\RequirePackage{fix-cm}
\documentclass[pdftex,onecolumn,epjc3,final]{svjour3}

\RequirePackage{enumitem}
\RequirePackage{graphicx}
\RequirePackage[T1]{fontenc}
\RequirePackage{subfig}
\RequirePackage{mathptmx}
\RequirePackage{helvet}
\RequirePackage{color}
\RequirePackage{url}
\RequirePackage{xspace}
\RequirePackage{booktabs}
\RequirePackage{amsmath}
\RequirePackage{algorithm}
\RequirePackage[noend]{algpseudocode}
\RequirePackage{pifont}
\RequirePackage{fancyvrb}
\RequirePackage{gensymb}

\DefineVerbatimEnvironment%
            {verbatimprog}%
                {Verbatim}%
  {fontsize=\footnotesize}%

\journalname{Eur. Phys. J. C}

\begin{document}
	\title{The Pandora multi-algorithm approach to automated pattern recognition of cosmic-ray muon and neutrino events in the MicroBooNE detector} 


	\author{R.~Acciarri\thanksref{g}
	\and
	C.~Adams\thanksref{bb,harvard}
	\and
	R.~An\thanksref{h}
	\and
	J.~Anthony\thanksref{c}
	\and
	J.~Asaadi\thanksref{y}
	\and
	M.~Auger\thanksref{a}
	\and
	L.~Bagby\thanksref{g}
	\and
	S.~Balasubramanian\thanksref{bb}
	\and
	B.~Baller\thanksref{g}
	\and
	C.~Barnes\thanksref{n}
% ...
	}

	\institute{Universit{\"a}t Bern, Bern CH-3012, Switzerland\label{a}
% ...
\and
	University of Cambridge, Cambridge CB3 0HE, United Kingdom\label{c}
% ...
\and
	Fermi National Accelerator Laboratory (FNAL), Batavia, IL 60510, USA\label{g}
	\and
        Harvard University, Cambridge, MA 02138, USA\label{harvard}
        \and
	Illinois Institute of Technology (IIT), Chicago, IL 60616, USA\label{h}
% ...
\and
	University of Michigan, Ann Arbor, MI, 48109, USA\label{n}
% ...
\and
	University of Texas, Arlington, TX, 76019, USA\label{y}
% ...
\and
	Yale University, New Haven, CT, 06520, USA\label{bb}
	}

    \maketitle

    \begin{abstract}
The development and operation of Liquid-Argon Time-Projection Chambers for neutrino physics has created a need for new approaches to pattern recognition in order to fully exploit the imaging capabilities offered by this technology. Whereas the human brain can excel at identifying features in the recorded events, it is a significant challenge to develop an automated, algorithmic solution. The Pandora Software Development Kit provides functionality to aid the design and implementation of pattern-recognition algorithms. It promotes the use of a multi-algorithm approach to pattern recognition, in which individual algorithms each address a specific task in a particular topology. Many tens of algorithms then carefully build up a picture of the event and, together, provide a robust automated pattern-recognition solution. This paper describes details of the chain of over one hundred Pandora algorithms and tools used to reconstruct cosmic-ray muon and neutrino events in the MicroBooNE detector. Metrics that assess the current pattern-recognition performance are presented for simulated MicroBooNE events, using a selection of final-state event topologies.
    \end{abstract}


\section{Introduction}
\label{sec::Introduction}
The MicroBooNE detector\,\cite{bib::Detector} has been operating in the Booster Neutrino Beam (BNB) at Fermilab since October 2015. It is an important step towards the realisation of large-scale Liquid-Argon Time-Projection Chambers (LArTPCs) for future long and short baseline neutrino experiments. It provides an opportunity to hone the development of automated algorithms for LArTPC event reconstruction and to test the algorithms using real data. The MicroBooNE physics goals are to perform measurements of neutrino cross-sections on argon in the 1\,GeV range and to probe the excess of low energy events observed in the MiniBooNE search for short-baseline neutrino oscillations\,\cite{bib::MiniBooNE}. One of the main reconstruction tools used by MicroBooNE is the Pandora Software Development Kit (SDK)\,\cite{bib::PandoraSDK}. This paper presents details of the Pandora pattern-recognition algorithms and characterises their current performance, using a selection of final-state event topologies.

The Pandora SDK was created to address the problem of identifying energy deposits from individual particles in fine-granularity detectors. It has been used to design and optimise the detectors proposed for use at future $e^{+}e^{-}$ collider experiments\,\cite{bib::ILC}\cite{bib::CLIC}. It specifically promotes the idea of a multi-algorithm approach to solving pattern-recognition problems. In this approach, the input building blocks (hits) describing the pattern-recognition problem are considered by large numbers of decoupled algorithms. Each algorithm targets a specific event topology and controls operations such as collecting hits together in clusters, merging or splitting clusters, or collecting clusters in order to build a representation of particles in the detector. Each algorithm aims only to perform pattern-recognition operations when it is deemed ``safe'', deferring complex topologies to later algorithms. In this way, the algorithms can remain decoupled and there is little inter-algorithm tension. Some algorithms are complex, whilst others are simple. The algorithms gradually build up a picture of the underlying events and collectively provide a robust reconstruction.

The Pandora algorithms are designed to be generic, to allow use by multiple experiments, but this paper describes their specific application to MicroBooNE. Section \ref{sec::Detector} describes the MicroBooNE detector and Section \ref{sec::Inputs} discusses the inputs to Pandora and the output pattern-recognition information. Section \ref{sec::Overview} describes the Pandora pattern-recognition algorithms and Section \ref{sec::Metrics} introduces metrics for assessing the pattern-recognition performance. Section \ref{sec::Performance} presents results for simulated BNB interactions, with a selection of exclusive final states, and Section \ref{sec::CR} presents the results for BNB interactions with overlaid cosmic-ray muon backgrounds. The principal focus of this paper is the reconstruction of $\nu_{\mu}$ charged-current (CC) and neutral-current (NC) interactions, supporting MicroBooNE's programme of cross-section measurements.


\section{The MicroBooNE detector}
\label{sec::Detector}
The MicroBooNE detector and its associated systems are described in detail in \cite{bib::Detector}. In this Section, just the key features pertaining to the pattern recognition are introduced.

The detector is a single-phase LArTPC with a rectangular active volume of the following dimensions: 2.6\,m (horizontal), 2.3\,m (vertical) and 10.4\,m (longitudinal)\footnote{MicroBooNE uses a right-handed Cartesian coordinate system where $x$ ranges from 0.0\,m at the innermost anode plane to +2.6\,m at the cathode, $y$ ranges from -1.15\,m at the bottom of the active volume to +1.15\,m at the top of the active volume, and $z$ ranges from 0.0\,m at the upstream end of the active volume to +10.4\,m at the downstream end.}. The TPC has an active mass of 85\,tonnes of argon and is immersed within a cryostat of 170\,tonne capacity. Charged particles passing through the liquid argon leave trails of ionisation electrons, which are transported through the highly-purified argon under the influence of a uniform electric field, here of strength 273\,V\,/\,cm. The anode and cathode planes are parallel to the BNB direction. At the anode plane, there are three planes of wires, with a 3\,mm pitch, held at specific bias voltages. The ionisation electrons induce signals on the first two planes of wires, which are oriented at $\pm60\degree$ to the vertical (here labelled the $u$ and $v$ planes). The electrons induce a signal on the third plane before being collected there. The wires in this third plane (here labelled $w$) are oriented vertically. Three separate two-dimensional (2D) images are formed, using the known positions of the wires and the recorded drift times; the times at which the ionisation signals are recorded, relative to the event trigger time. The waveforms observed for each wire are examined, detector effects are removed and a hit-finding algorithm searches for local maxima and minima. A Gaussian distribution is fitted to each peak and hit objects are created, forming the input to the pattern recognition.

A particular challenge for the reconstruction of neutrino interactions in MicroBooNE is the high level of cosmic-ray muon background inherent in a surface-based LArTPC, which has a long exposure per event due to lengthy drift times (up to a few milliseconds). Further complications in any operating LArTPC such as MicroBooNE include the presence of partially-correlated noise, unresponsive readout channels, and residual inefficiencies or imperfections in the input hits, which may affect the fine detail of the pattern recognition. The characterisation and mitigation of the observed noise in the MicroBooNE detector is discussed in \cite{bib::Noise}.


\section{Inputs and outputs}
\label{sec::Inputs}
The Pandora reconstruction is integrated into the LArSoft\,\cite{bib::LArSoft} framework via the LArPandora translation module. This module is required to translate the input pattern-recognition building blocks from the LArSoft Event Data Model (EDM) to the Pandora EDM, initiate and apply the Pandora algorithms, and then translate the output Pandora pattern-recognition results back to the LArSoft EDM. Translation modules are ultimately responsible for controlling a reconstruction using Pandora and are described in detail in \cite{bib::PandoraSDK}.

In its initialisation step, the translation module uses Application Programming Interfaces (APIs) to:
\begin{itemize}
    \item Create a Pandora instance. For MicroBooNE, all input hits are given to this single instance and the event is reconstructed using a single thread.
    \item Provide simple detector geometry information, including wire pitches and wire angles to the vertical. These details are used by the plugin that provides coordinate transformations between readout planes.
    \item Register factories, which create the instances of the algorithms, algorithm tools and plugins used in the multi-algorithm reconstruction.
    \item Provide a user-defined PandoraSettings configuration file. This specifies which algorithms will run over each event, in which order, and provides their configuration details.
\end{itemize}

On a per-event basis, the translation module uses Pandora APIs to:
\begin{itemize}
    \item Translate input hits from the LArSoft EDM to Pandora hits, which provide a self-describing input to the Pandora pattern-recognition algorithms.
    \item Translate records of the true, generated particles in simulated data to create Pandora MCParticles. These are not used to influence the pattern recognition, but enable evaluation of performance metrics. MCParticles can have parent-daughter relationship hierarchies and links to the Pandora hits.
    \item Instruct the Pandora instance to process the event. The thread is passed to the Pandora instance and the algorithms are applied to the input hits, as specified in the PandoraSettings configuration file.
    \item Extract the list of reconstructed particles, which represent the pattern-recognition solution. These particles are translated to the LArSoft EDM and written to the event record.
    \item Reset the Pandora instance, so that it is ready to receive new input objects for the next event.
\end{itemize}

Each input hit represents a signal detected on a single wire at a definite drift time. The Pandora hits are placed in the $x-$wire plane, with $x$ representing the drift time coordinate, converted to a position, and the second coordinate representing the wire number, converted to a position. The Pandora hits have a width in the $x$ coordinate defined by the Gaussian distribution fitted by the hit-finding algorithm: hits extend across positions corresponding to drift times one standard deviation below and above the peak time. In the wire coordinate, the hits have extent equal to the wire pitch. The readout plane is specified for each Pandora hit, so three 2D images (the $u$, $v$ and $w$ ``views'') are provided of events within the active volume of the MicroBooNE detector. The $x$ coordinate is common to all three images and so can be exploited by the pattern-recognition algorithms to correlate features in the different images and perform three-dimensional (3D) reconstruction.

\begin{figure}[]
  \begin{center}
     \includegraphics[width=0.6\textwidth]{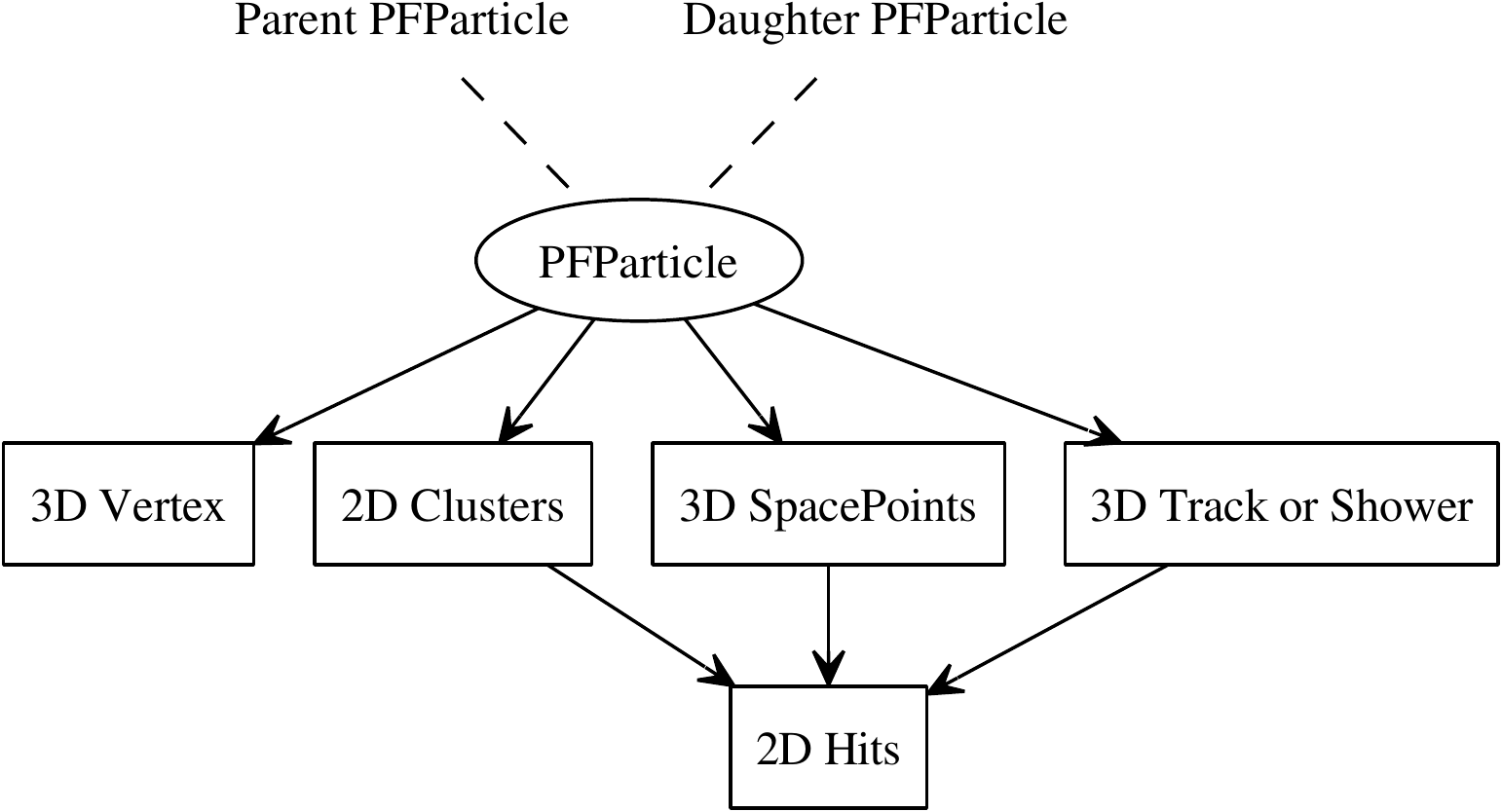}
     \caption{The Pandora output data products, as persisted in the LArSoft Event Data Model. Navigation along PFParticle hierarchies is achieved using the PFParticle interface, represented by dashed lines. Navigation from PFParticles to their associated objects is represented by solid arrows.\label{fig::LArSoftOutput}}
  \end{center}
\end{figure}

The pattern-recognition output is illustrated in Figure \ref{fig::LArSoftOutput}. The most important output is the list of reconstructed ``PFParticles'' (PF stands for Particle Flow). Each PFParticle corresponds to a distinct track or shower, and is associated with a list of 2D clusters. The 2D clusters group together the relevant hits from each readout plane. Each PFParticle is also associated with a set of reconstructed 3D positions (termed SpacePoints) and with a reconstructed vertex position, which defines its interaction point or first energy deposit. The PFParticles are placed in a hierarchy, which identifies parent-daughter relationships and describes the particle flow in the observed interactions. A neutrino PFParticle can be created as part of the hierarchy and can form the primary parent particle for a neutrino interaction. The type of each particle is not currently reconstructed, but they are instead identified as track-like or shower-like. Track and shower objects carry additional metadata, such as position and momentum information for tracks or principal-axis information for showers.

\section{Algorithm overview}
\label{sec::Overview}
Two Pandora multi-algorithm reconstruction paths have been created for use in the analysis of MicroBooNE data. One option, PandoraCosmic, is optimised for the reconstruction of cosmic-ray muons and their daughter delta rays. The second option, PandoraNu, is optimised for the reconstruction of neutrino interactions. Many algorithms are shared between the PandoraCosmic and PandoraNu reconstruction paths, but the overall algorithm selection results in the following key features:

\begin{itemize}
    \item \textbf{PandoraCosmic:} This reconstruction is more strongly track-oriented, producing primary particles that represent cosmic-ray muons. Showers are assumed to be delta rays and are added as daughter particles of the most appropriate cosmic-ray muon. The reconstructed vertex/start-point for the cosmic-ray muon is the high-$y$ coordinate of the muon track.
    \item \textbf{PandoraNu:} This reconstruction identifies a neutrino interaction vertex and uses it to aid the reconstruction of all particles emerging from the vertex position. There is careful treatment to reconstruct tracks and showers. A parent neutrino particle is created and the reconstructed visible particles are added as daughters of the neutrino.
\end{itemize}

\begin{figure}[]
  \begin{center}
     \includegraphics[width=0.9\textwidth]{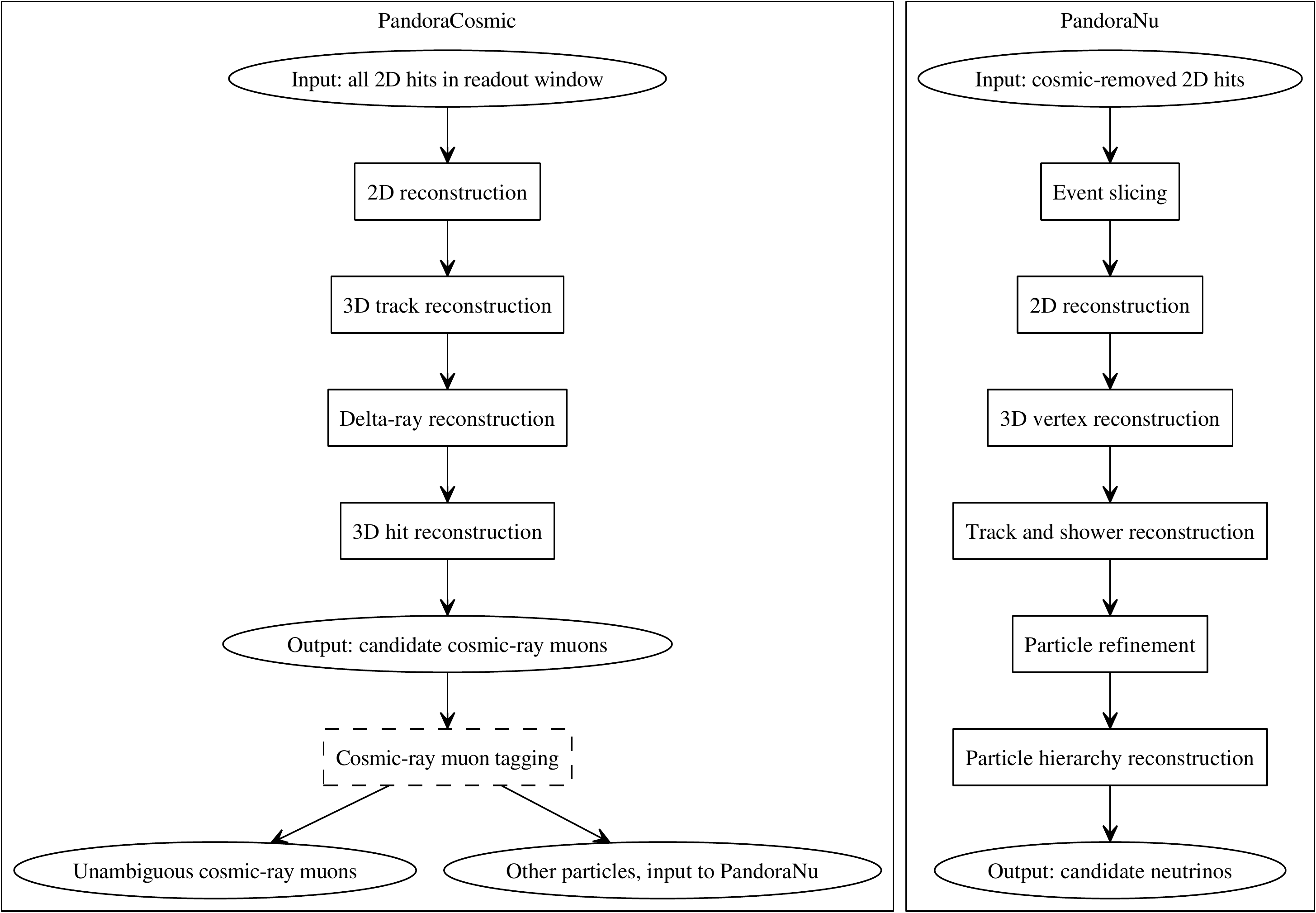}
     \caption{A simple representation of the two multi-algorithm reconstruction paths created for use in MicroBooNE. Particles formed by the PandoraCosmic reconstruction are examined by a cosmic-ray tagging module, external to Pandora. Hits associated with unambiguous cosmic-ray muons are flagged and a new cosmic-removed hit collection provides the input to the PandoraNu reconstruction.\label{fig::AlgorithmChain}}
  \end{center}
\end{figure}

The PandoraCosmic and PandoraNu reconstructions are applied to the MicroBooNE data in two passes. The PandoraCosmic reconstruction is first used to process all hits identified during a specified readout window and to provide a list of candidate cosmic-ray particles. This list of particles is then examined by a cosmic-ray tagging module, implemented within LArSoft, which identifies unambiguous cosmic-ray muons, based on their start and end positions and associated hits. Hits associated with particles flagged as cosmic-ray muons are removed from the input hit collection and a new cosmic-removed hit collection is created. This second hit collection provides the input to the PandoraNu reconstruction, which outputs a list of candidate neutrinos. The overall chain of Pandora algorithms is illustrated in Figure \ref{fig::AlgorithmChain}.

\subsection{Cosmic-ray muon reconstruction}
\label{sec::Cosmic}

The PandoraCosmic reconstruction proceeds in four main stages, each of which uses multiple algorithms and algorithm tools, as described in this Section.

\subsubsection{Two-dimensional reconstruction}
\label{sec::TwoDRec}
The first step is to separate the input hits into three separate lists, corresponding to the three readout planes ($u$, $v$ and $w$). This operation is performed by the EventPreparation algorithm\footnote{For expediency, algorithms are referred to by their self-describing names throughout this paper.}. For each wire plane, the TrackClusterCreation algorithm then produces a list of 2D clusters that represent continuous, unambiguous lines of hits. Separate clusters are created for each structure in the input hit image, with clusters starting/stopping at each branch feature or any time there is any bifurcation or ambiguity. This initial clustering provides clusters of high \textit{purity}, representing energy deposits from exactly one true particle, even if this means that the clusters are initially of low \textit{completeness}, containing only a small fraction of the total hits associated with the true particle. The clusters are then examined by a series of topological algorithms.

Cluster-merging algorithms identify associations between multiple 2D clusters and look to grow the clusters to improve completeness, without compromising purity. The typical approach used by cluster-merging algorithms is to identify pairs of clusters that are either in close proximity, or which point towards each other. The challenge for the algorithms is to make cluster-merging decisions in the context of the entire event, rather than just by considering individual pairs of clusters in isolation. The ClusterAssociation and ClusterExtension algorithms are reusable base classes, which allow different definitions of cluster association to be provided. They provide a common implementation that evaluates association for all cluster combinations and identifies chains of associated clusters, allowing decisions to be made based upon an understanding of the overall event topology. Algorithms inheriting from these base classes are used to extend 2D clusters in both the longitudinal (beam) and transverse directions. They are also used to merge clusters across registered gaps in the detector instrumentation.

\begin{figure}[]
  \begin{center}
     \subfloat[][]{\includegraphics[height=0.4\textwidth]{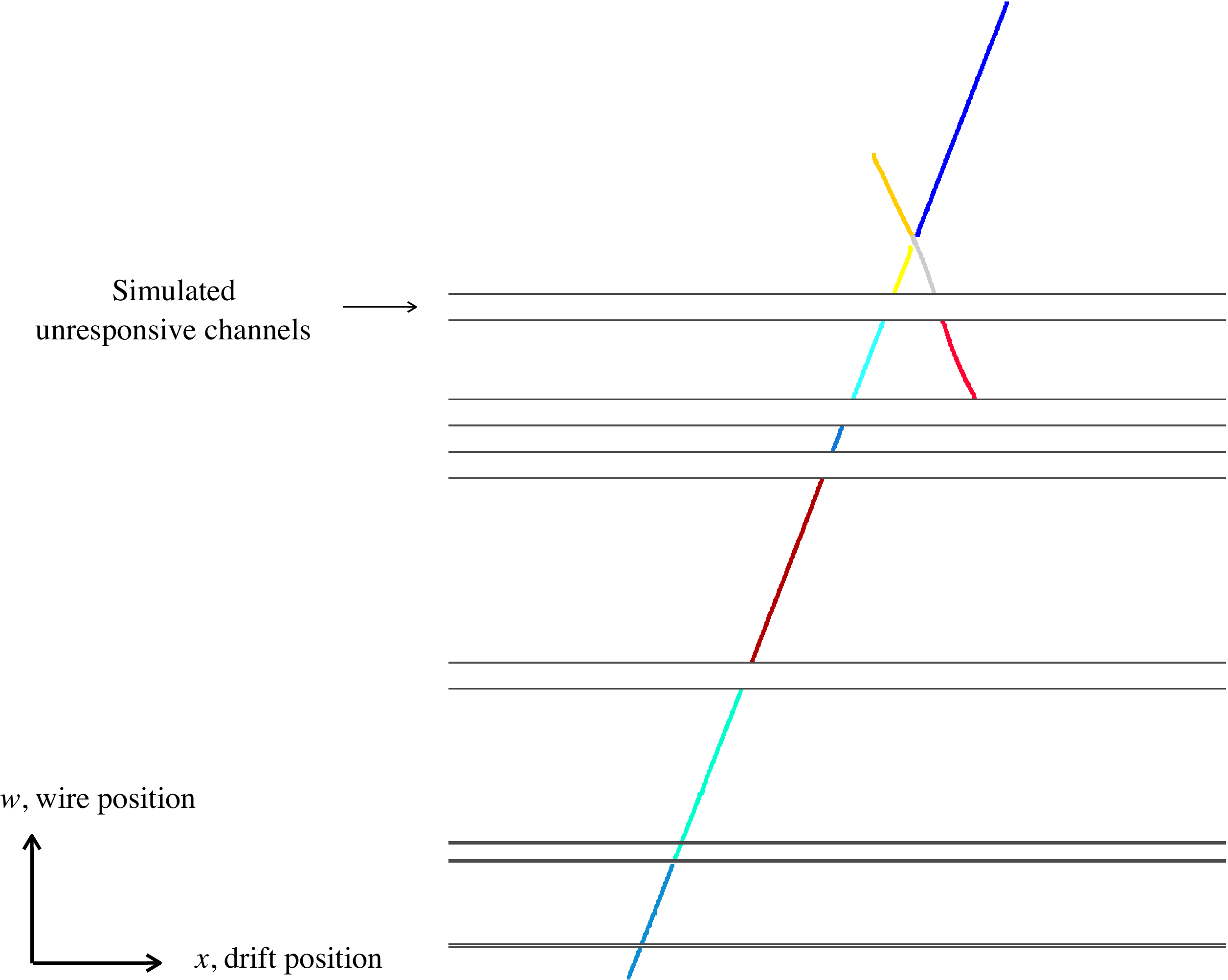}\label{fig::2DRecoPre}}\hspace{1.2cm}
     \subfloat[][]{\includegraphics[height=0.4\textwidth]{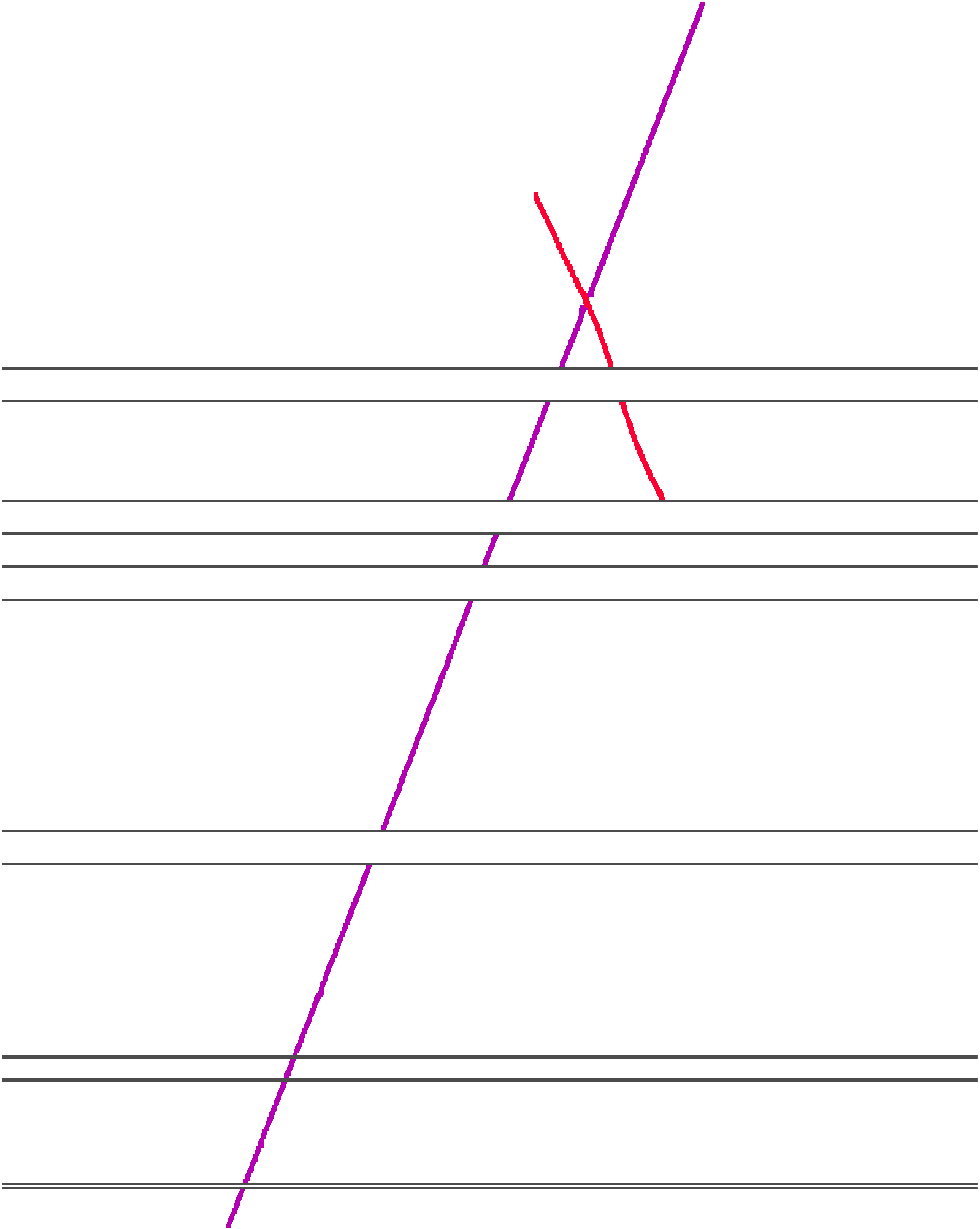}\label{fig::2DRecoPost}}
     \caption{(a) Clusters produced by the TrackClusterCreation algorithm for two crossing cosmic-ray muons in the MicroBooNE detector simulation. Separate clusters are formed where the tracks cross and on either side of unresponsive regions of the detector. (b) The refined clusters formed by the series of topological algorithms, now with one cluster for each cosmic-ray muon track. Each coloured track corresponds to a separately reconstructed cluster of hits and the gaps indicate possible unresponsive portions of the TPC.}
  \end{center}
\end{figure}

To improve purity, cluster-splitting algorithms refine the hit selection by breaking single clusters into two parts if topological features indicate the inclusion of hits from multiple particles. Clusters are split if there is a significant discontinuity in the cluster direction, or if multiple clusters intersect or point towards a common position along the length of an existing cluster. Figure \ref{fig::2DRecoPre} shows initial clusters formed for simulated cosmic-ray muons in MicroBooNE. These clusters form the input to the series of topological algorithms, in which multiple cluster-merging and cluster-splitting procedures are interspersed. Processing by these algorithms results in the refined clusters shown in Figure \ref{fig::2DRecoPost}. The final 2D clusters provide the input to the process used to ``match'' features reconstructed in multiple readout planes, and to construct particles.

\subsubsection{Three-dimensional track reconstruction}
\label{sec::ThreeDTracks}
The aim of the 3D track reconstruction is to collect the 2D clusters from the three readout planes that represent individual, track-like particles. The clusters can be assigned as daughter objects of new Pandora particles. The challenge for the algorithms is to identify consistent groupings of clusters from the different views. The 3D track reconstruction is primarily performed by the ThreeDTransverseTracks algorithm. This algorithm considers the suitability of all combinations of clusters from the three readout planes and stores the results in a three-dimensional array, hereafter loosely referred to as a rank-three tensor. The three tensor indices are the clusters in the $u$, $v$ and $w$ views and, for each combination of clusters, a detailed TransverseOverlapResult is stored. The information in the tensor is examined in order to identify cluster-matching ambiguities. If ambiguities are discovered, the information can be used to motivate changes to the 2D reconstruction that would ensure that only unambiguous combinations of clusters emerge. This procedure is often loosely referred to as ``diagonalising'' the tensor.

To populate the TransverseOverlapResult for three clusters (one from each of the $u$, $v$ and $w$ views), a number of sampling points are defined in the $x$ (drift time) region common to all three clusters. Sliding linear fits to each cluster are then constructed. These record the results of a series of linear fits, each using only hits from a local region of the cluster. For a sampling point at a given $x$ coordinate, the sliding-fit position can be extracted for a pair of clusters, e.g. in the $u$ and $v$ views. These positions, together with the coordinate transformation plugin, can be used to predict the position of the third cluster, e.g. in the $w$ view, at the same $x$ coordinate. This prediction can be compared to the sliding-fit position for the third cluster and, by considering all combinations ($u,v\rightarrow w$; $v,w\rightarrow u$; $u,w\rightarrow v$), a quantity approximating to a $\chi^{2}$ can be calculated. The $\chi^{2}$-like value, together with the common $x$-overlap span, the number of sampling points and the number of consistent sampling points, is stored in the TransverseOverlapResult in the tensor.

Crucially, the results stored in the tensor do not just provide isolated information about the consistency of groups of three clusters. The results also provide detailed information about the connections between multiple clusters and their matching ambiguities. For instance, starting from a given cluster, it is possible to navigate between multiple tensor elements, each of which indicate good cluster matching but share, or ``re-use'', one or two clusters. In this way, a complete set of connected clusters can be extracted. If this set contains more than one cluster from any single view, an ambiguity is identified. The exact form of the ambiguity can often indicate the mechanism by which it may be addressed and can identify the specific clusters that require modication. This detailed information about cluster connections is queried by a series of algorithm tools, which can create particles or modify the 2D pattern recognition. The algorithm tools have a specific ordering and, if any tool makes a change, the tensor is updated and the full list of tools runs again. The tensor is processed until no tool can perform any further operations.

\begin{figure}[!ht]
  \begin{center}
     \subfloat[][]{\includegraphics[height=0.44\textwidth]{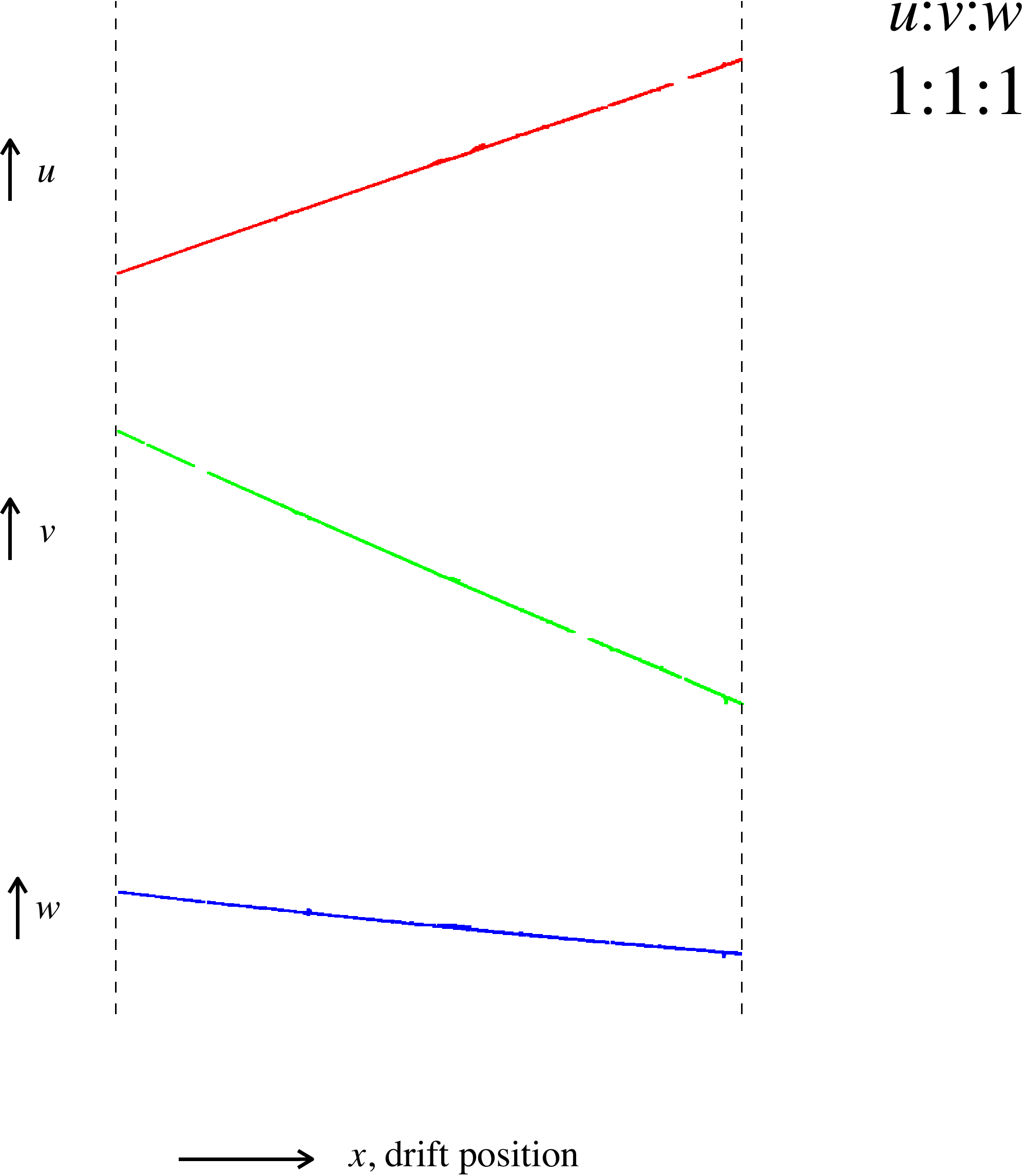}\label{fig::ClearTracksTool}}\hspace{2cm}
     \subfloat[][]{\includegraphics[height=0.44\textwidth]{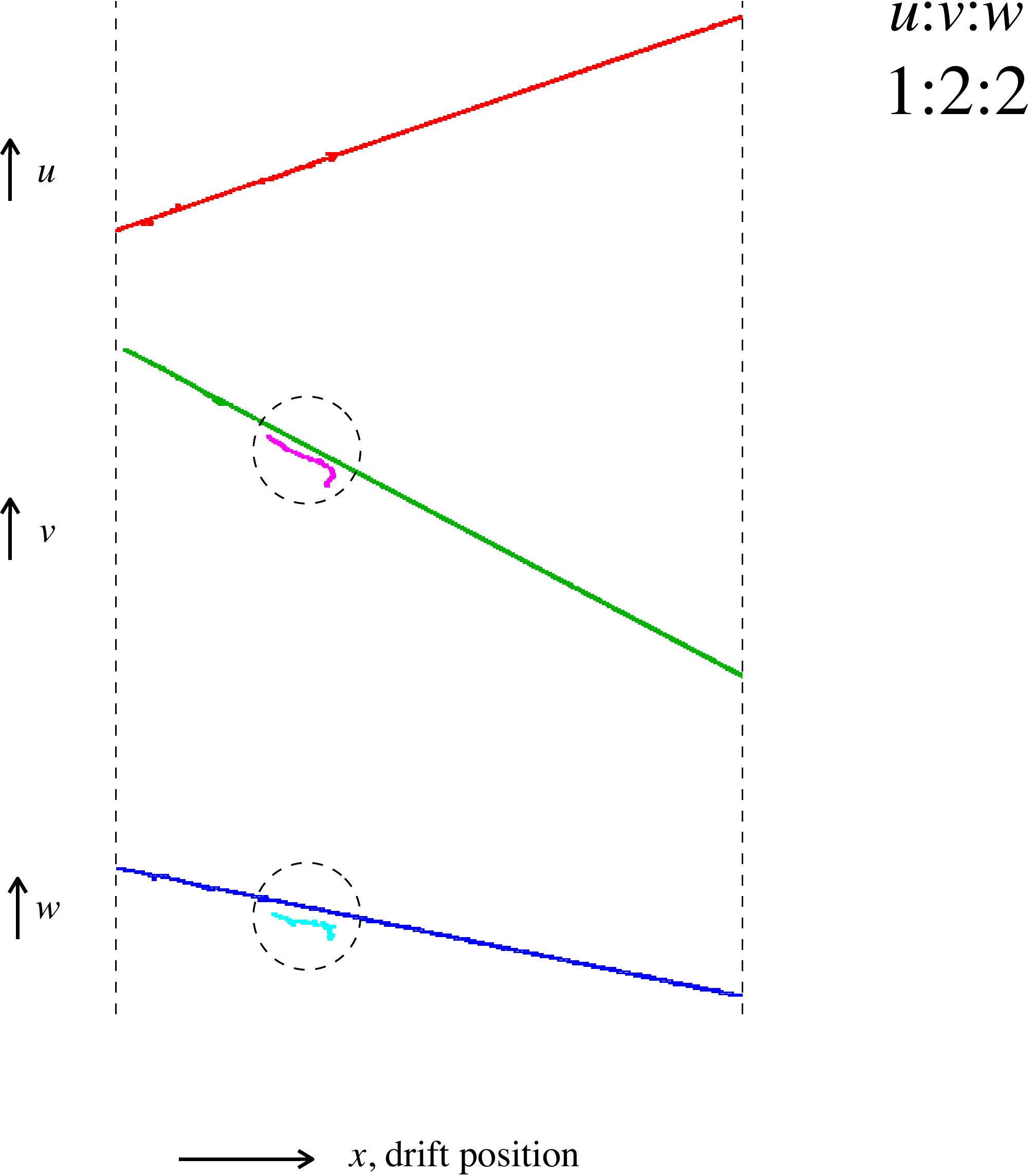}\label{fig::LongTracksTool}}\\
     \subfloat[][]{\includegraphics[height=0.44\textwidth]{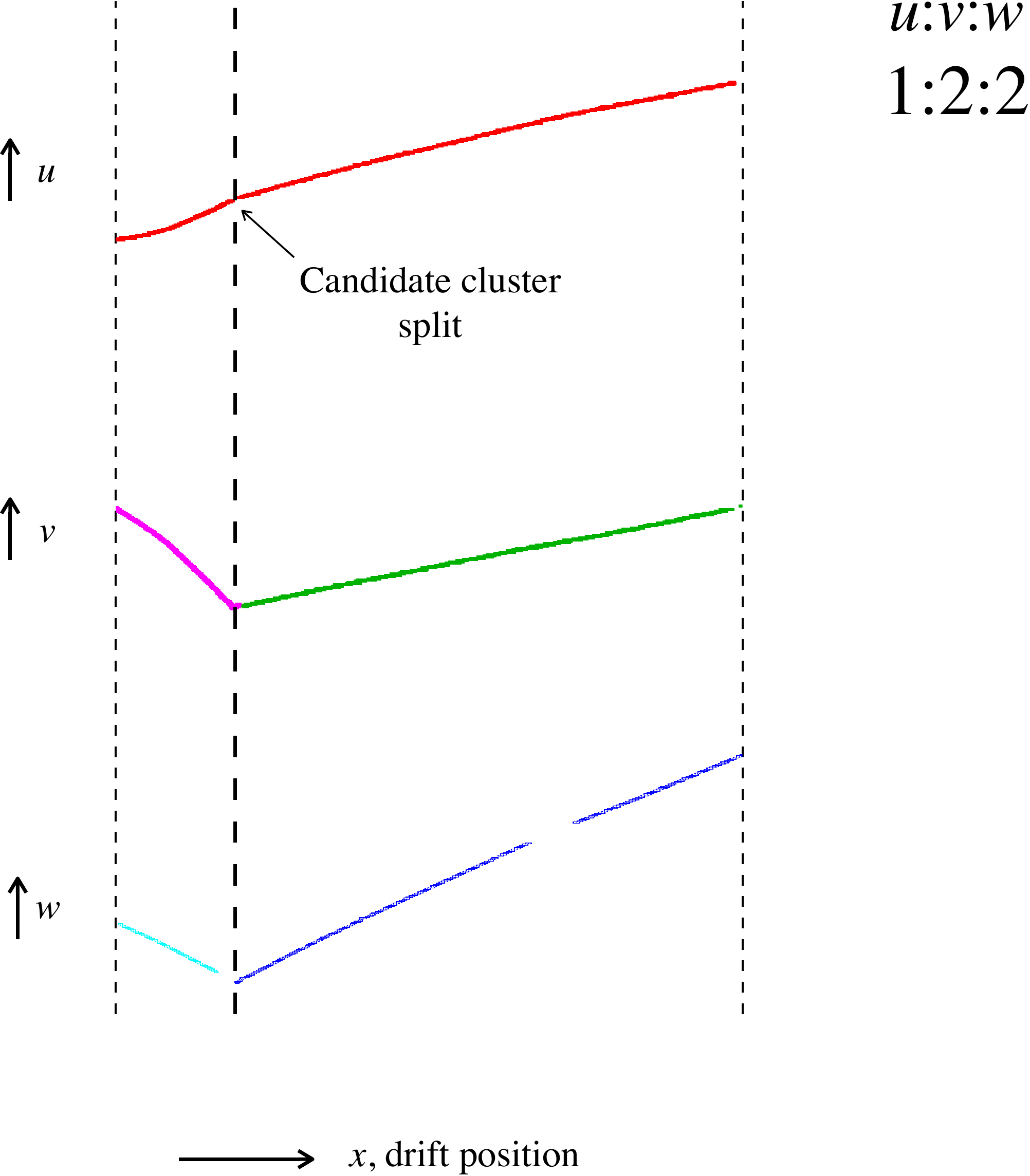}\label{fig::OvershootTracksTool}}\hspace{2cm}
     \subfloat[][]{\includegraphics[height=0.44\textwidth]{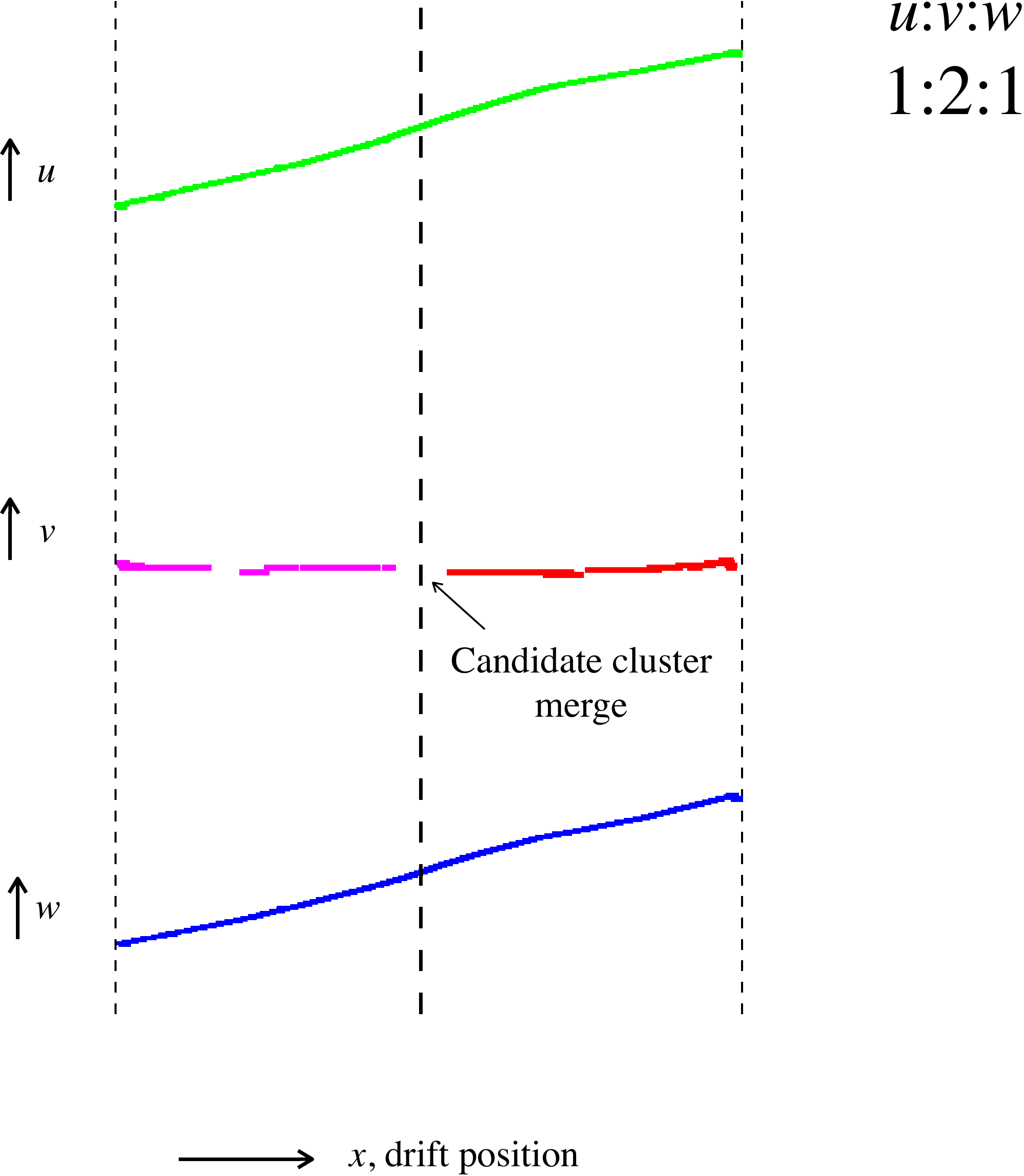}\label{fig::UndershootTracksTool}}
     \caption{Example topologies considered by the 3D track reconstruction, which aims to identify unambiguous groupings of 2D clusters, one from each readout plane. (a) An unambiguous grouping of clusters, with complete overlap in the common $x$ coordinate. (b) The presence of two small delta-ray clusters (circled), near the muon tracks, means that the cluster matching is ambiguous, but the most appropriate grouping of 2D clusters can be identified. (c) An overshoot in the clustering in the $u$ view leads to ambiguous cluster matching, which can be resolved by splitting the $u$ cluster at the indicated position. (d) An undershoot in the clustering in the $v$ view leads to ambiguous cluster matching, which can be resolved by merging the two $v$ cluster fragments.}
  \end{center}
\end{figure}

The algorithm tools, in the order that they are run, are:
\begin{itemize}
\item ClearTracks tool, which looks to create particles from unambiguous groupings of three clusters. It examines the tensor to find regions where only three clusters are connected, one from each of the $u$, $v$ and $w$ views, as illustrated in Figure \ref{fig::ClearTracksTool}. Quality cuts are  applied to the TransverseOverlapResult and, if passed (the common $x$-overlap must be >90\% of the $x$-extent for all clusters at this stage), a new particle is created.

\item LongTracks tool, which aims to resolve obvious ambiguities. In the example in Figure \ref{fig::LongTracksTool}, the presence of small delta-ray clusters near long muon tracks means that clusters are matched in multiple configurations and the tensor is not diagonal. One of the combinations of clusters is, however, better than the others (with larger $x$-overlap and a larger number of consistent sampling points) and is used to create a particle. The common $x$-overlap threshold remains >90\% of the $x$-extent for all clusters.

\item OvershootTracks tool, which addresses cluster-matching ambiguities of the form 1:2:2 (one cluster in the $u$ view, matched to two clusters in the $v$ view and two clusters in the $w$ view). In the example in Figure \ref{fig::OvershootTracksTool}, the pairs of clusters in the $v$ and $w$ views connect at a common $x$ coordinate, but there is a single, common cluster in the $u$ view, which spans the full $x$-extent. The tool considers all clusters and decides whether they represent a kinked topology in 3D. If a 3D kink is identified, the $u$ cluster can be split at the relevant position and two new $u$ clusters fed back into the tensor. The initial ClearTracks tool will then be able to identify two unambiguous groupings of three clusters and create two particles.

\item UndershootTracks tool, which examines the tensor to find cluster-matching ambiguities of the form 1:2:1. In the example in Figure \ref{fig::UndershootTracksTool}, two clusters in the $v$ view are matched to common clusters in the $u$ and $w$ views, leading to two conflicting TransverseOverlapResults in the tensor. The tool examines all the clusters to assess whether they represent a kinked topology in 3D. If a 3D kink is not found, the two $v$ clusters can be merged and a single $v$ cluster fed-back into the tensor, removing the ambiguity.

\item MissingTracks tool, which understands that particle features may be obscured in one view, with a single cluster representing multiple overlapping particles. If this tool identifies appropriate cluster overlap, using the cluster-relationship information available from the tensor, the tool can create two-cluster particles.

\item TrackSplitting tool, which looks to split clusters if the matching between views is unambiguous, but there is a significant discrepancy between the cluster $x$-extents and evidence of gaps in a cluster.

\item MissingTrackSegment tool, which looks to add missing hits to the end of a cluster if the matching between views is unambiguous, but there is a significant discrepancy between the cluster $x$-extents.

\item LongTracks tool, which is used again with the common $x$-overlap threshold reduced to >75\% of the $x$-extent for all clusters.
\end{itemize}

In addition to the ThreeDTransverseTracks algorithm, there are other algorithms that form a cluster-association tensor, and query it using algorithm tools. These algorithms target different topologies and store different information in the tensor. The ThreeDLongitudinalTracks algorithm examines the case where the $x$-extent of a cluster grouping is small. In this case, there are too many ambiguities when trying to sample the clusters at fixed $x$ coordinates. The ThreeDTrackFragments algorithm is optimised to look for situations where there are single, clean clusters in two views, associated with multiple fragment clusters in a third view.


\subsubsection{Delta-ray reconstruction}
Following 3D track reconstruction, the PandoraCosmic reconstruction dissolves any 2D clusters that have not been included in a reconstructed particle. The assumption is that these clusters likely represent fragments of delta-ray showers. The relevant hits are reclustered using the SimpleClusterCreation algorithm, which is a proximity-based clustering algorithm. A number of topological algorithms, which re-use implementation from the earlier 2D reconstruction,  refine the clusters to provide a more complete delta-ray reconstruction. The DeltaRayMatching algorithm subsequently matches the delta-ray clusters between views, creates new shower-like particles and identifies the appropriate parent cosmic-ray particles. The cluster matching is simple and assesses the $x$-overlap between clusters in multiple views. Parent cosmic-ray particles are identified via simple comparison of inter-cluster distances.


\subsubsection{Three-dimensional hit reconstruction}
\label{sec::ThreeDHits}
At this point, the assignment of hits to particles is complete and the particles contain 2D clusters from one, two or usually all three readout planes. For each input (2D) hit in a particle, a new 3D hit is created. The mechanics differ depending upon the cluster topology, with separate approaches for: hits on transverse tracks (significant extent in $x$ coordinate) with clusters in all views; hits on longitudinal tracks (small extent in $x$ coordinate) with clusters in all views; hits on tracks that are multi-valued at specific $x$ coordinates; hits on tracks with clusters in only two views; and hits in shower-like particles. Only two such approaches are described here:
\begin{itemize}
\item For transverse tracks with clusters in all three views, a 2D hit in one view, e.g. $u$, is considered and sliding linear fit positions are evaluated for the two other clusters, e.g. $v$ and $w$, at the same $x$ coordinate. An analytic $\chi^{2}$ minimisation is used to extract the optimal $y$ and $z$ coordinates at the given $x$ coordinate. It is also possible to run in a mode whereby the chosen $y$ and $z$ coordinates ensure that the 3D hit can be projected precisely onto the specific wire associated with the input 2D hit.

\item For a 2D hit in a shower-like particle (a delta ray, e.g. in the $u$ view), all combinations of hits (e.g. in the $v$ and $w$ views) located in a narrow region around the hit $x$ coordinate are considered. For a given combination of hit $u$, $v$ and $w$ values, the most appropriate $y$ and $z$ coordinates can be calculated. The position yielding the best $\chi^{2}$ value is identified and a $\chi^{2}$ cut is applied to help ensure that only satisfactory positions emerge.
\end{itemize}

After 3D hit creation, the PandoraCosmic reconstruction is completed by the placement of vertices/start-positions at the high-$y$ coordinates of the cosmic-ray muon particles. Vertices are also reconstructed for delta-ray particles and are placed at the 3D point of closest approach between the parent cosmic-ray muon and daughter delta ray.


\subsection{Neutrino reconstruction}
\label{sec::Neutrino}

A key requirement for the PandoraNu reconstruction path is that it must be able to deal with the presence of any cosmic-ray muon remnants that remain in the input, cosmic-removed hit collection. The approach is to begin by running the 2D reconstruction, 3D track reconstruction and 3D hit reconstruction algorithms described in Section \ref{sec::Cosmic}. The 3D hits are then divided into \textit{slices} (separate lists of hits), using proximity and direction-based metrics. The intent is to isolate neutrino interactions and cosmic-ray muon remnants in individual slices. The original 2D hits associated with each slice are then used as an input to the dedicated neutrino reconstruction described in this Section. Each slice (including those containing cosmic-ray muon remnants) is processed in isolation and results in one candidate reconstructed neutrino.

The dedicated neutrino reconstruction begins with a track-oriented clustering algorithm and series of topological algorithms, as described in Section \ref{sec::TwoDRec}. The lists of 2D clusters for the different readout planes are then used to identify the neutrino interaction vertex. The vertex reconstruction is a key difference between the PandoraCosmic and PandoraNu reconstruction paths, and the 3D vertex position plays an important role throughout the subsequent algorithms. Correct identification of the neutrino interaction vertex helps algorithms to identify individual primary particles and to ensure that they each result in separate reconstructed particles.


\subsubsection{Three-dimensional vertex reconstruction}
\begin{figure}[]
  \begin{center}
     \includegraphics[height=0.4\textwidth]{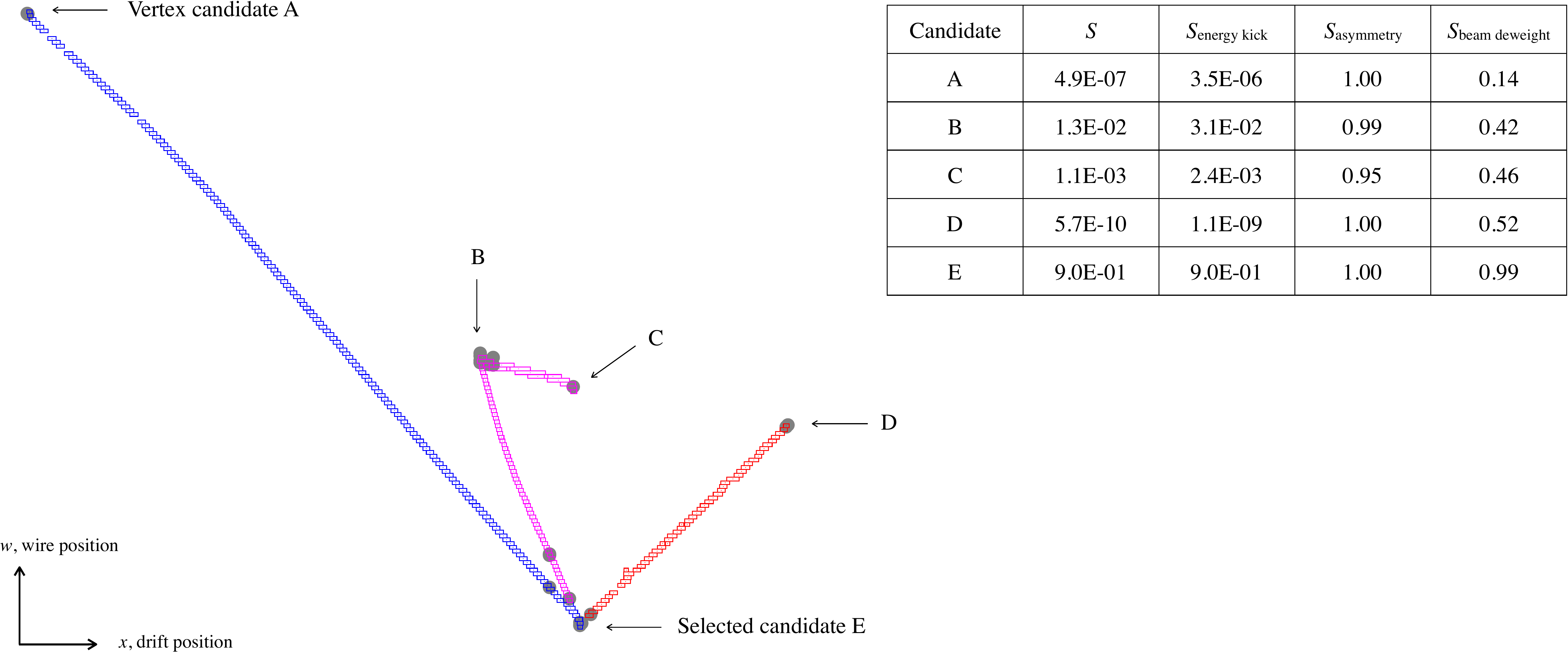}
     \caption{The positions of 3D neutrino interaction vertex candidates, as projected into the $w$ view. A comprehensive list of candidates is produced for each event, identifying all the key features in the event topology. To select the neutrino interaction vertex, each candidate is assigned a score. The scores are indicated for a number of candidates and a breakdown of each score into its component parts is provided.\label{fig::Vertexing}}
  \end{center}
\end{figure}

Reconstruction of the neutrino interaction vertex begins with creation of a list of possible vertex positions. The CandidateVertexCreation algorithm compares pairs of 2D clusters, ensuring that the two clusters are from different readout planes and have some overlap in the common $x$ coordinate. The endpoints of the two clusters are then compared. For instance, the low-$x$ endpoint of one cluster can be identified. The same $x$ coordinate will not necessarily correspond to an endpoint of the second cluster, but the position of the second cluster at this $x$ coordinate can be evaluated, using a sliding linear fit and allowing some extrapolation of cluster positions. The two cluster positions, from two views, are sufficient to provide a candidate 3D position. Using all of the cluster endpoints allows four candidate vertices to be created for each cluster pairing. Figure \ref{fig::Vertexing} shows the candidate vertex positions created for a typical simulated CC $\nu_{\mu}$ event in MicroBooNE.

Having identified an extensive list of candidate vertex positions, it is necessary to select one as the most likely neutrino interaction vertex. There are a large number of candidates, so each is required to pass a simple quality cut before being put forward for assessment: candidates are required to sit on or near a hit, or in a registered detector gap, in all three views. The EnergyKickVertexSelection algorithm then assigns a score to each remaining candidate  and the candidate with the highest score is selected.

There are three components to the score:

\begin{equation}
S = S_\text{energy kick} \cdot S_\text{asymmetry} \cdot S_\text{beam deweight} \\
\label{equ::TotalVertexScore}
\end{equation}

\begin{itemize}
\item \textbf{Energy kick score:} Each 3D vertex candidate is projected into the $u$, $v$ and $w$ views. A parameter, $E^{T'}_{ij}$, is then calculated to assess whether the candidate is consistent with observed cluster $j$ in view $i$. This parameter is closely related to the transverse energy, but has additional degrees of freedom that introduce a dependence on the displacement between the cluster and vertex projection. Candidates are suppressed if the sum of $E^{T'}_{ij}$, over all clusters, is large. This reflects the fact that primary particles produced in the interaction should point back towards the true interaction vertex, whilst downstream secondary particles may not, but are expected to be less energetic:

\begin{equation}
S_\text{energy kick} =  \exp\left\{-\sum_{\text{view } i}\,\,\,\,\sum_{\text{cluster } j}{\frac{E^{T'}_{ij}}{\epsilon}}\right\}
\label{equ::EnergyKick1}
\end{equation}
\begin{equation}
{E^{T'}_{ij} = \frac{E_j\times(x_{ij} + \delta_x)}{(d_{ij} + \delta_d)}}
\label{equ::EnergyKick2}
\end{equation}
\noindent where $x_{ij}$ is the transverse impact parameter between the vertex and a linear fit to cluster $j$ in view $i$, $d_{ij}$ is the closest distance between the vertex and cluster and $E_{j}$ is the cluster energy, taken as the integral of the hit waveforms converted to a modified GeV scale. The parameters $\epsilon$, $\delta_d$ and $\delta_x$ are tunable constants: $\epsilon$ determines the relative importance of the energy kick score, $\delta_d$ protects against cases where $d_{ij}$ is zero and controls weighting as a function of $d_{ij}$, and $\delta_x$ controls weighting as a function of $x_{ij}$.\\

\item \textbf{Asymmetry score:} This suppresses candidates incorrectly placed along single, straight clusters, by counting the numbers of hits deemed upstream and downstream of the candidate position. For the true vertex, the expectation is that there should be a large asymmetry. In each view, a 2D principal axis is determined and used to define which hits are upstream or downstream of the projected vertex candidate. The difference between the numbers of hits is used to calculate a fractional asymmetry, $A_{i}$ for view $i$:
\begin{equation}
S_\text{asymmetry} = \exp\left\{{\sum_{\text{view } i}\frac{A_i}{\alpha}}\right\}
\label{equ::Asymmetry}
\end{equation}
\begin{equation}
{A_{i} = \frac{|N^{\uparrow}_{i} - N^{\downarrow}_{i}|}{N^{\uparrow}_{i} + N^{\downarrow}_{i}}}
\label{equ::Asymmetry2}
\end{equation}
\noindent where $\alpha$ is a tunable constant that determines the relative importance of the asymmetry score and $N^{\uparrow}_{i}$ and $N^{\downarrow}_{i}$ are the numbers of hits deemed upstream and downstream of the projected vertex candidate in view $i$.\\

\item \textbf{Beam deweighting score:} For the reconstruction of beam neutrinos, knowledge of the beam direction can be used to preferentially select vertex candidates with low $z$ coordinates:

\begin{equation}
S_\text{beam deweight} = \exp\left\{-Z/\zeta\right\}
\label{equ::BeamDeweight}
\end{equation}
\begin{equation}
{Z = \frac{z - z_\mathrm{min}}{z_\mathrm{max} - z_\mathrm{min}}}
\label{equ::BeamDeweight2}
\end{equation}
\noindent where $\zeta$ is a tunable constant that determines the relative importance of the beam deweighting score and $z_{\mathrm{min}}$ and $z_{\mathrm{max}}$ are the lowest and highest $z$ positions from the list of candidate vertices.\\
\end{itemize}

Figure \ref{fig::Vertexing} shows the scores assigned to a number of vertex candidates in a typical simulated CC $\nu_{\mu}$ event in MicroBooNE, including a breakdown of each score into its component parts. Following selection of the neutrino interaction vertex, any 2D clusters crossing the vertex are split into two pieces, one on either side of the projected vertex position.


\subsubsection{Track and shower reconstruction}
\label{sec::ShowerReco}
The PandoraNu 3D track reconstruction proceeds as described in Section \ref{sec::ThreeDTracks}. Unlike the cosmic-ray muon reconstruction, PandoraNu also attempts to reconstruct primary electromagnetic showers, from electrons and photons. An example of the typical topologies under investigation is shown in Figure \ref{fig::TrackShwId}. PandoraNu performs 2D shower reconstruction by adding branches to any long clusters that represent ``shower spines''. This  procedure uses the following steps:

\begin{figure}[]
  \begin{center}
     \includegraphics[width=0.7\textwidth]{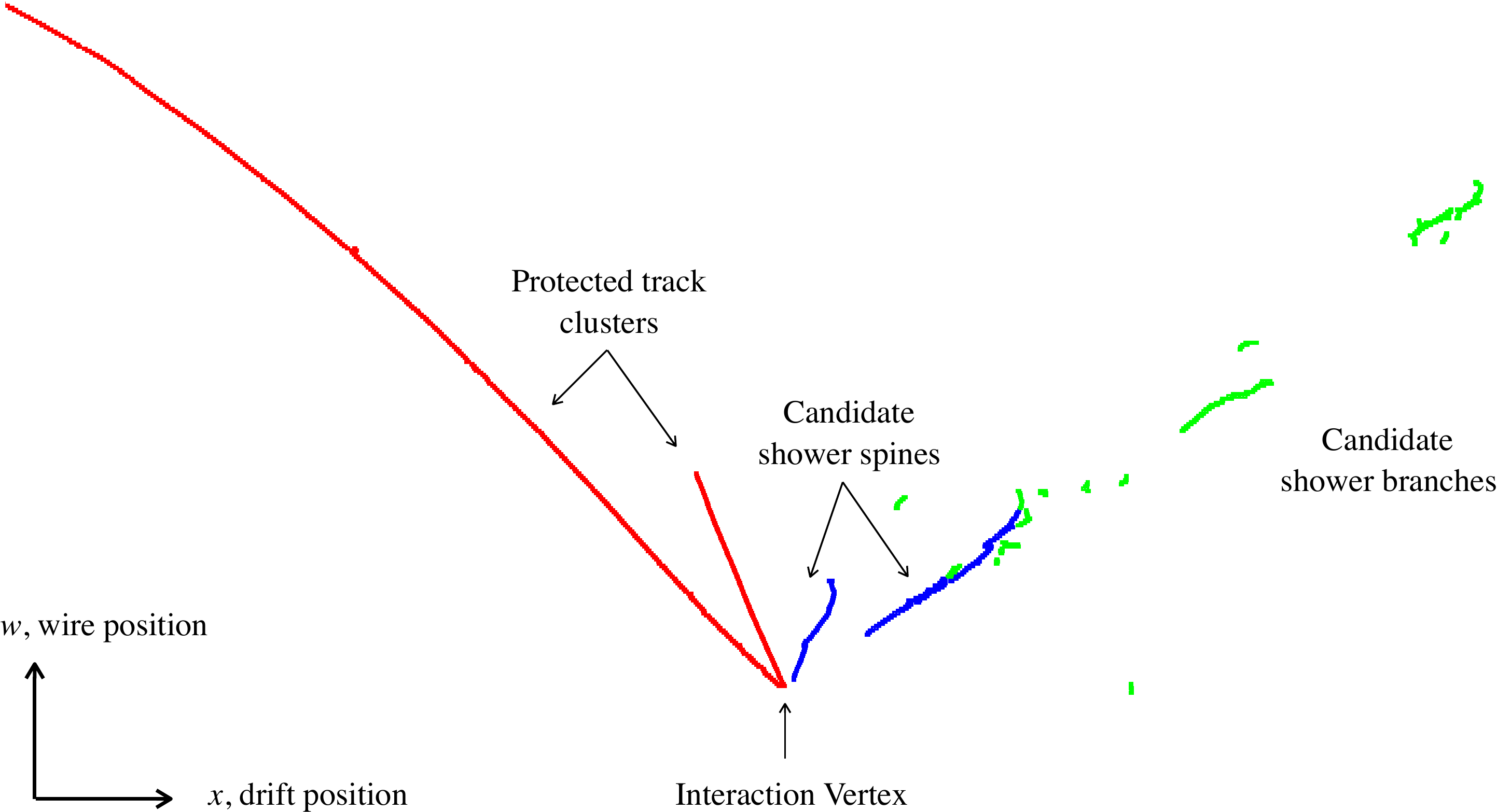}
     \caption{Cluster labels used by the shower reconstruction algorithms. Clusters identified as track-like (red) are excluded from the shower reconstruction. Long, typically vertex-associated, shower-like clusters (blue) are identified as possible shower spines. The ShowerGrowing algorithm looks to add shower-like branch clusters (green) to the most appropriate shower spines, providing 2D shower-like clusters of high completeness.\label{fig::TrackShwId}}
  \end{center}
\end{figure}

\begin{itemize}
    \item The 2D clusters are characterised as track-like or shower-like, based on length, variations in sliding-fit direction along the length of the cluster, an assessment of the extent of the cluster transverse to its linear-fit direction, and the closest approach to the projected neutrino interaction vertex.
    \item Any existing track particles that are now deemed to be shower-like are dissolved to allow assessment of the clusters as shower candidates.
    \item Long, shower-like 2D clusters that could represent shower spines are identified. The shower spines will typically point back towards the interaction vertex.
    \item Short, shower-like 2D branch clusters are added to shower spines. The ShowerGrowing algorithm operates recursively, finding branches on a candidate shower spine, then branches on branches. For every branch, a strength of association to each spine is recorded. Branch addition decisions are then made in the context of the overall event topology.
\end{itemize}

\begin{figure}[]
  \begin{center}
     \includegraphics[width=0.6\textwidth]{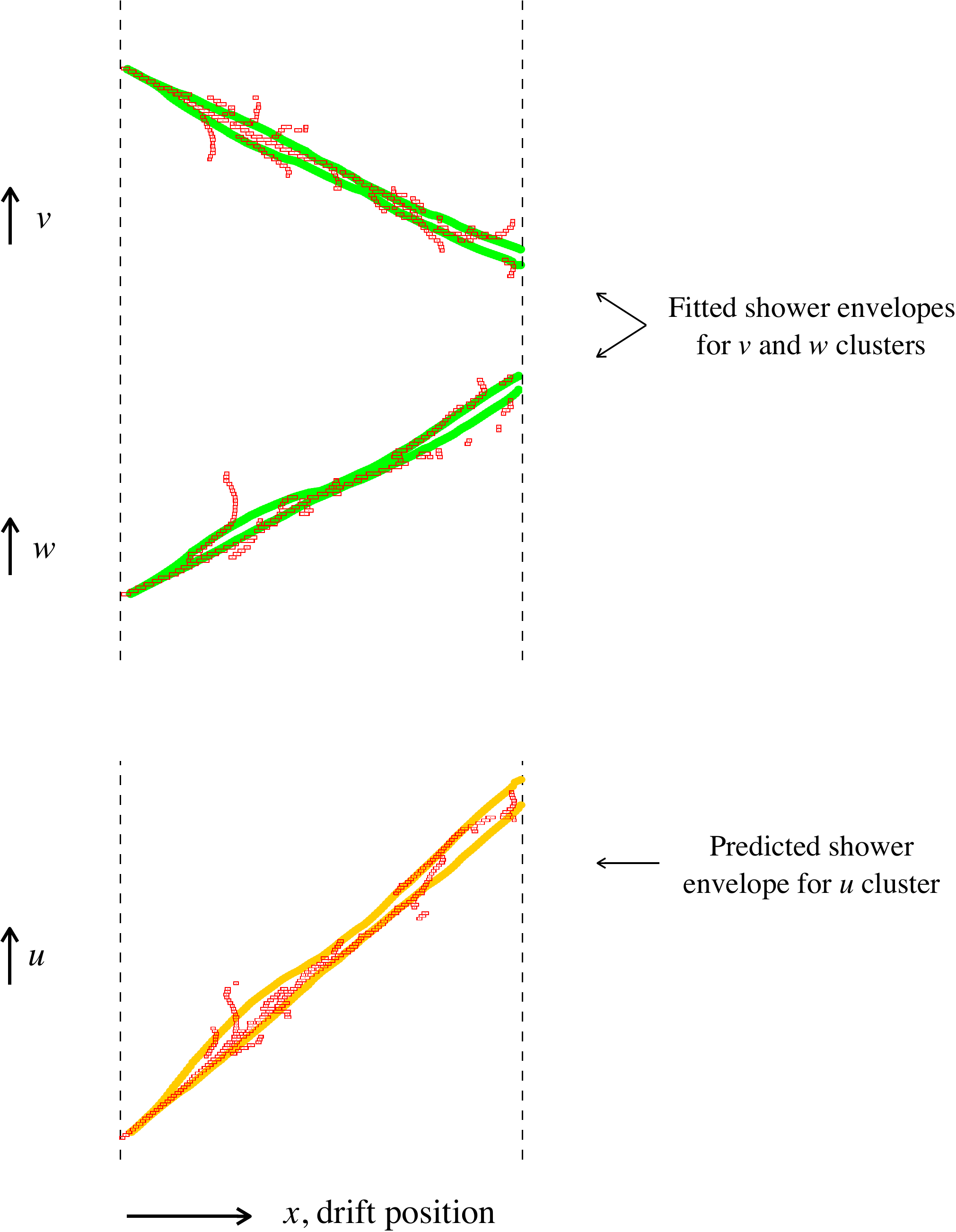}
     \caption{The 3D shower reconstruction aims to identify the clusters representing the same shower in each of the three readout planes. The hits in candidate 2D clusters are shown as red boxes. Fits to the hit positions are used to characterise the spatial extent of the clusters. The fitted shower envelopes (green markers) from two clusters are then used to predict a shower envelope (orange markers) for the third cluster. The fraction of hits in the third cluster enclosed by the predicted envelope is calculated. Predictions made using all cluster combinations ($u,v\rightarrow w$; $v,w\rightarrow u$; $u,w\rightarrow v$) are used to decide whether to add the three clusters to a new shower particle.\label{fig::ShowerOverlap}}
  \end{center}
\end{figure}

Following 2D shower reconstruction, the 2D shower-like clusters are matched between readout planes in order to form 3D shower particles. The ideas described in Section \ref{sec::ThreeDTracks} are re-used for this process. The ThreeDShowers algorithm builds a rank-three tensor to store cluster-overlap and relationship information, then a series of algorithm tools examine the tensor. Iterative changes are made to the 2D reconstruction to diagonalise the tensor and ensure that 3D shower particles can be formed without ambiguity. Fits to the hit positions in 2D shower-like clusters are used to characterise the spatial extent of the shower. In order to calculate a ShowerOverlapResult for a group of three clusters, the shower edges from two are used to predict a shower envelope for the third cluster. The fraction of hits in the third cluster contained within the envelope is then stored, alongside details of the common cluster $x$-overlap. This procedure is illustrated in Figure \ref{fig::ShowerOverlap}.

The shower tensor is first queried by the ClearShowers tool, which looks to form shower particles from any unambiguous associations between three clusters. The association between the clusters must satisfy quality cuts on the common $x$-overlap and fraction of hits enclosed in the predicted shower envelopes. The SplitShowers tool then looks to resolve ambiguities associated with splitting of sparse showers into multiple 2D clusters. This tool searches for 2D clusters that can be merged in order to ensure that each electromagnetic shower is represented by a single cluster from each readout plane.

After 3D shower reconstruction, a second pass of the 3D track reconstruction is applied, to recover any inefficiencies associated with dissolving track particles to examine their potential as showers. The ParticleRecovery algorithm then examines any groups of clusters that previously failed to satisfy the quality cuts for particle creation, due to problems with the hit-finding or 2D clustering, or due to significant detector gaps. Ideas from the earlier 3D track and shower reconstruction are re-used, but the thresholds for matching clusters between views are reduced. Finally, the ParticleCharacterisation algorithm classifies each particle as being either track-like or shower-like.


\subsubsection{Particle refinement}
\label{sec::Refinement}
The list of 3D track-like and shower-like particles can be examined and refined, to provide the final assignment of hits to particles. For MicroBooNE, the primary issue to address at this stage is the completeness of sparse showers, which can frequently be represented as multiple, separate reconstructed particles. A number of distinct algorithms are used:
\begin{itemize}
\item The ClusterMopUp algorithms consider 2D clusters that have been assigned to shower-like particles. They use parameterisations of the 2D cluster extents, including cone fits and sliding linear fits to the edges of the showers, to pick up any remaining, unassociated 2D clusters that are either bounded by the assigned clusters, or in close proximity.

\item The SlidingConeParticleMopUp algorithm uses sliding linear fits to the 3D hits for shower-like particles. Local 3D cone axes and apices are defined and cone opening angles can be specified as algorithm parameters or derived from the topology of the 3D shower hits. The 3D cones are extrapolated and downstream particles deemed fragments of the same shower are collected and merged into the parent particle.

\item The SlidingConeClusterMopUp algorithm projects fitted 3D cones into each view and searches for any remaining 2D clusters (not added to any particle) that are bounded by the projections.

\item The IsolatedClusterMopUp algorithm dissolves any remaining unassociated 2D clusters and looks to add their hits to nearby shower-like particles.
\end{itemize}

\subsubsection{Particle hierarchy reconstruction}
The final step in the PandoraNu reconstruction is to organise the reconstructed particles into a hierarchy. The procedure used is:
\begin{itemize}
    \item A neutrino particle is created and the 3D neutrino interaction vertex is added to this particle.
    \item The 3D hits associated with the reconstructed particles are considered and any particles deemed to be associated to the interaction vertex are added as primary daughters of the neutrino particle.
    \item Algorithm tools look to add subsequent daughter particles to the existing primary daughters of the neutrino, for example a decay electron may be added as a daughter of a primary muon particle.
    \item If the primary daughter particle with the largest number of hits is flagged as track-like or shower-like, the reconstructed neutrino will be labelled as a $\nu_{\mu}$ or a $\nu_{e}$ respectively.
    \item 3D vertex positions are calculated for each of the particles in the neutrino hierarchy. The vertex positions are the points of closest approach to their parent particles, or to the neutrino interaction vertex.
\end{itemize}

\begin{figure}[]
  \begin{center}
     \includegraphics[width=0.64\textwidth]{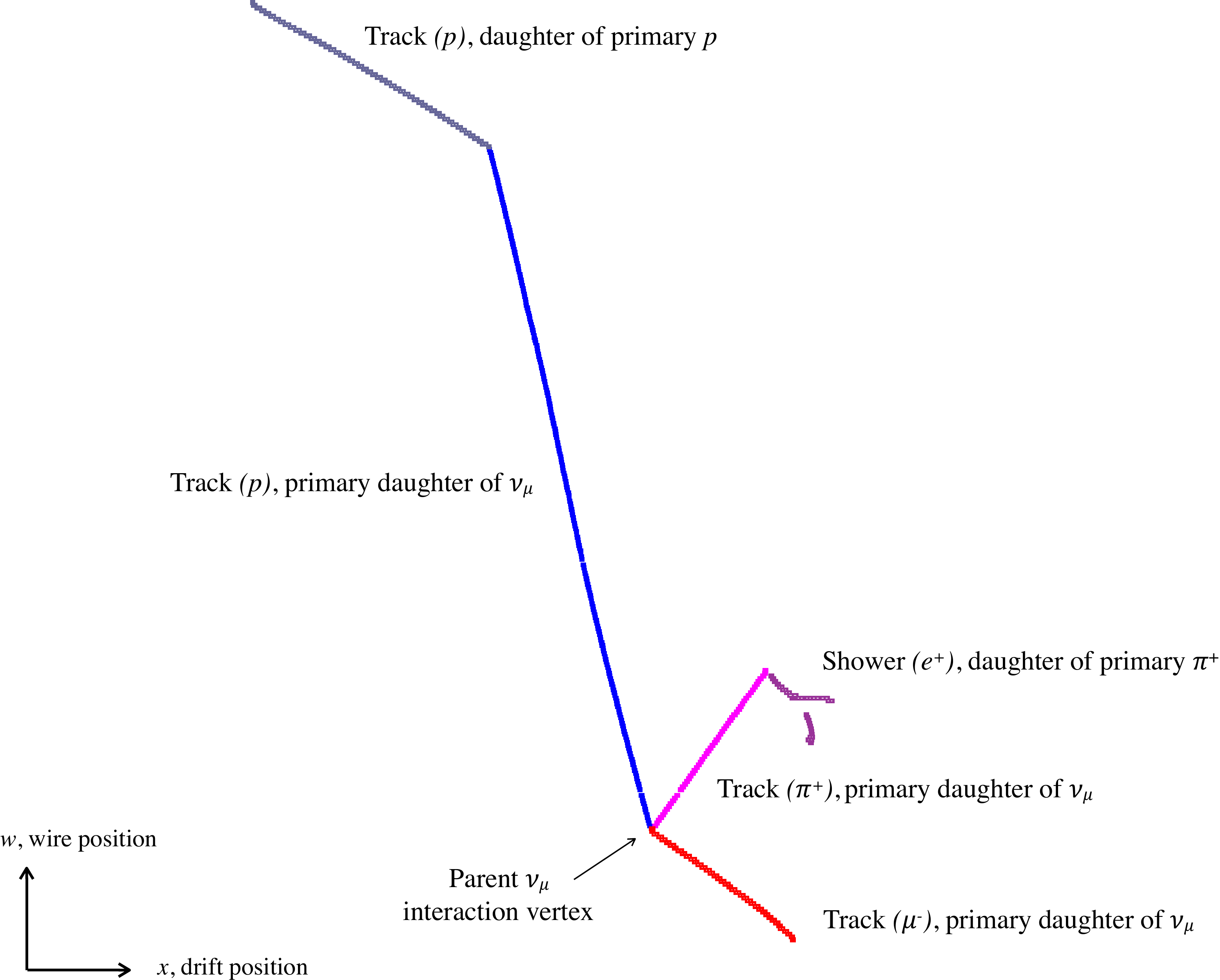}
     \caption{The hierarchy of particles reconstructed for a simulated CC $\nu_{\mu}$ event in MicroBooNE with a muon, proton and charged pion in the visible final state. Each reconstructed visible particle is shown in a separate colour. The neutrino particle has a reconstructed interaction vertex and three track-like primary daughter particles. The charged-pion decays into a $\mu^{+}$, which rapidly decays into a $e^{+}$ and is reconstructed as a shower-like secondary daughter particle. The proton scatters off a nucleus, resulting in a track-like secondary daughter particle. Pandora identifies each particle as track-like or shower-like and the explicit particle types were identified using information from the simulation.\label{fig::Hierarchy}}
  \end{center}
\end{figure}

Each slice results in a single reconstructed neutrino particle, with a hierarchy of reconstructed daughter particles. The particles reconstructed for a typical simulated CC $\nu_{\mu}$ event in MicroBooNE are illustrated in Figure \ref{fig::Hierarchy}.

\section{Performance metrics}
\label{sec::Metrics}
There are many ways in which to define and interpret performance metrics for pattern recognition, and each must be fully qualified. The performance metrics presented in this paper are based on the sharing of hits between the true, generated particles (MCParticles) and the reconstructed particles. A list of target MCParticles is selected by examining the MCParticle hierarchy. This hierarchy comprises the incident neutrino, the final-state particles emerging from the neutrino interaction, and cascades of daughter particles produced by subsequent decays or interactions. Starting with the neutrino and considering each daughter MCParticle in turn, the first visible particles (defined as one of $e^{\pm}, \mu^{\pm}, \gamma, K^{\pm}, \pi^{\pm}, p$) are identified as targets for the reconstruction. Each reconstructed 2D hit is matched to the target MCParticle responsible for the largest deposit of energy in the region of space and time covered by the hit, and the list of 2D hits matched to each MCParticle is known as its collection of ``true hits''. Any hits associated with downstream MCParticles in the hierarchy are folded into the relevant target MCParticle.

\begin{figure}[]
  \begin{center}
     \includegraphics[width=0.6\textwidth]{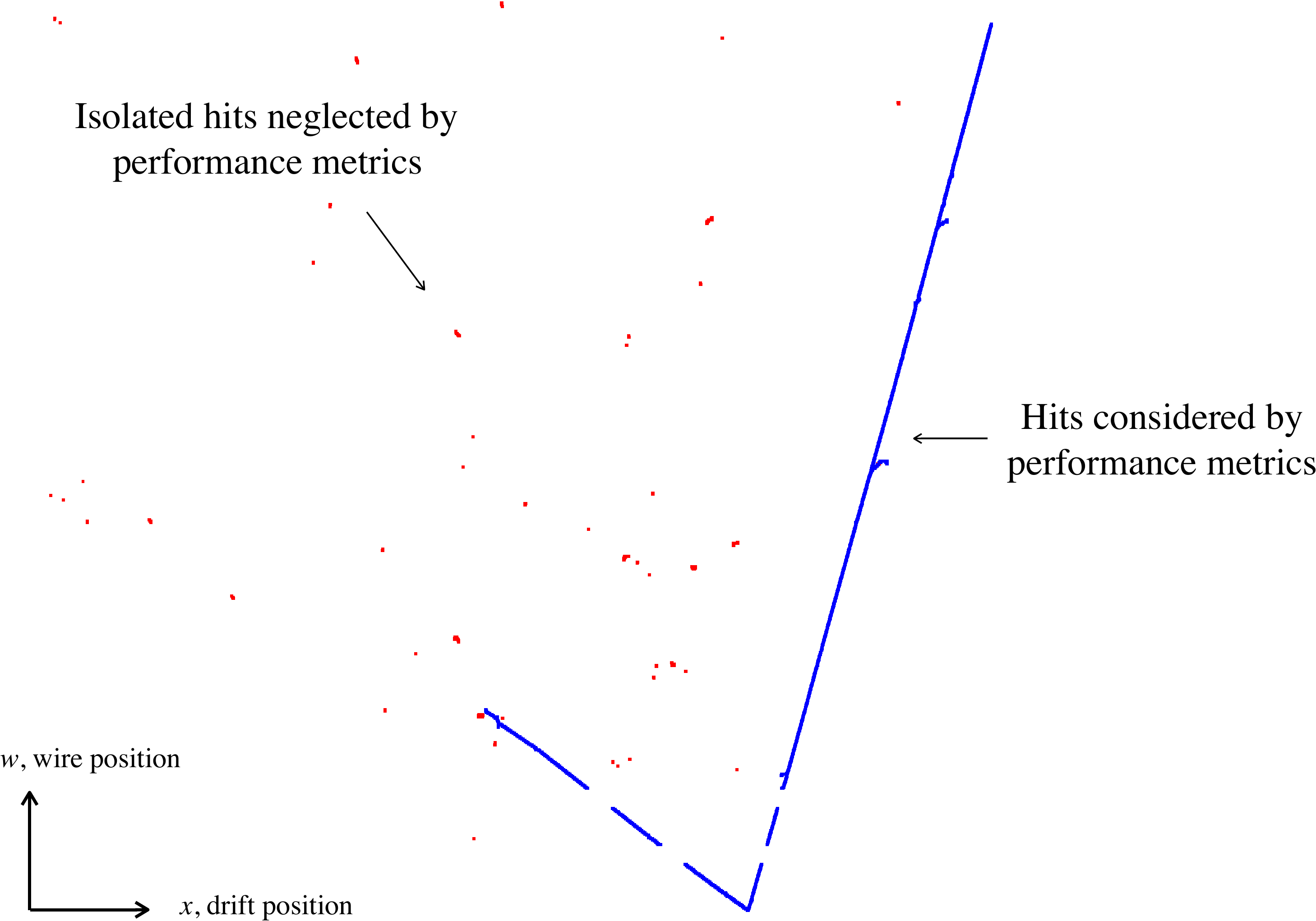}
     \caption{The hits that are considered (blue) and neglected (red) in the construction of pattern-recognition performance metrics for a typical simulated CC $\nu_{\mu}$ event in MicroBooNE. By considering the MCParticle hierarchy, hits that will likely form part of an isolated and diffuse topology are not used to identify or characterise the reconstructable target MCParticles in an event.\label{fig::IsolatedHits}}
  \end{center}
\end{figure}

In practice, some MCParticles will not be reconstructable and should not be considered as viable targets for the reconstruction. This may be because the MCParticle does not have sufficient true hits, or because its true hits form an isolated and diffuse topology, following a decay or interaction. For this reason, hits are neglected in the performance evaluation if the hierarchy shows they are associated to MCParticles downstream of a far-travelling neutron, or, if the primary MCParticle is track-like, a far-travelling photon (this avoids cases of capture of low energy particles, followed by nuclear excitation and decay, producing photons and neutrons). An example of the hits removed by this selection procedure is shown, for a typical simulated CC $\nu_{\mu}$ event in MicroBooNE, in Figure \ref{fig::IsolatedHits}. Target MCParticles are then only considered viable if they are associated to at least 15 hits passing the selection, including at least five hits in at least two views. When counting hits associated with a target MCParticle, the relevant MCParticle must be responsible for at least 90\% of the true energy deposition recorded for the hit. This selection corresponds to true momentum thresholds of approximately 60\,MeV for muons and 250\,MeV for protons in the MicroBooNE simulation.

Reconstructed particles are then matched to the target MCParticles. A matrix of associations is constructed, recording the number of hits shared between each target MCParticle and each reconstructed particle. As with the MCParticle hierarchy, the reconstructed particle hierarchy is used to fold hit associations with reconstructed daughter particles into the parent visible particles (the primary daughters of the reconstructed neutrino). The following performance metrics can then be defined:
\begin{itemize}
    \item \textbf{Efficiency}, for a type of target MCParticle, is the fraction of such target MCParticles with at least one matched reconstructed particle 
    \item \textbf{Completeness}, for a given pairing of reconstructed particle and target MCParticle, is the fraction of the MCParticle true hits that are shared with the reconstructed particle
    \item \textbf{Purity}, for a given pairing of reconstructed particle and target MCParticle, is the fraction of hits in the reconstructed particle that are shared with the target MCParticle
\end{itemize}

The information collected in the matching process is comprehensive, but single reconstructed particles can contain hits from multiple target MCParticles and some interpretation of the information can clarify the reconstruction performance. For instance, a distinction can be made between the case where a few hits are incorrectly assigned in regions where several target MCParticles meet, and the case where a single reconstructed particle incorporates a significant fraction of true hits from multiple target MCParticles. Matches between target MCParticles and reconstructed particles are only considered if there are at least five hits shared between the two. The reconstructed particle must also match the target MCParticle with a purity of 50\%, so that it is more strongly associated to the given MCParticle than to any other. Matches must also have a completeness of at least 10\%, which is a low threshold designed to remove low-quality matches between target MCParticles and small, fragment reconstructed particles\footnote{The completeness and purity thresholds are omitted from plots of completeness and purity themselves, where they would present direct selection cuts.}. The procedure below is used to provide a final, human interpretation of the reconstruction output:
\begin{enumerate}
    \item Identify the single strongest match, with the largest number of shared hits, between any of the \textit{available} target MCParticles and reconstructed particles.
    \item Repeat step 1 until no further matches are possible, ensuring that each target MCParticle and reconstructed particle can only be matched at most once, and are then subsequently \textit{unavailable}.
    \item Assign any remaining available, unmatched reconstructed particles to the target MCParticle with which they share most hits, even if the target MCParticle already has reported matches.
\end{enumerate}

In step 3 of the interpretation, the number of reconstructed particles matched to a target MCParticle can increase from one to e.g. two or three, but can never increase if it is zero upon the completion of step 2 (this target MCParticle must have been lost). An event is deemed to have a ``correct'' overall reconstruction if there is exactly one reconstructed particle for each target MCParticle at the end of this procedure. The fraction of events deemed correct provides a useful, and highly sensitive, picture of the pattern-recognition performance.


\section{Performance}
\label{sec::Performance}
The performance of the PandoraNu reconstruction is considered separately for specific neutrino interaction types and a selection of exclusive final states in generated BNB events in the MicroBooNE detector simulation. Only neutrino interactions in the fiducial volume of the LArTPC are considered. The fiducial volume is the active volume excluding the region within 10\,cm of the detector edges in $x$ and $z$, and within 20\,cm of the edges in $y$. In Sections \ref{sec::CCQEL_MU_P}, \ref{sec::CCRES_MU_P_PIPLUS} and \ref{sec::CCRES_MU_P_PIZERO}, the performance of the neutrino reconstruction is tested using three specific topologies: two-track, three-track, and two-track plus two-shower $\nu_{\mu}$ CC interactions in argon. In Section \ref{sec::BNB_SUMMARY}, the performance is assessed for more complex final states. A combined reconstruction chain containing both PandoraCosmic and PandoraNu is then studied in Section \ref{sec::CR}, using simulated BNB interactions overlaid with simulated cosmic-ray muon interactions.

The event generation and detector simulation steps use LArSoft v04.36.00.03, which includes v2.8.6 of the GENIE\,\cite{bib::GENIE} neutrino Monte Carlo event generator, and v7.4003 of the CORSIKA\,\cite{bib::CORSIKA} Monte Carlo simulation of air showers initiated by cosmic-ray particles. The simulation of the MicroBooNE detector geometry incorporates unresponsive parts of the readout, but does not include a full description of detector noise. Signal processing, including hit finding, uses LArSoft v05.08.00.05 and the Pandora pattern recognition uses v03.02.00 of the LArPandoraContent library, which contains the Pandora algorithms and tools implemented for LArTPC event reconstruction and requires v03.00.00 of the Pandora SDK. The cosmic-ray tagging and hit removal modules of LArSoft v06.15.01 were used. For each LArSoft version, the corresponding version of uboonecode\,\cite{bib::uboonecode} was used to provide MicroBooNE-specific additions to the LArSoft functionality.


\subsection{BNB CC quasi-elastic events: $\nu_{\mu} + Ar  \rightarrow  \mu^{-} + p$}
\label{sec::CCQEL_MU_P}
\begin{figure}[!h]
  \begin{center}
     \includegraphics[width=0.6\textwidth]{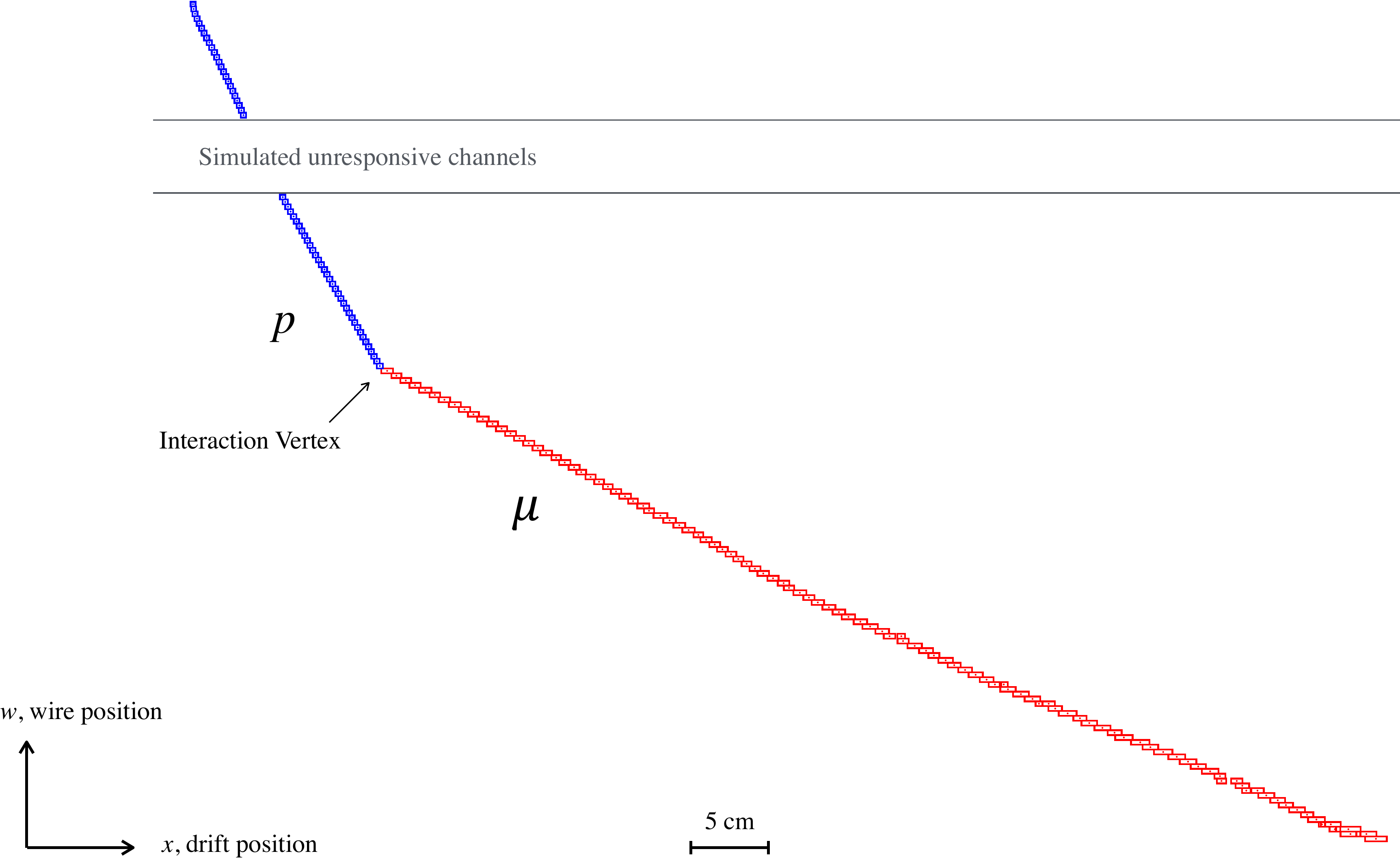}
     \caption{The reconstruction of a simulated 500-MeV CC $\nu_{\mu}$ quasi-elastic interaction. The target particles for the reconstruction are the muon and proton. A gap in the reconstructed proton track is observed due to the presence of unresponsive channels, which are included in the simulation.\label{fig::CCQEL_MU_P_Example}}
  \end{center}
\end{figure}

Quasi-elastic CC interactions with exactly one reconstructable muon and one reconstructable proton in the visible final state provide a clean topology to evaluate pattern-recognition performance. This clean topology represents only a small subset of the possible final states produced by quasi-elastic CC interactions in argon. The true momentum distributions for muons and protons in selected BNB events both peak at approximately 400\,MeV; an example event topology is displayed in Figure \ref{fig::CCQEL_MU_P_Example}. Table \ref{tab::CCQEL_MU_P} provides a thorough assessment of the pattern-recognition performance for this kind of interaction, showing the distribution of numbers of reconstructed particles matched to each target MCParticle. Events with a correct reconstruction should match exactly one reconstructed particle to the muon and exactly one to the proton. The Table shows that 95.8\% of target muons and 87.3\% of target protons are matched to exactly one reconstructed particle; 86.0\% of events are deemed to be reconstructed correctly. A small fraction of muons (1.3\%) are not reconstructed and a more significant fraction (8.9\%) of protons also have no matched reconstructed particle. This is predominantly due to merging of the muon and proton into a single reconstructed particle. Some muons and protons are split into two (or more) reconstructed particles. One mechanism for splitting target MCParticles is failure to reconstruct all the required parent-daughter links when true daughter MCParticles are present: reconstruction of a decay electron as a separate primary particle, for example. Another mechanism is incomplete reclamation of target MCParticles that are split across gaps in the detector instrumentation.

\begin{table}[!h]
  \begin{center}
    \begin{tabular}{ c | c  c  c  c }
      \toprule
      \#Matched Particles  & 0                  & 1            & 2                 & 3+              \\
      \midrule
      $\mu$                & \,\,\,$1.3\%$      & $95.8\%$     & \,\,\,$2.9\%$     & \,\,\,$0.1\%$   \\
      $p$                  & \,\,\,$8.9\%$      & $87.3\%$     & \,\,\,$3.6\%$     & \,\,\,$0.2\%$   \\
      \bottomrule
    \end{tabular}
  \end{center}
  \caption{Pattern-recognition performance for the target muon and proton in simulated BNB CC $\nu_{\mu}$ quasi-elastic interactions. The total number of events was 53,168 and 86.0\% were deemed to have exactly one reconstructed particle matched to each target.\label{tab::CCQEL_MU_P}}
\end{table}

\begin{figure}[!h]
  \begin{center}
     \subfloat[][]{\includegraphics[width=0.34\textwidth]{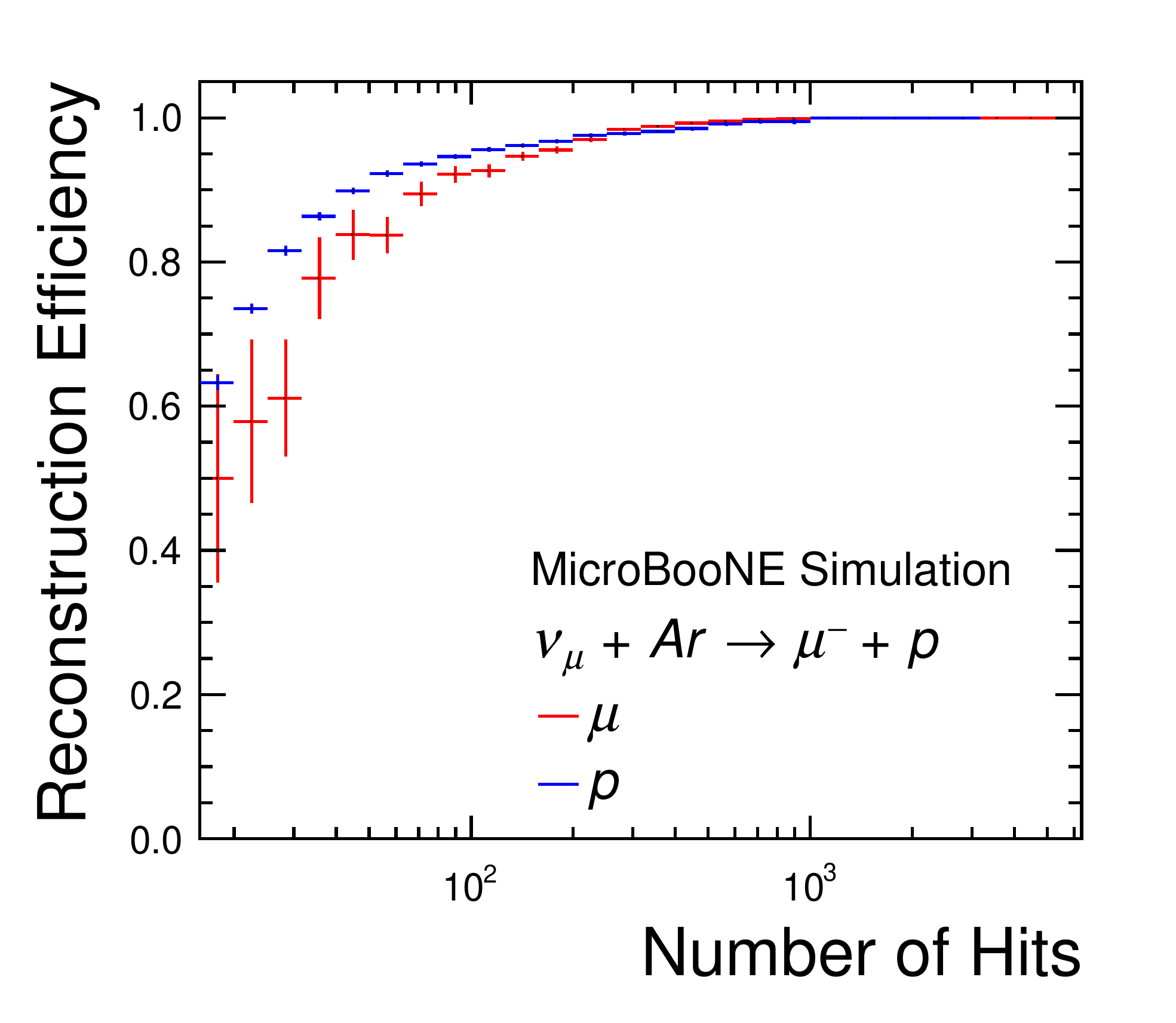}\label{fig::CCQEL_MU_P_EffHits}}
     \subfloat[][]{\includegraphics[width=0.34\textwidth]{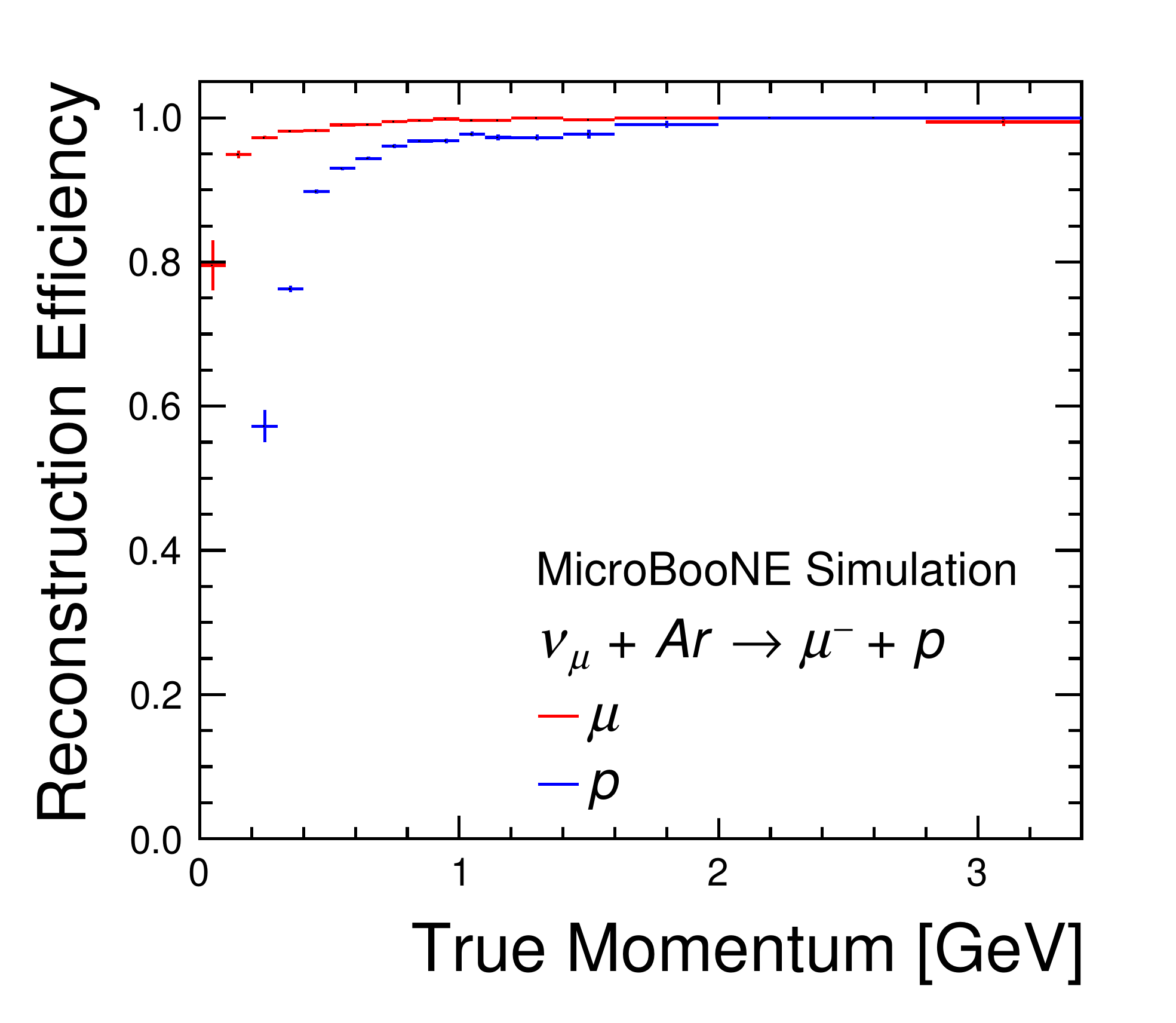}\label{fig::CCQEL_MU_P_EffMom}}
     \subfloat[][]{\includegraphics[width=0.34\textwidth]{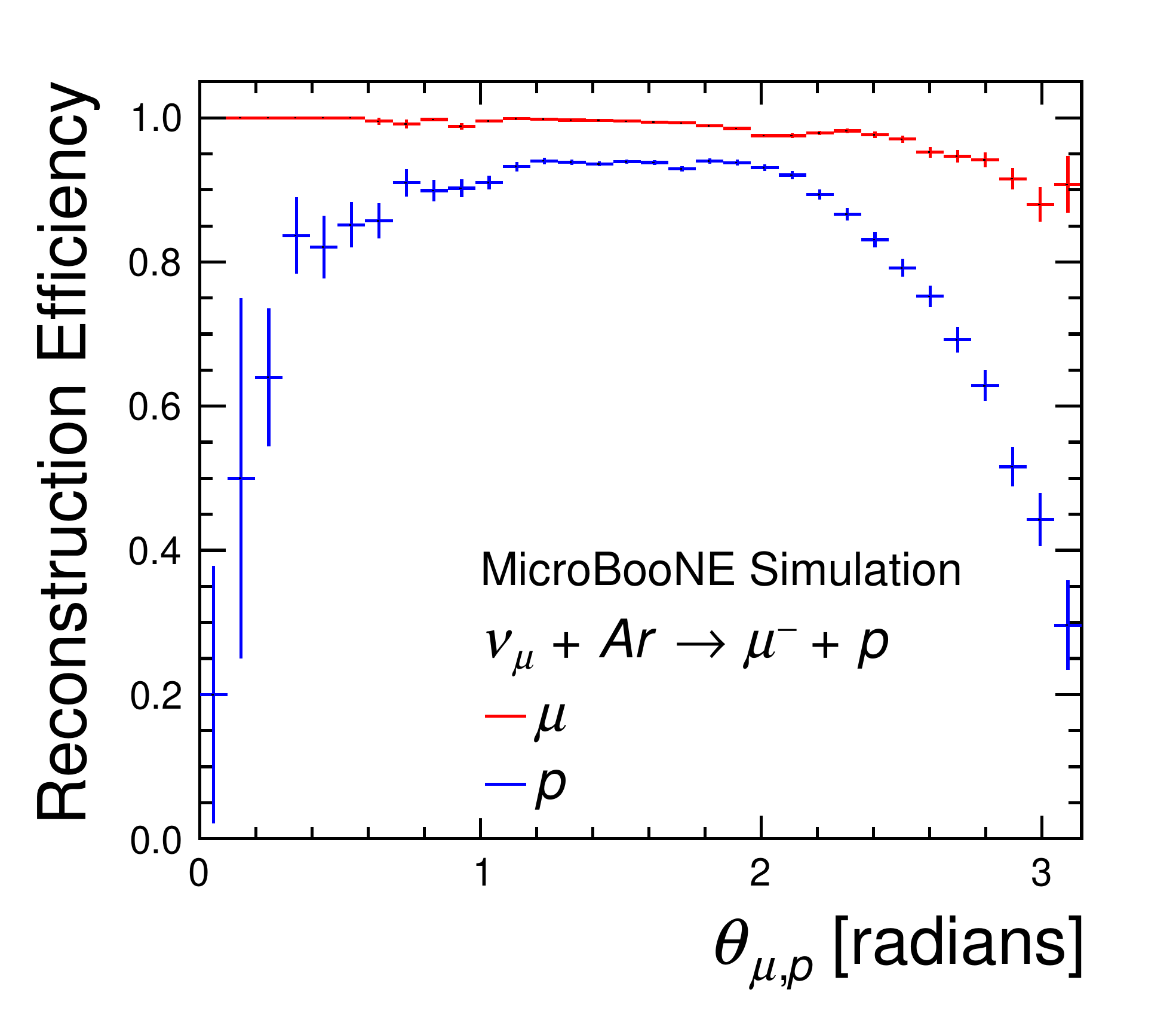}\label{fig::CCQEL_MU_P_EffAngle}}
     \caption{Reconstruction efficiencies for the target muon and proton in simulated BNB CC $\nu_{\mu}$ quasi-elastic interactions, (a) as a function of the numbers of true hits, (b) as a function of true momenta and (c) as a function of the true opening angle between the muon and proton.\label{fig::CCQEL_MU_P_Efficiencies}}
  \end{center}
\end{figure}

Figure \ref{fig::CCQEL_MU_P_Efficiencies} displays the reconstruction efficiencies for the target muon and proton as a function of the numbers of true hits, as a function of true momenta and as a function of the true opening angle between the muon and proton. The proton reconstruction efficiency is lower than the muon reconstruction efficiency across the full range of momenta, with the most common failure mechanism being merging of the muon and proton into a single particle. The efficiency in Figure \ref{fig::CCQEL_MU_P_EffHits} is better for protons with small numbers of hits than for muons with the same numbers of hits, because of their respective $\mathrm{d}E/\mathrm{d}x$ distributions. Figure \ref{fig::CCQEL_MU_P_EffAngle} shows that the muon and proton are most likely to be merged into a single particle when the two tracks are close to collinear. The single reconstructed particle will be matched to the target with which it shares most hits, which will preferentially be the muon. When the muon and proton are collinear, use of $\mathrm{d}E/\mathrm{d}x$ information might allow the individual particles to be resolved. This information is not yet exploited by the pattern recognition, but is expected to yield improvements in the future.

\begin{figure}[!h]
  \begin{center}
     \subfloat[][]{\includegraphics[width=0.34\textwidth]{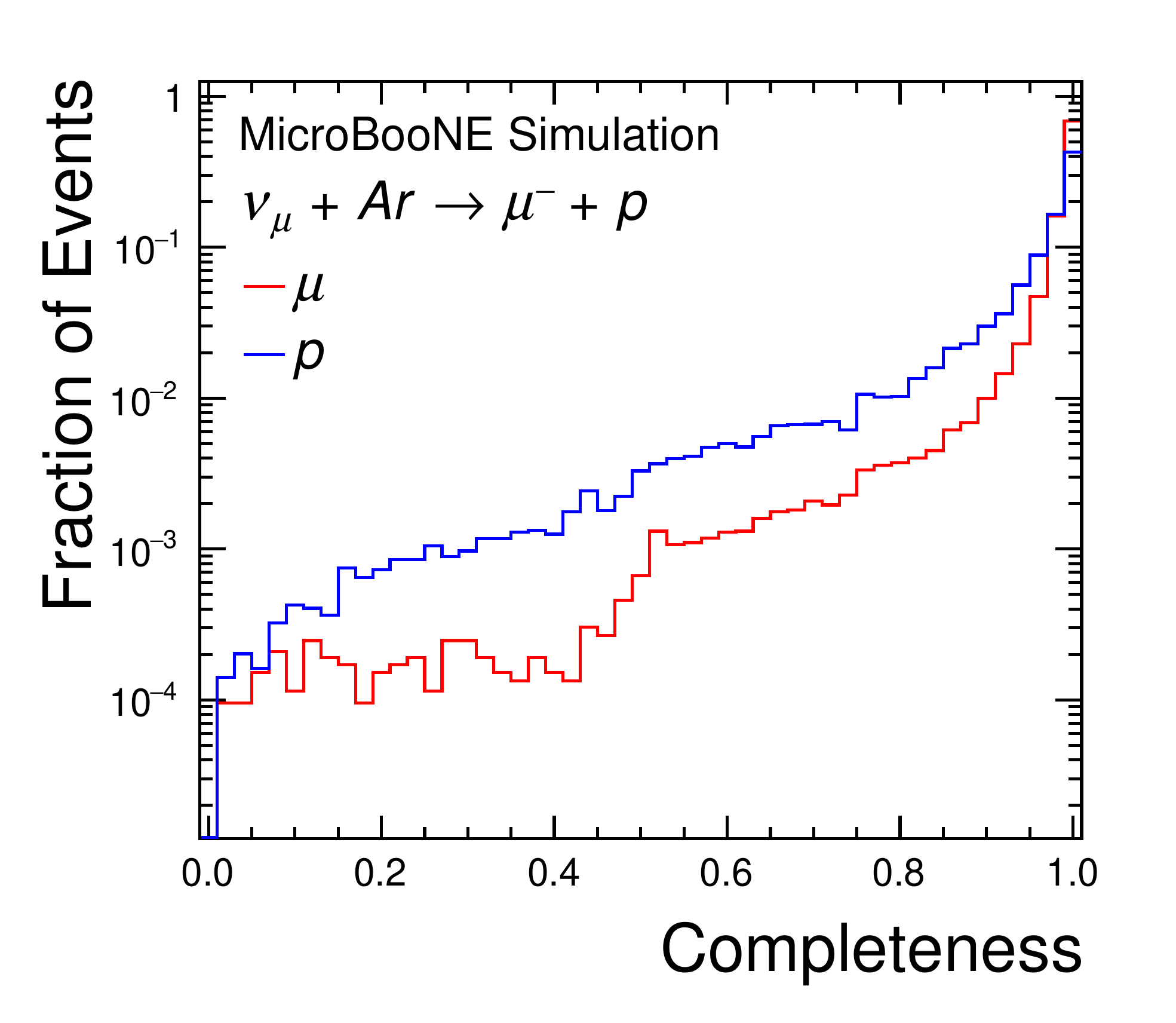}\label{fig::CCQEL_MU_P_Comp}}
     \subfloat[][]{\includegraphics[width=0.34\textwidth]{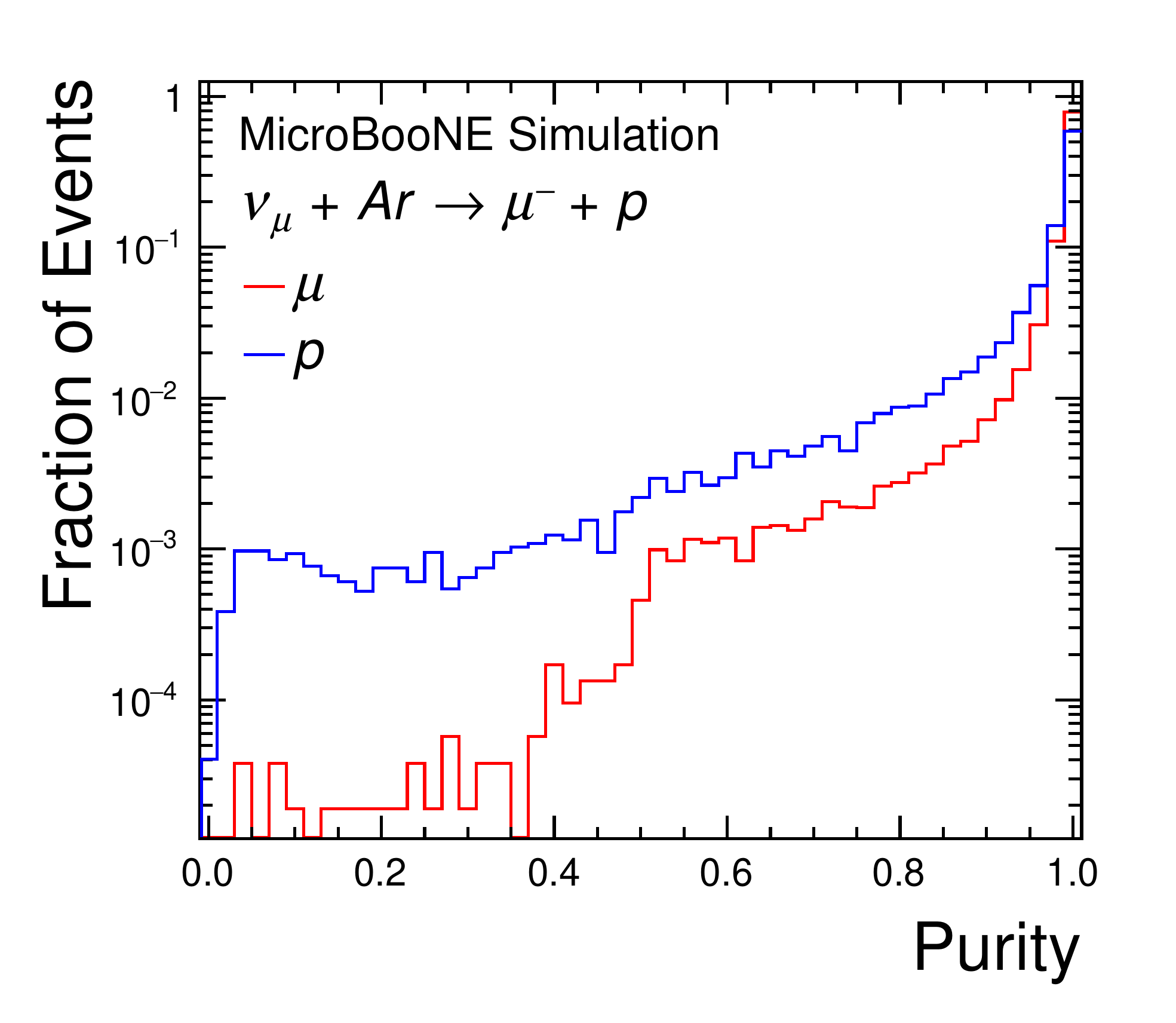}\label{fig::CCQEL_MU_P_Pur}}
     \subfloat[][]{\includegraphics[width=0.34\textwidth]{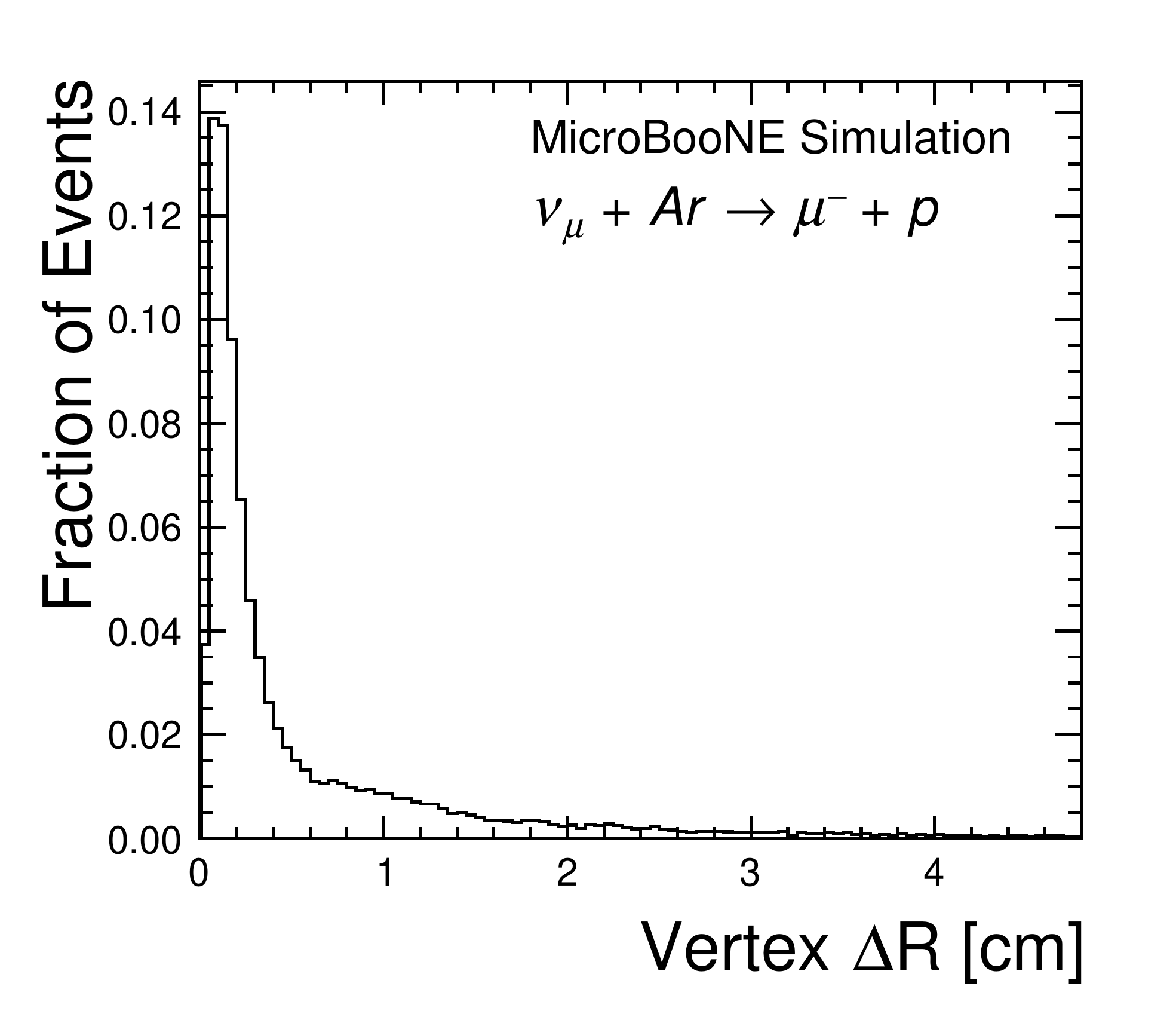}\label{fig::CCQEL_MU_P_Vtx}}
     \caption{Completeness (a) and purity (b) of the reconstructed particles with the strongest matches to the target muon and proton in simulated BNB CC $\nu_{\mu}$ quasi-elastic interactions and (c) the distance between generated and reconstructed 3D vertex positions.\label{fig::CCQEL_MU_P_PurCompVtx}}
  \end{center}
\end{figure}

Figure \ref{fig::CCQEL_MU_P_PurCompVtx} shows the completeness and purity of the reconstructed particles with the strongest matches to the target muon and proton; the distributions strongly peak at one. Figure \ref{fig::CCQEL_MU_P_Comp} shows that it is more difficult to achieve high reconstructed completeness for protons than for muons, as this can require collection of all hits in complex hadronic shower topologies downstream of the main proton track. Figure \ref{fig::CCQEL_MU_P_Pur} shows that there is a notable population of low purity protons, which are those that just satisfy the requirements to be matched to the target proton, but which also track significantly into the nearby muon. 

Figure \ref{fig::CCQEL_MU_P_Vtx} shows the displacement of the reconstructed neutrino interaction vertex from the true, generated position. It is found that 68\% of events have a displacement below 0.74\,cm. The 10.4\% of events with a displacement above 5\,cm are mainly due to placement of the vertex at the incorrect end of one of the particle tracks. This typically happens when there is a track of significant length with direction back towards the beam source. The presence of decay electrons can also yield topologies where multiple, distinct particles are associated with a specific point and can make the downstream end of the muon track appear to be a strong vertex candidate.

\subsection{BNB CC resonance events: $\nu_{\mu} + Ar  \rightarrow  \mu^{-} + p + \pi^{+}$}
\label{sec::CCRES_MU_P_PIPLUS}
\begin{figure}[!h]
  \begin{center}
     \includegraphics[width=0.5\textwidth]{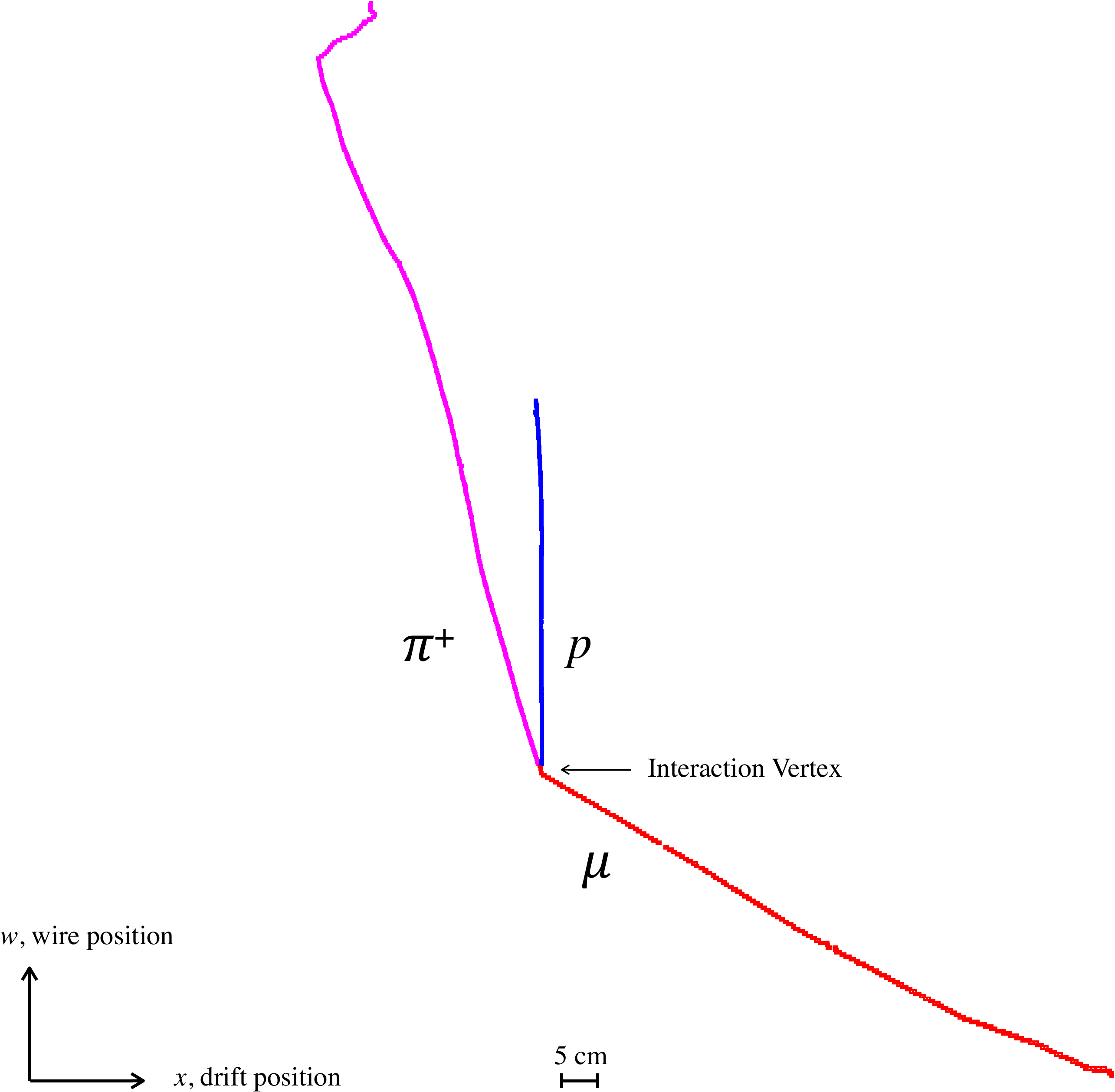}
     \caption{The reconstruction of a simulated 1.1-GeV CC $\nu_{\mu}$ interaction with resonant charged-pion production. Target particles for the reconstruction are the muon, proton and charged pion.\label{fig::CCRES_MU_P_PIPLUS_Example}}
  \end{center}
\end{figure}

The performance for three-track final states is studied using simulated BNB CC $\nu_{\mu}$ interactions with resonant charged-pion production. A specific subset of events is selected: those with one reconstructable muon, one reconstructable proton and one reconstructable charged pion in the visible final state. The true momentum distributions for particles in selected BNB events peak at approximately 300\,MeV for muons, 400\,MeV for protons and 200\,MeV for charged pions. An example event topology is shown in Figure \ref{fig::CCRES_MU_P_PIPLUS_Example}.

Table \ref{tab::CCRES_MU_P_PIPLUS} shows that 95.1\% of target muons, 86.8\% of target protons and 80.9\% of target pions result in a single reconstructed particle; 70.5\% of events are deemed correct, matching exactly one reconstructed particle to each target MCParticle. The performance for muons and protons is similar to that observed for the quasi-elastic events considered in Section \ref{sec::CCQEL_MU_P}. The fraction of muons with no matched reconstructed particles is higher than for quasi-elastic events, because the muon and pion tracks can be merged into a single particle. The pions will sometimes interact, leading to a MCParticle hierarchy of a parent and one or more daughter, and this explains the frequency at which the target pion is matched to more than one reconstructed particle: if the parent and daughter are reconstructed as separate particles, with no corresponding reconstructed parent-daughter links, multiple matches to the target pion will be recorded.

\begin{table}[!h]
  \begin{center}
    \begin{tabular}{ c | c  c  c  c }
      \toprule
      \#Matched Particles  & 0                  & 1             & 2                  & 3+             \\
      \midrule
      $\mu$                & \,\,\,$3.5\%$      & $95.1\%$      & \,\,\,$1.4\%$      & \,\,\,$0.0\%$  \\
      $p$                  & \,\,\,$9.0\%$      & $86.8\%$      & \,\,\,$4.0\%$      & \,\,\,$0.3\%$  \\
      $\pi^{+}$            & \,\,\,$6.9\%$      & $80.9\%$      & $11.4\%$           & \,\,\,$0.8\%$  \\
      \bottomrule
    \end{tabular}
  \end{center}
  \caption{Pattern-recognition performance for the target muon, proton and charged pion in simulated BNB CC $\nu_{\mu}$ interactions with resonant pion production. The total number of events was 47,754 and 70.5\% were deemed to have exactly one reconstructed particle matched to each target.\label{tab::CCRES_MU_P_PIPLUS}}
\end{table}

\begin{figure}[!h]
  \begin{center}
     \subfloat[][]{\includegraphics[width=0.34\textwidth]{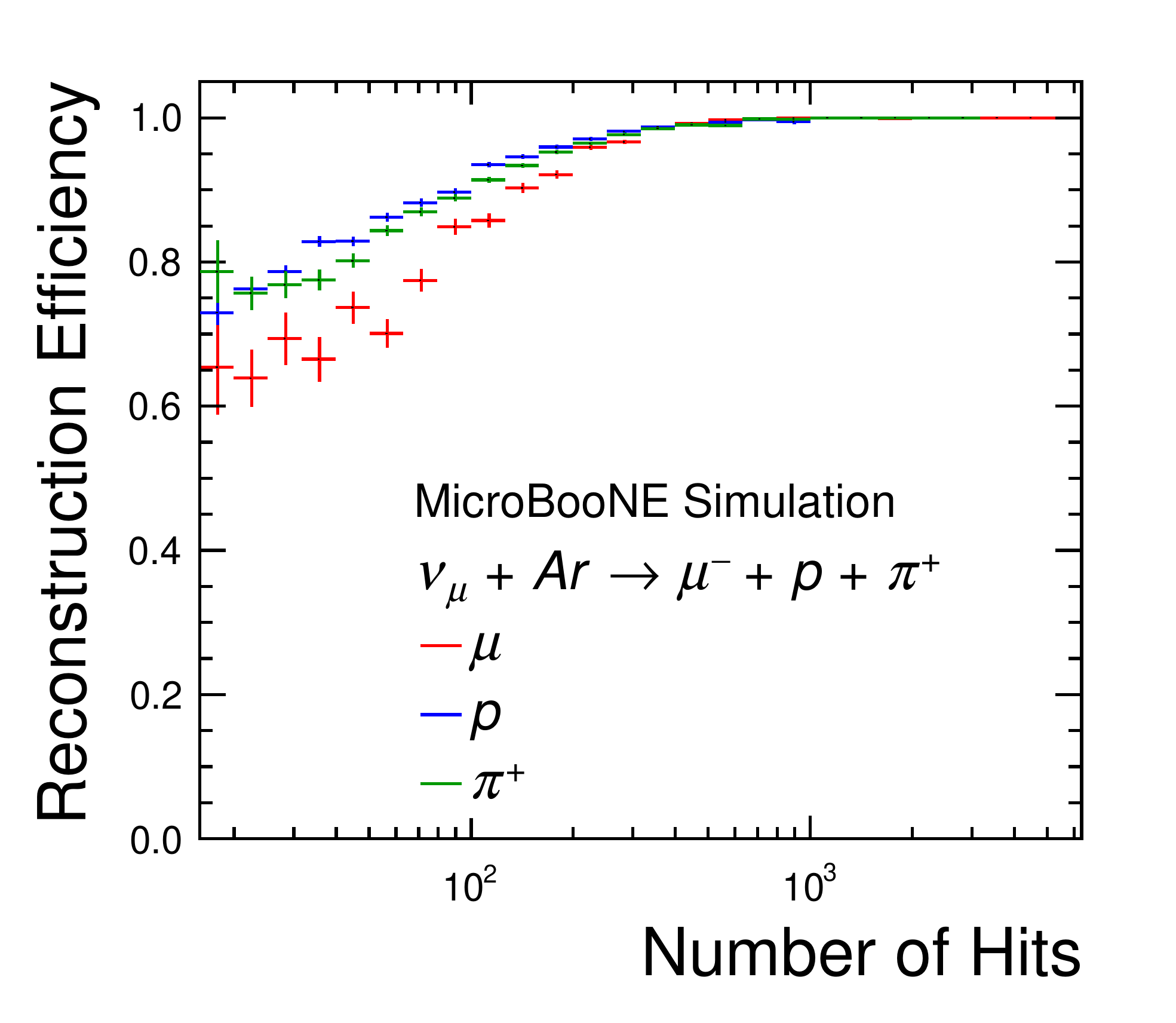}}
     \subfloat[][]{\includegraphics[width=0.34\textwidth]{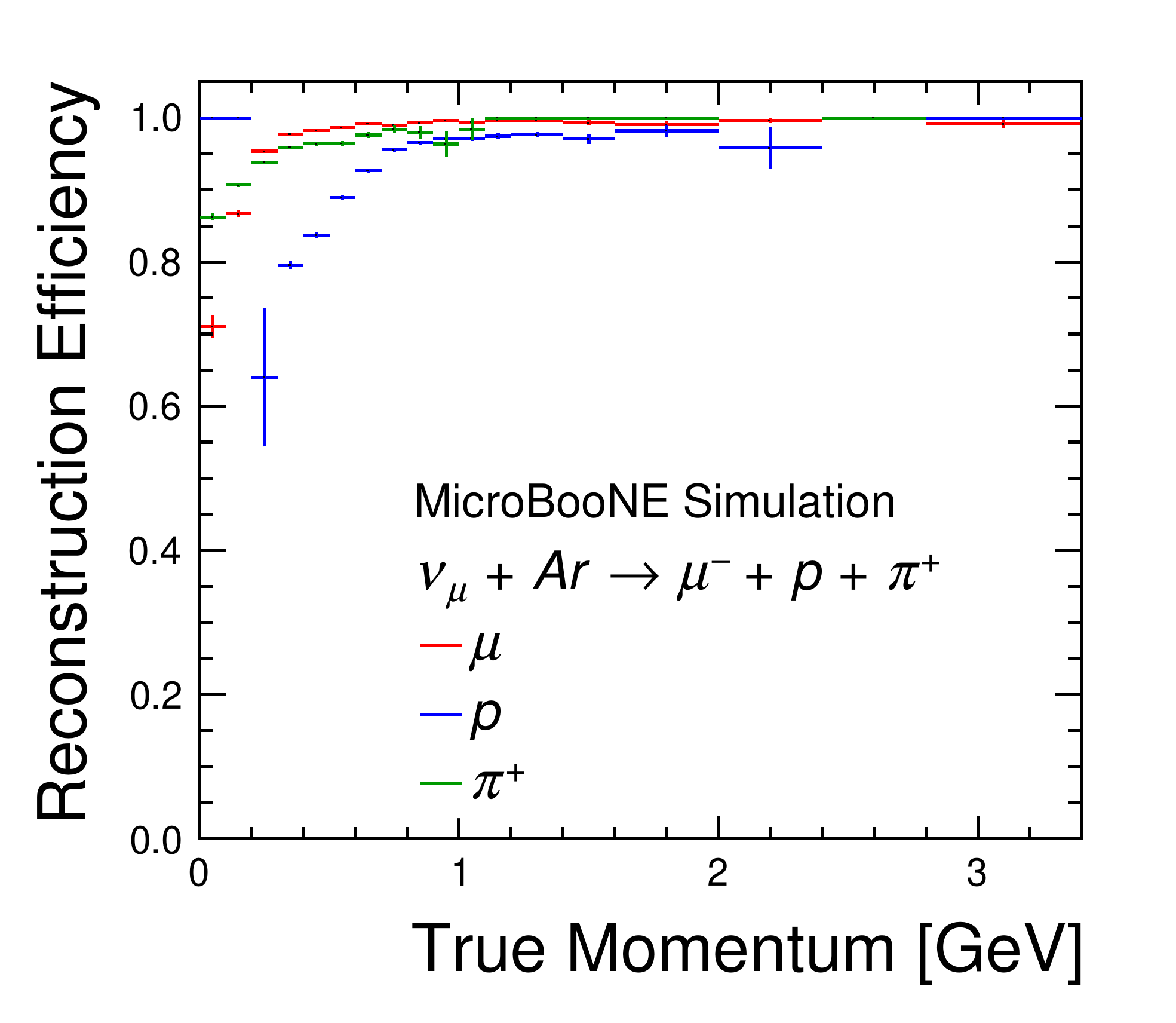}}
     \subfloat[][]{\includegraphics[width=0.34\textwidth]{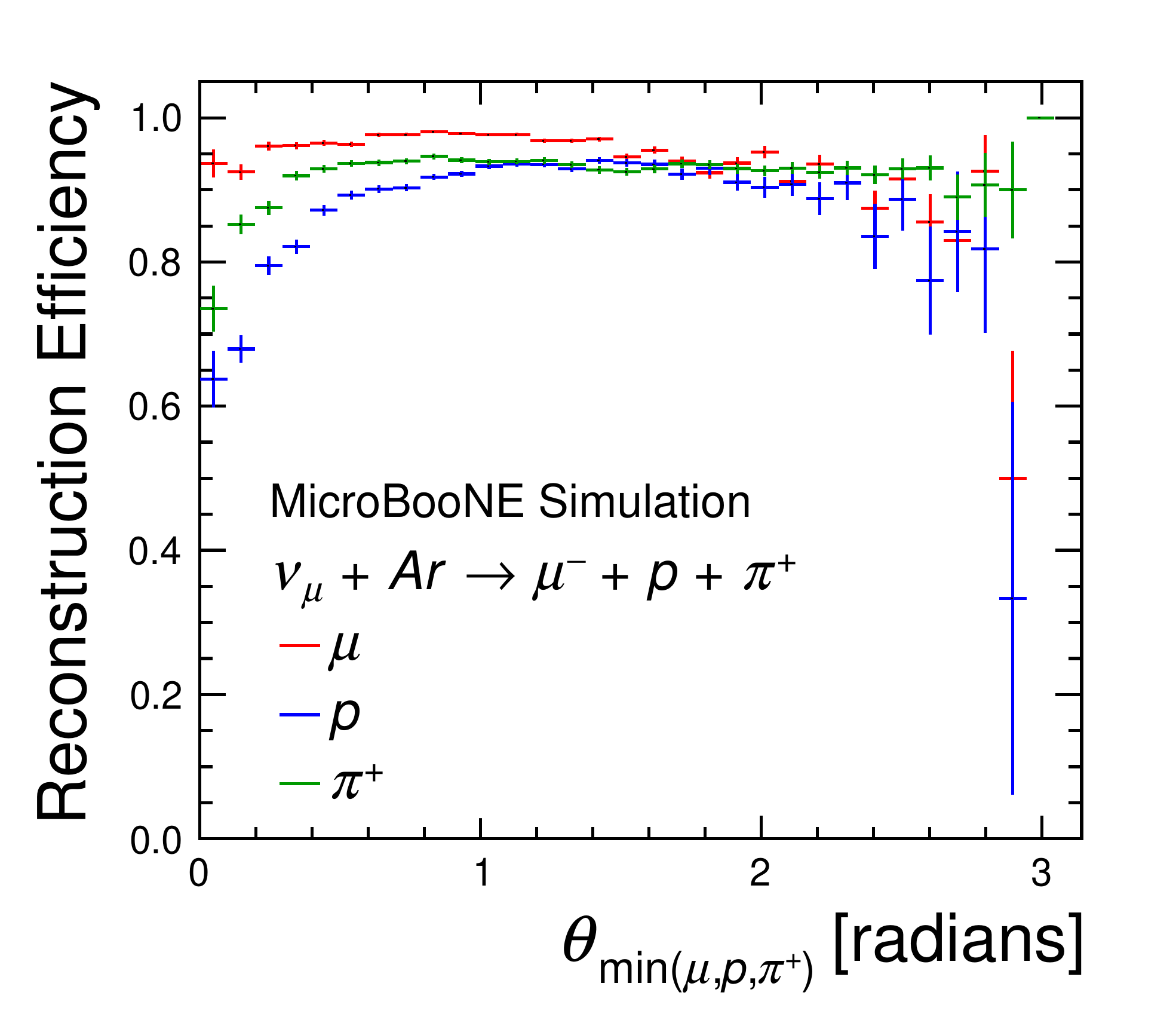}}
     \caption{Reconstruction efficiencies for the target muon, proton and charged pion in simulated BNB CC $\nu_{\mu}$ interactions with resonant pion production, (a) as a function of the numbers of true hits, (b) as a function of true momenta and (c) as a function of the true opening angles to the nearest-neighbour target MCParticle. For instance, for the muon in a given event, this would be the smaller of its true opening angles to the proton and the charged pion.\label{fig::CCRES_MU_P_PIPLUS_Efficiencies}}
  \end{center}
\end{figure}

\begin{figure}[!h]
  \begin{center}
     \subfloat[][]{\includegraphics[width=0.34\textwidth]{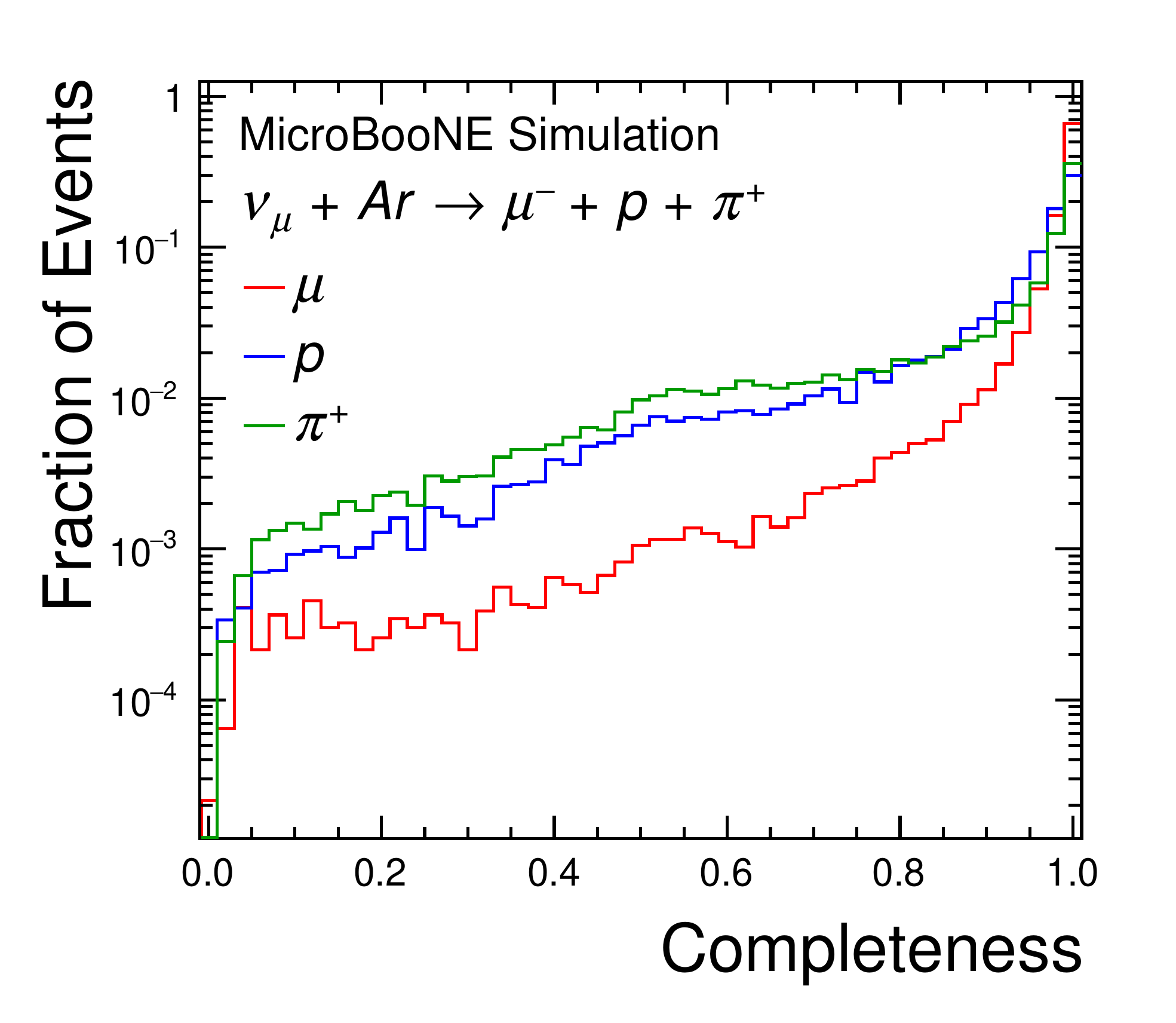}}
     \subfloat[][]{\includegraphics[width=0.34\textwidth]{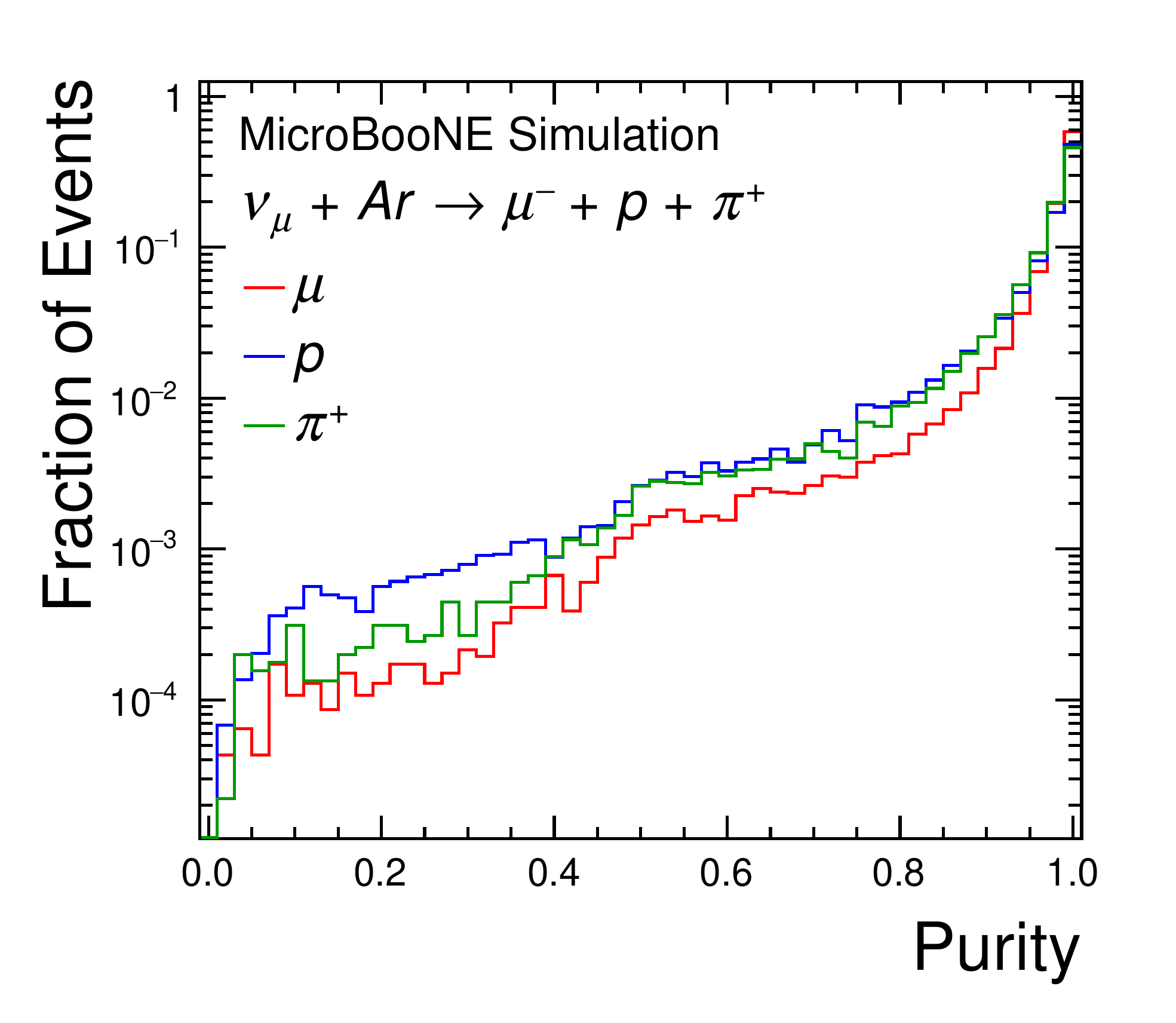}}
     \subfloat[][]{\includegraphics[width=0.34\textwidth]{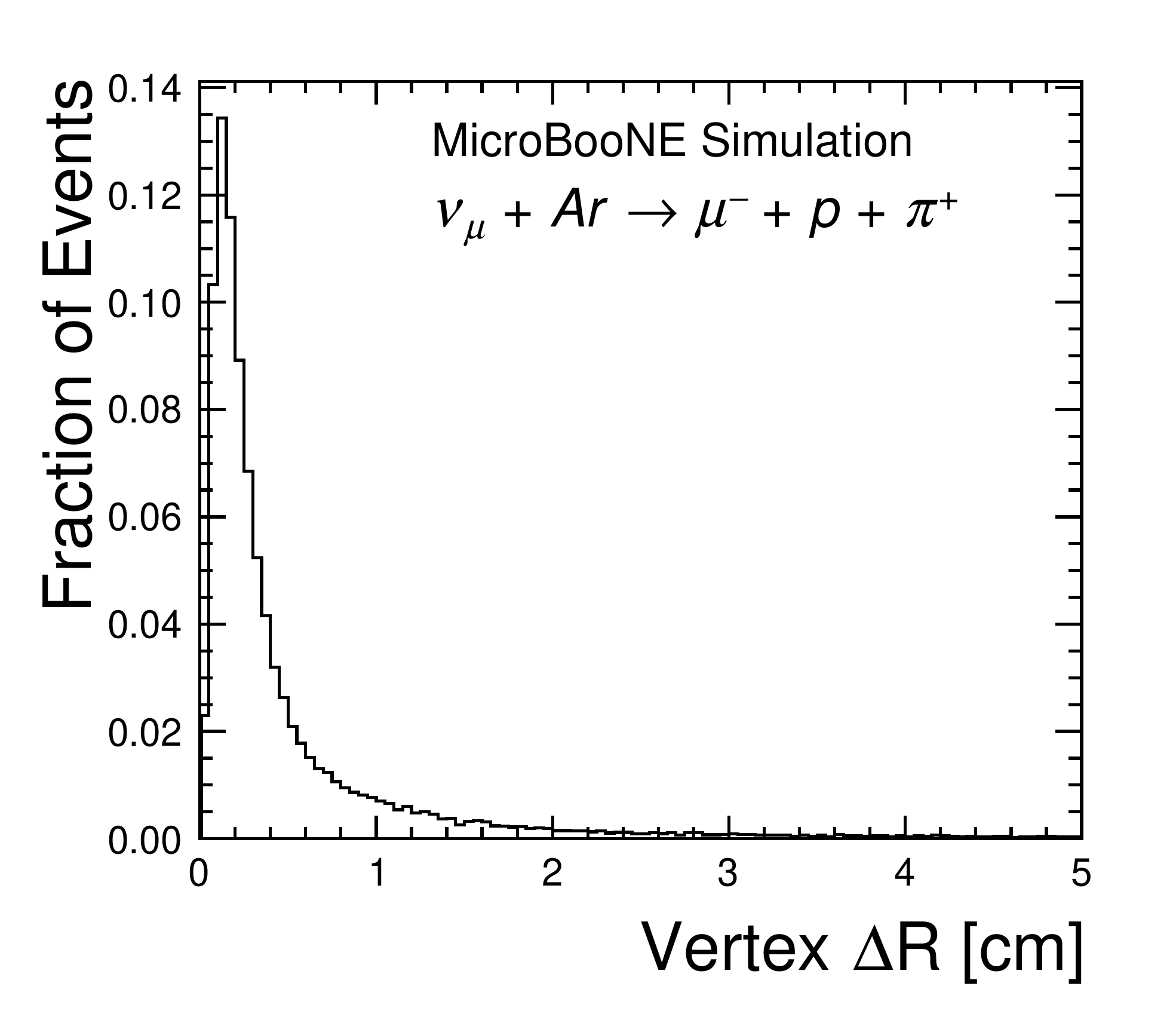}\label{fig::CCRES_MU_P_PIPLUS_Vtx}}
     \caption{Completeness (a) and purity (b) of the reconstructed particles with the strongest matches to the target muon, proton and charged pion in simulated BNB CC $\nu_{\mu}$ interactions with resonant pion production and (c) the distance between generated and reconstructed 3D vertex positions.\label{fig::CCRES_MU_P_PIPLUS_PurCompVtx}}
  \end{center}
\end{figure}

Figure \ref{fig::CCRES_MU_P_PIPLUS_Efficiencies} displays the reconstruction efficiencies for the target muon, proton and pion as a function of the numbers of true hits, true momenta and the true opening angles to their nearest-neighbour target MCParticle. As expected, target MCParticles are most likely to be merged into single reconstructed particles when the targets are collinear. Figure \ref{fig::CCRES_MU_P_PIPLUS_PurCompVtx} shows the completenesses and purities of the reconstructed particles with the strongest matches to the target muon, proton and pion. The reported completeness is lowest for the target pions because of the difficulty inherent in fully reconstructing the hierarchy of daughter particles, even when all the separate particles are reconstructed.

Figure \ref{fig::CCRES_MU_P_PIPLUS_Vtx} shows the displacement of the reconstructed neutrino interaction vertex from the true, generated position. It is found that 68\% of events have a displacement below 0.48\,cm, whilst 7.3\% of events have a displacement above 5\,cm. The vertex reconstruction performance is better than for the quasi-elastic events considered in Section \ref{sec::CCQEL_MU_P}. The presence of the pion track, whilst adding to the complexity of the events, provides additional pointing information indicating the position of the interaction vertex.

\subsection{BNB CC resonance events: $\nu_{\mu} + Ar  \rightarrow  \mu^{-} + p + \pi^{0}$}
\label{sec::CCRES_MU_P_PIZERO}
\begin{figure}[!h]
  \begin{center}
     \includegraphics[width=0.7\textwidth]{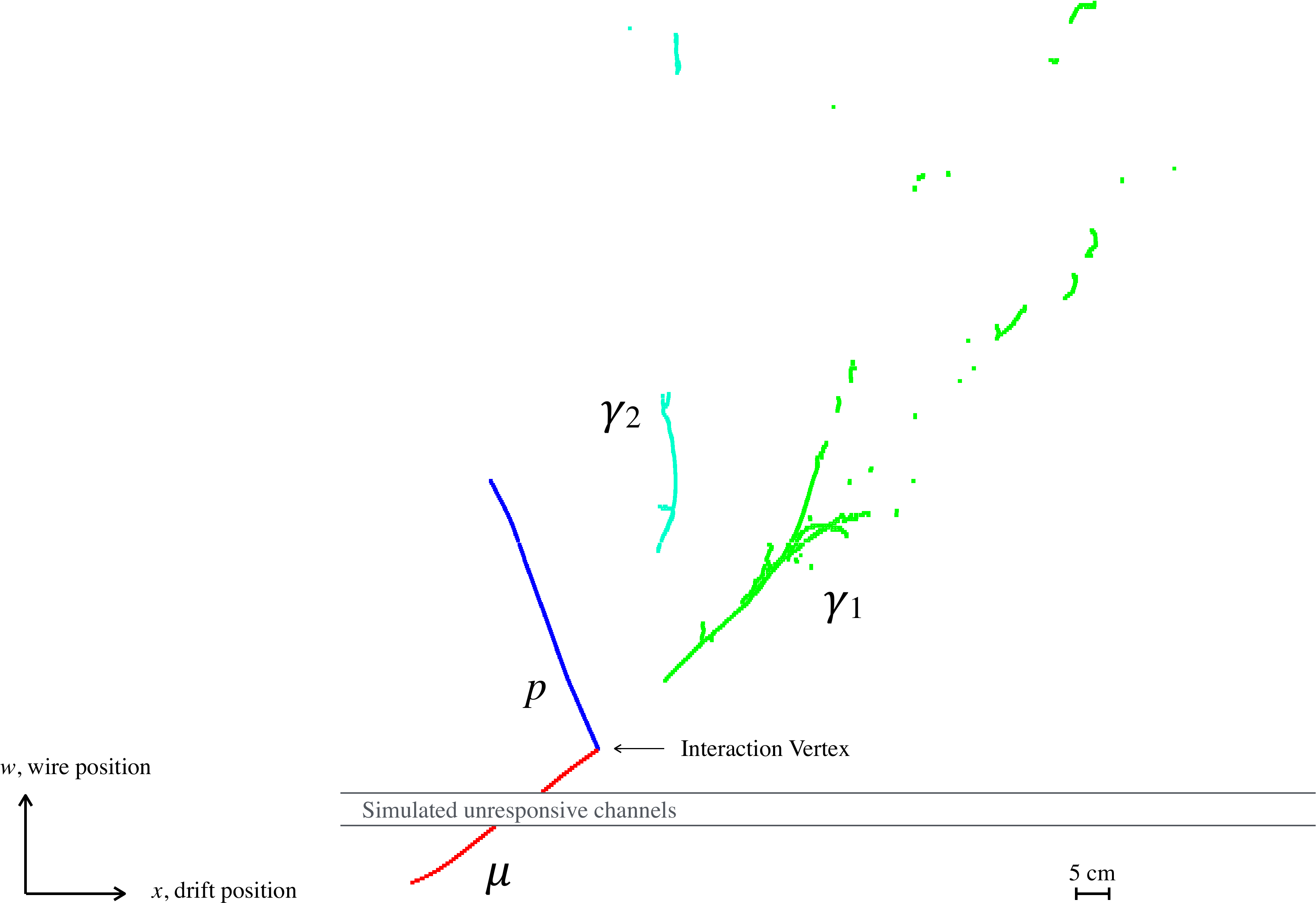}
     \caption{The reconstruction of a simulated 1.4-GeV CC $\nu_{\mu}$ interaction with resonant neutral-pion production. Target particles for the reconstruction are the muon, proton and two photons from $\pi^{0}$ decay. The label $\gamma_{1}$ identifies the target photon with the largest number of true hits, whilst $\gamma_{2}$ identifies the photon with fewer true hits.\label{fig::CCRES_MU_P_PIZERO_Example}}
  \end{center}
\end{figure}

The reconstruction of photons from $\pi^{0}$ decays is challenging, but the ability to distinguish a $\pi^{0}$ from a single electromagnetic shower is of direct relevance to the MicroBooNE physics goals. Here, the quality of reconstruction is benchmarked using simulated BNB CC $\nu_{\mu}$ interactions with resonant neutral-pion production. Events are considered if they produce exactly one reconstructable muon, one reconstructable proton and two reconstructable photons in the visible final state. The true momentum distributions for particles in selected BNB events peak at approximately 300\,MeV for muons and 400\,MeV for protons. The true energy distributions peak at approximately 150\,MeV for the larger photon ($\gamma_{1}$), with most associated hits, and 60\,MeV for the smaller photon ($\gamma_{2}$). An example event topology is shown in Figure \ref{fig::CCRES_MU_P_PIZERO_Example}. The presence of two photon-induced showers presents a different reconstruction challenge, compared to the track-only topologies considered in Sections \ref{sec::CCQEL_MU_P} and \ref{sec::CCRES_MU_P_PIPLUS}. Small opening angles between the two showers can cause them to be merged into a single reconstructed particle, whilst sparse shower topologies can result in single showers being split into multiple reconstructed particles.

Table \ref{tab::CCRES_MU_P_PIZERO} shows that the performance for muons and protons remains similar to that seen in Sections \ref{sec::CCQEL_MU_P} and \ref{sec::CCRES_MU_P_PIPLUS}. Exactly one reconstructed particle is matched to 94.8\% of target muons and to 85.5\% of target protons. The slightly larger fractions of lost muons or protons is associated with a new failure mechanism, whereby the tracks are merged into nearby showers. As anticipated, the diverse and complex shower topologies lead to problems with both merging and splitting of particles. $\gamma_{1}$ is matched to exactly one reconstructed particle in 88.0\% of events. In 6.8\% of events, no particle is matched to $\gamma_{1}$ and this failure is typically associated with small showers being absorbed into a nearby track particle. Sparse shower topologies can mean that $\gamma_{1}$ is reconstructed as multiple, distinct particles. $\gamma_{2}$ is matched to exactly one reconstructed particle in 66.4\% of events. $\gamma_{2}$ can be split into multiple reconstructed particles, but the dominant failure mechanism for this target shower is the lack of any matched reconstructed particle. This can be due to accidental merging into a nearby particle, typically that associated with the larger shower, or due to an inability to reconstruct the small 2D shower clusters or to match these clusters between views.

\begin{table}[!h]
  \begin{center}
    \begin{tabular}{ c | c  c  c  c }
      \toprule
      \#Matched Particles  & 0                  & 1             & 2                  & 3+              \\
      \midrule
      $\mu$                & \,\,\,$3.7\%$      & $94.8\%$      & \,\,\,$1.5\%$      & \,\,\,$0.0\%$   \\
      $p$                  & \,\,\,$9.9\%$      & $85.5\%$      & \,\,\,$4.3\%$      & \,\,\,$0.3\%$   \\
      $\gamma_{1}$         & \,\,\,$6.8\%$      & $88.0\%$      & \,\,\,$4.8\%$      & \,\,\,$0.4\%$   \\
      $\gamma_{2}$         & $29.9\%$           & $66.4\%$      & \,\,\,$3.6\%$      & \,\,\,$0.2\%$   \\
      \bottomrule
    \end{tabular}
  \end{center}
  \caption{Pattern-recognition performance for the target muon, proton and two photons ($\gamma_{1}$ is the photon with the largest number of true hits, $\gamma_{2}$ has fewer true hits) in simulated BNB CC $\nu_{\mu}$ interactions with resonant neutral-pion production. The total number of events was 17,939 and 49.9\% had exactly one reconstructed particle matched to each target.\label{tab::CCRES_MU_P_PIZERO}}
\end{table}

Events with a $\mu + p + \gamma_{1} + \gamma_{2}$ topology, from CC $\nu_{\mu}$ resonance interactions, represent a significant challenge and 49.9\% of events are deemed correct, matching exactly one reconstructed particle to each target MCParticle. To reconstruct these events, there are fundamental tensions in the pattern recognition. Algorithms need to be inclusive to avoid splitting true showers into multiple reconstructed particles, but they also need to avoid merging together hits from separate, nearby target MCParticles. Algorithm thresholds for individual particle creation also need to be sufficiently low to enable efficient reconstruction of small showers, without leading to the creation of excessive numbers of separate fragment particles. Aggressive searches for small particles associated with the reconstructed neutrino interaction vertex can help to address this second source of tension.

\begin{figure}[!h]
  \begin{center}
     \subfloat[][]{\includegraphics[width=0.34\textwidth]{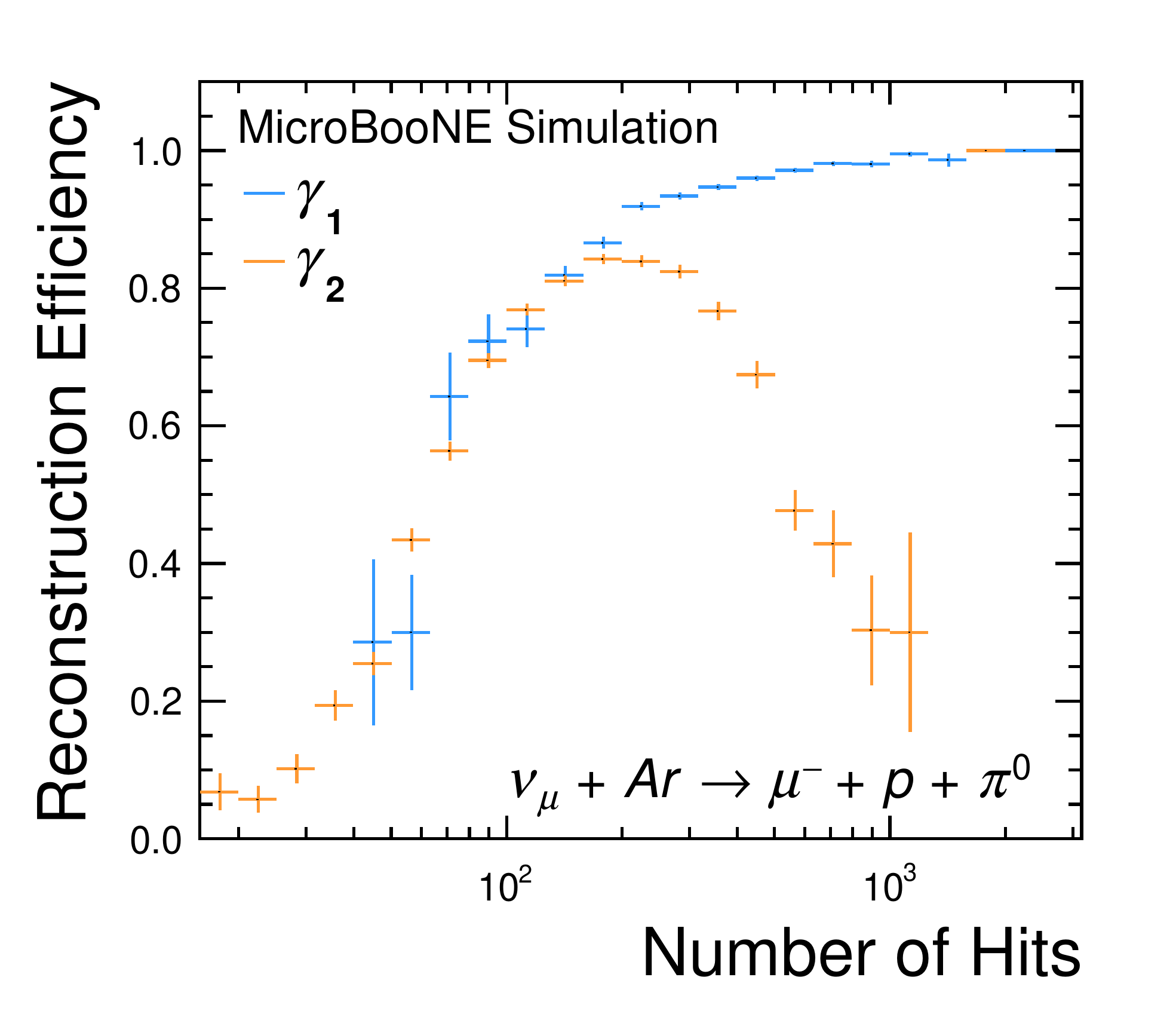}}
     \subfloat[][]{\includegraphics[width=0.34\textwidth]{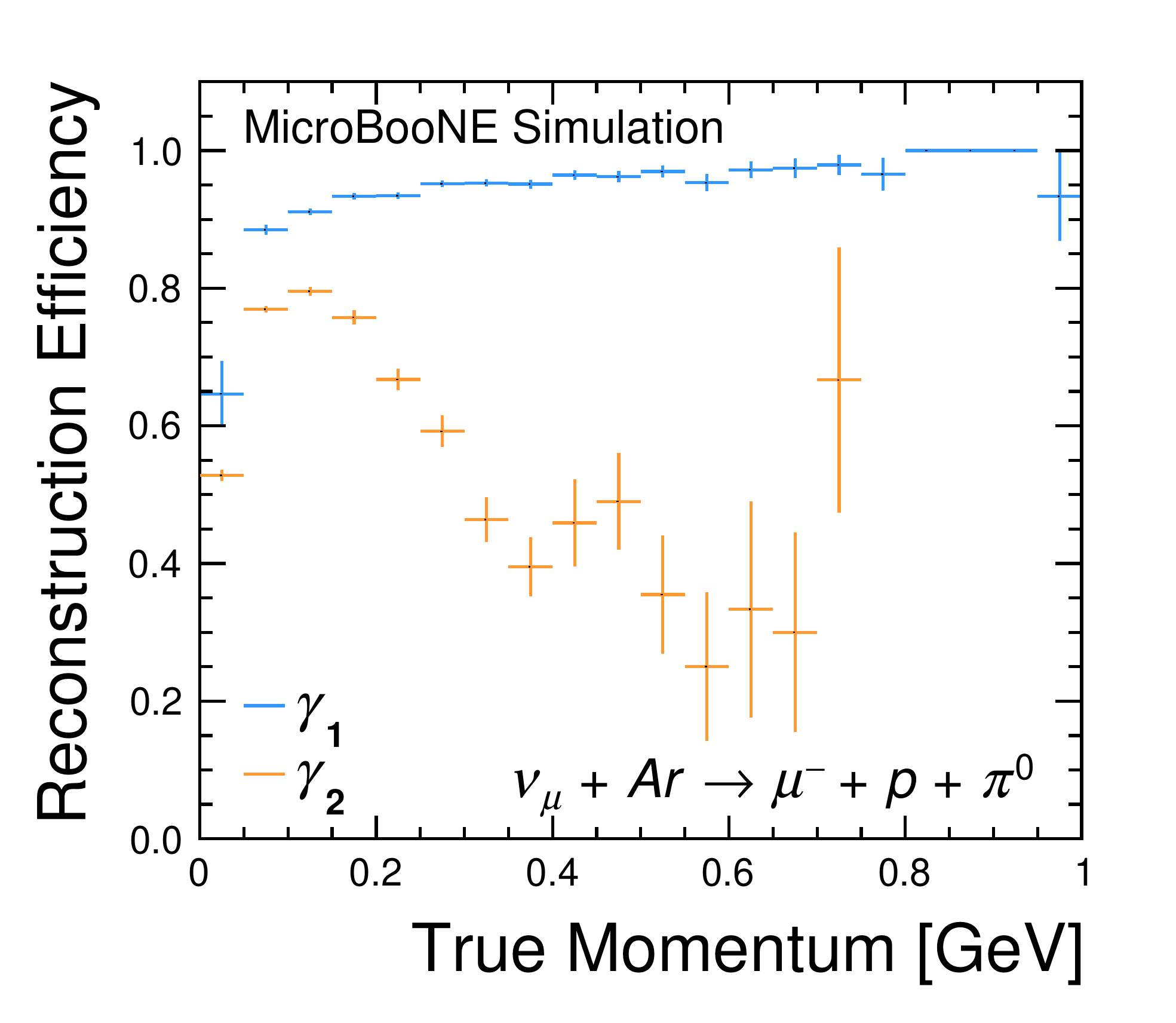}}
     \subfloat[][]{\includegraphics[width=0.34\textwidth]{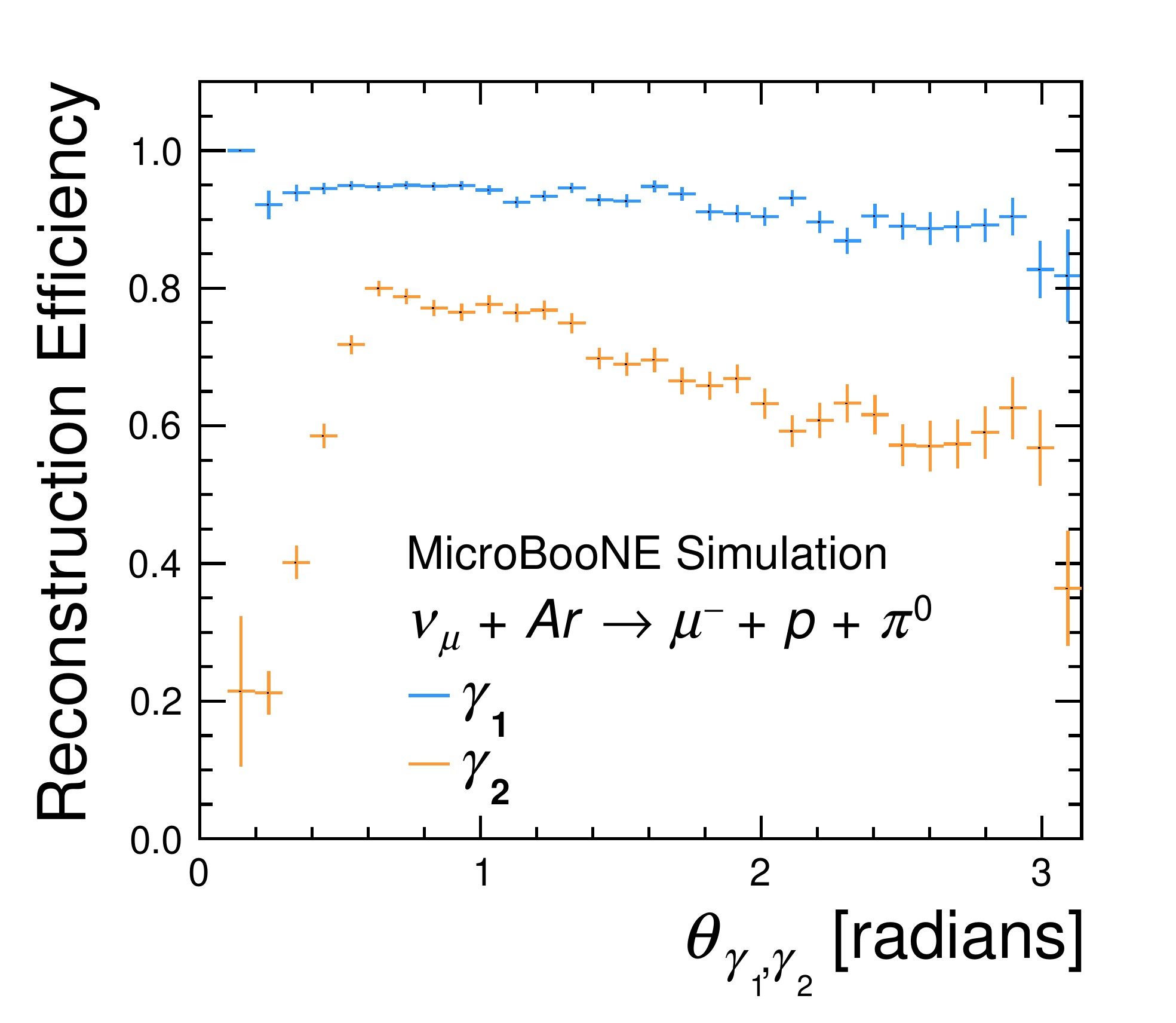}\label{fig::CCRES_MU_P_PIZERO_EffAngle}}
     \caption{Reconstruction efficiencies for the target photons ($\gamma_{1}$ is the photon with the largest number of true hits, $\gamma_{2}$ has fewer true hits) in simulated BNB CC $\nu_{\mu}$ interactions with resonant neutral-pion production, (a) as a function of the numbers of true hits, (b) as a function of true momenta and (c) as a function of the true opening angle.\label{fig::CCRES_MU_P_PIZERO_Efficiencies}}
  \end{center}
\end{figure}

The reconstruction efficiencies, purities and completenesses for the target muon and proton are essentially unchanged from those reported for the event topologies considered in Sections \ref{sec::CCQEL_MU_P} and \ref{sec::CCRES_MU_P_PIPLUS}. Figure \ref{fig::CCRES_MU_P_PIZERO_Efficiencies} therefore concentrates on the pattern-recognition performance for the two showers, showing reconstruction efficiencies as a function of the numbers of true hits, true momenta and the true opening angle between the two photons. The efficiency for $\gamma_{1}$ increases, almost monotonically, with the number of true hits. The efficiency for $\gamma_{2}$ initially displays the same rise with number of true hits, but then falls away as the two showers are more frequently merged into a single reconstructed particle that is associated to the larger target, $\gamma_{1}$.

Figure \ref{fig::CCRES_MU_P_PIZERO_EffAngle} shows that the efficiency for $\gamma_{2}$ is very low when the opening angle between the two photons is small and the two showers are coincident. The efficiency then rises as the opening angle increases and the two showers begin to appear as separate entities, reaching a maximum at a true opening angle of approximately $36\degree$. The efficiency then decreases slowly as the angle increases, before falling steeply as the two showers appear in a back-to-back topology. The efficiency for $\gamma_{2}$ is always lower than that for $\gamma_{1}$, as merged reconstructed particles will typically be associated to $\gamma_{1}$ and as more of the smaller showers do not cross the threshold for creation of a reconstructed particle.

\begin{figure}[!h]
  \begin{center}
     \subfloat[][]{\includegraphics[width=0.34\textwidth]{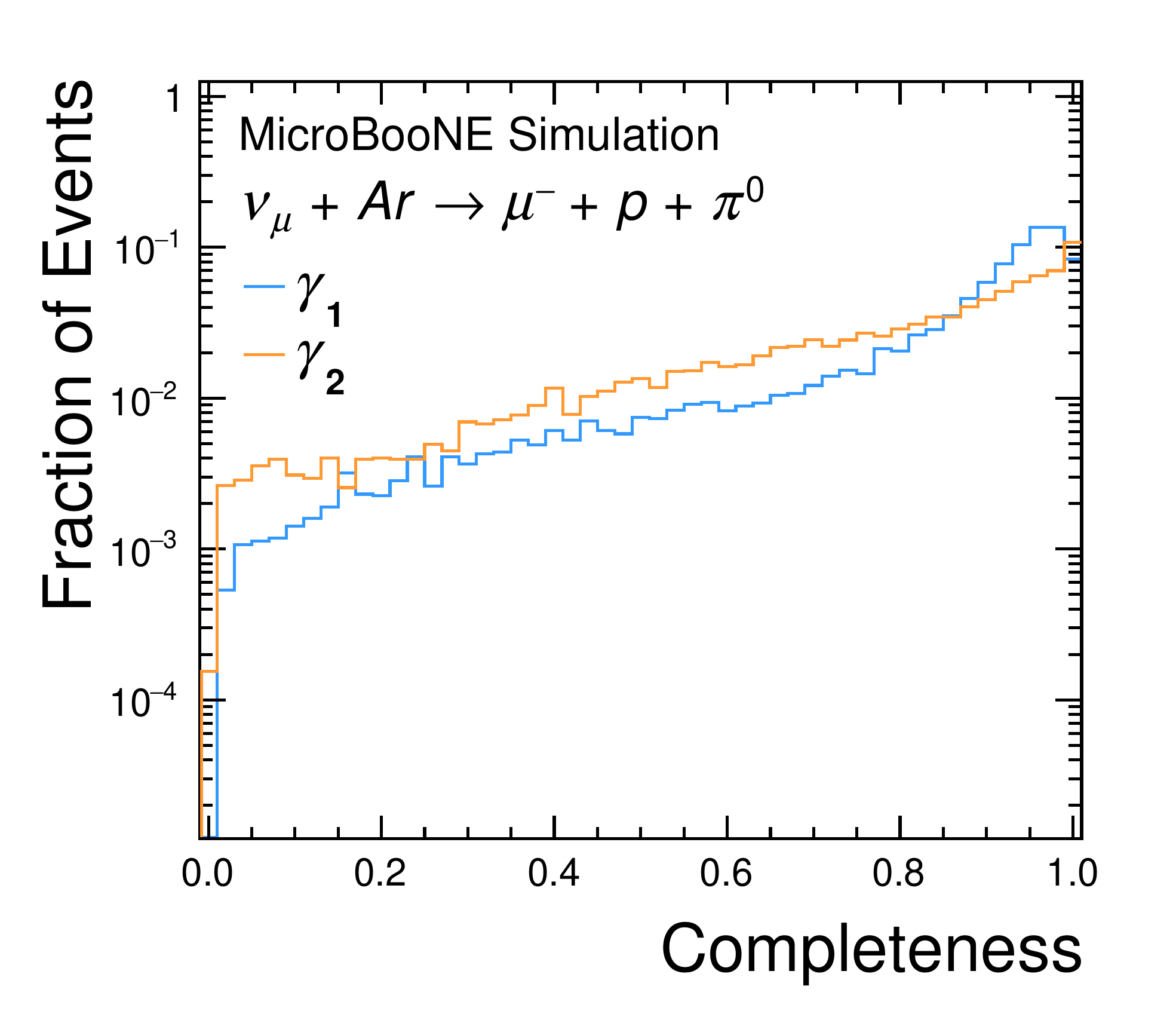}}
     \subfloat[][]{\includegraphics[width=0.34\textwidth]{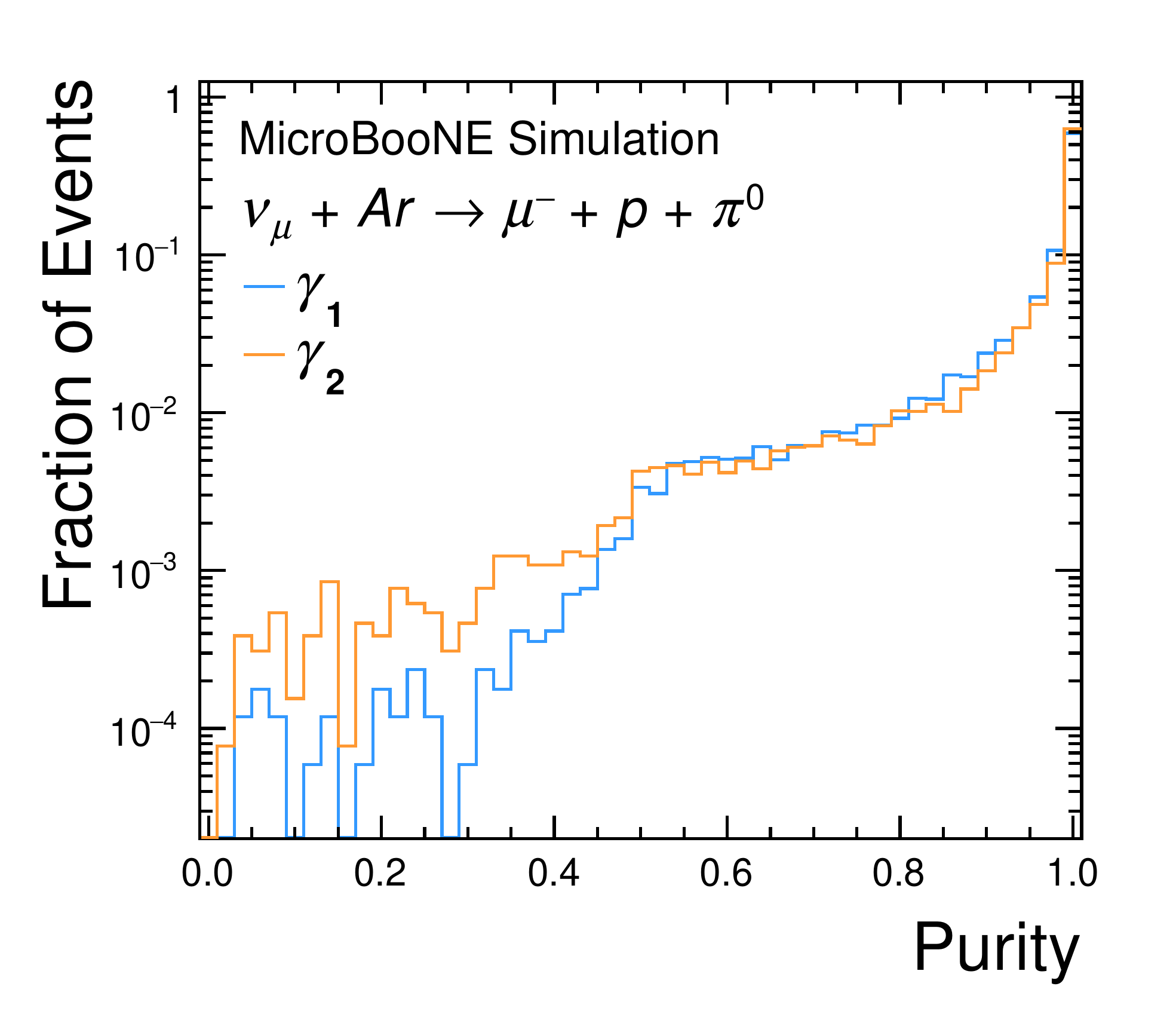}}
     \subfloat[][]{\includegraphics[width=0.34\textwidth]{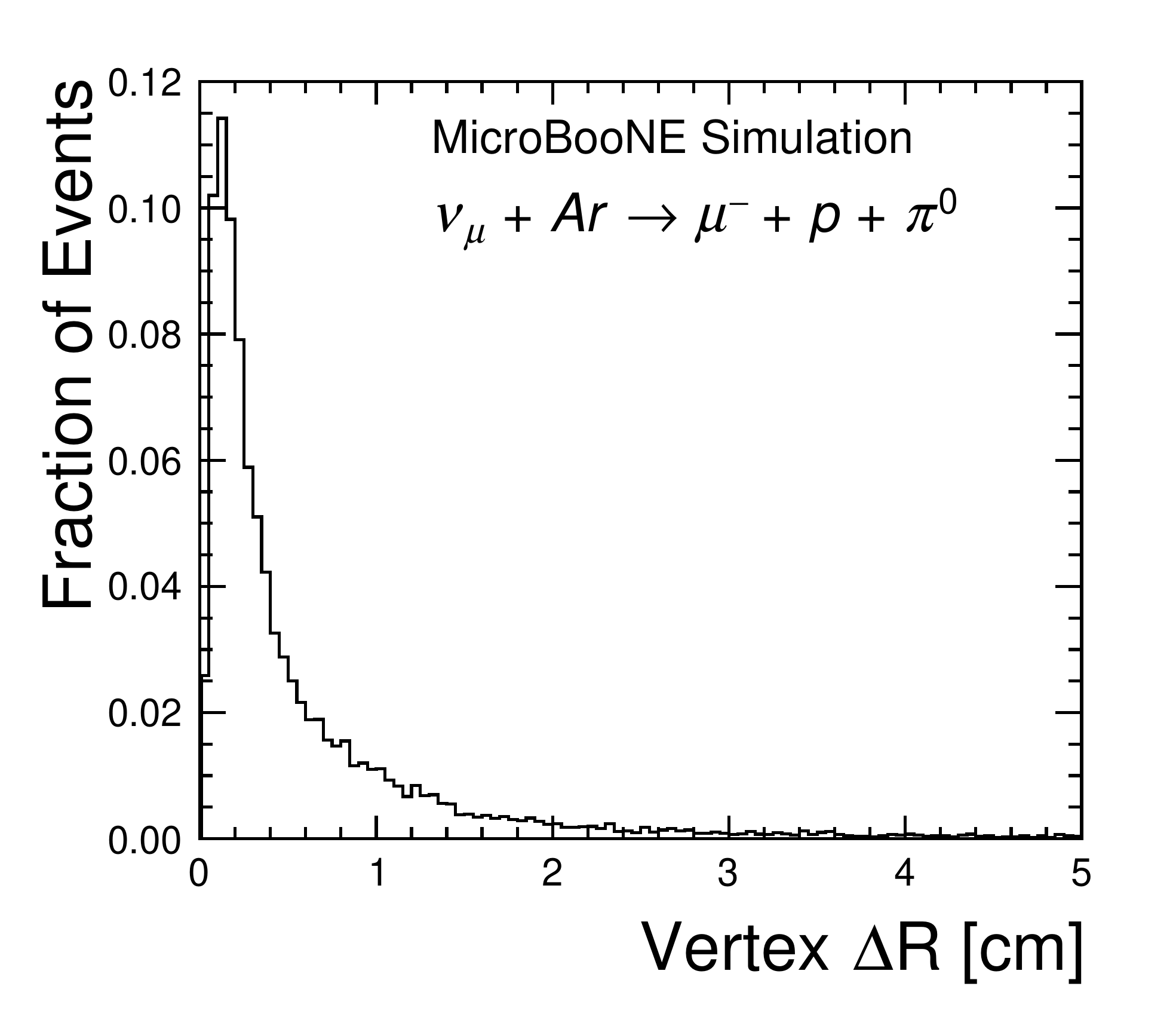}\label{fig::CCRES_MU_P_PIZERO_Vtx}}
     \caption{Completeness (a) and purity (b) of the reconstructed particles with the strongest matches to the target photons ($\gamma_{1}$ is the photon with the largest number of true hits, $\gamma_{2}$ has fewer true hits) in simulated BNB CC $\nu_{\mu}$ interactions with resonant neutral-pion production and (c) the distance between generated and reconstructed 3D vertex positions.\label{fig::CCRES_MU_P_PIZERO_PurCompVtx}}
  \end{center}
\end{figure}

Figure \ref{fig::CCRES_MU_P_PIZERO_PurCompVtx} shows the completenesses and purities of the reconstructed particles with the strongest matches to the two target showers. The completenesses are markedly lower than for target track-like particles in this event topology, and in the event topologies in Sections \ref{sec::CCQEL_MU_P} and \ref{sec::CCRES_MU_P_PIPLUS}. This is associated with the problems of splitting sparse showers into multiple reconstructed particles. The observed purities indicate that mixing of hits between the reconstructed shower particles is rather low.

Figure \ref{fig::CCRES_MU_P_PIZERO_Vtx} shows the displacement of the reconstructed neutrino interaction vertex from the true, generated position. It is found that 68\% of events have a displacement below 0.52\,cm, whilst 4.5\% of events have a displacement above 5\,cm. The distribution is not quite as sharp as for events with target muon, proton and charged pion, but there are fewer failures, with displacements above 5\,cm. This reflects the fact that there is more information available in these events, with a muon, proton and two showers emerging from the interaction position, but that the pointing information available from the two showers is typically not of the same quality as that provided by a charged-pion track.

\subsection{Selection of exclusive final states}
\label{sec::BNB_SUMMARY}

\begin{figure}[!h]
  \begin{center}
     \includegraphics[width=0.95\textwidth]{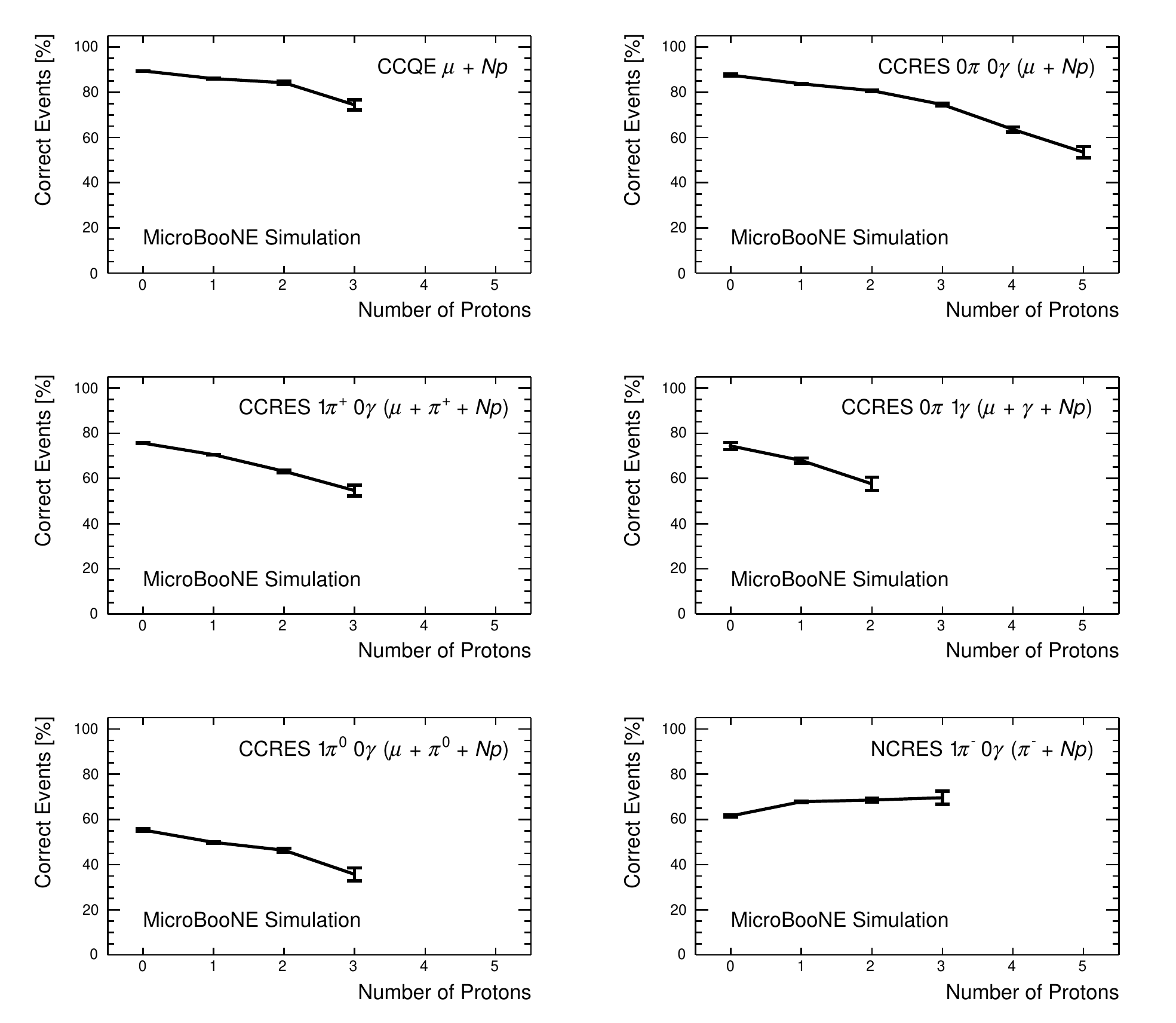}
     \caption{The fraction of events deemed to have correct pattern recognition, shown for a selection of different BNB interactions with exclusive final states. For each interaction type (and combination of final-state leptons, pions or photons), the correct event fraction is displayed as a function of the number of final-state protons. Correct events are those deemed to have exactly one reconstructed particle matched to each target MCParticle.\label{fig::BNB_Summary}}
  \end{center}
\end{figure}

Sections \ref{sec::CCQEL_MU_P}, \ref{sec::CCRES_MU_P_PIPLUS}, and \ref{sec::CCRES_MU_P_PIZERO} focused on three specific event topologies. In general, CC quasi-elastic and CC resonance interactions in argon are more complex and will produce other final states than just muon and single proton, or muon, single pion and single proton, respectively. Here, a somewhat larger selection of exclusive final states is considered for BNB interactions in the MicroBooNE detector. In each case, the pattern-recognition performance is characterised by the fraction of events deemed to be completely correct; i.e. those for which exactly one reconstructed particle is matched to each target MCParticle. This provides a single, highly-sensitive metric to indicate the quality of the pattern recognition. Figure \ref{fig::BNB_Summary} displays the fraction of correct events, for specific interaction types, as a function of the number of target protons in the final state. This includes CC quasi-elastic events with ($\mu + Np$) final states, where $N$ is the number of protons. It also includes CC resonance events with ($\mu + Np$), ($\mu + \pi^{+} + Np$), ($\mu + \gamma + Np$) and ($\mu + \pi^{0} + Np$) final states, and NC resonance events with ($\pi^{-} + Np$) final states. For the CC events, the correct event fraction decreases as the number of protons in the final state increases and the events become more complex. For CC resonance events, the correct event fraction ranges from 87.6\% for the $\mu$ final state, to 74.5\% for the ($\mu + 3p$) final state and 53.4\% for the ($\mu + 5p$) final state. For the NC events, the correct event fraction displays a small rise as a function of the number of protons in the final state. This is because the presence of additional protons aids the reconstruction of the neutrino interaction vertex. Once the vertex position has been determined, the algorithms are more efficient at reconstructing small particles and they are better at avoiding incorrect particle merges in the vertex region, thereby protecting individual target particles as single entities.

\section{Impact of cosmic-ray muon background}
\label{sec::CR}
Section \ref{sec::Performance} considered samples of pure neutrino interactions in the MicroBooNE detector. In practice, MicroBooNE is a surface-based experiment and each neutrino event is overlaid with cosmic-ray muons. In this Section, the simulated BNB neutrino interactions are overlaid with simulated cosmic-ray muon interactions. In the MicroBooNE simulation, there is exactly one neutrino interaction for each 3.2\,ms readout window, and the typical number of cosmic-ray muons (having at least 30 true hits) is $20.6\pm0.2$. The neutrino reconstruction is assessed using the full procedure of running PandoraCosmic, tagging and removing unambiguous cosmic-ray muon candidates, then running PandoraNu on a cosmic-removed hit collection. The cosmic-ray muon tagging takes place in a LArSoft module and is external to the Pandora pattern recognition. Particles are flagged as unambiguous cosmic-ray muons if some of the associated hits are placed outside the detector when the event time is taken to be the neutrino beam trigger time, or if the reconstructed trajectories are through-going, with the exception of particles that pass through both the upstream and downstream faces of the detector.

The presence and removal of cosmic-ray muons can degrade the neutrino reconstruction, due to:
\begin{itemize}
    \item Removal of key features of the neutrino interaction prior to the PandoraNu reconstruction. This could be due to an inability of the PandoraCosmic reconstruction to cleanly separate all neutrino-induced particles from nearby cosmic-ray muons, or due to incorrect tagging of (elements of) the neutrino interaction as a cosmic-ray muon.
    \item Confusion of the PandoraNu pattern recognition by the presence of cosmic-ray muon remnants. It is then the responsibility of the Pandora slicing algorithm to ensure that hits from the neutrino interaction and hits from cosmic-ray muon remnants are assigned to different slices, and so produce separate reconstructed candidate neutrinos.
\end{itemize}

To assess the fraction of neutrino interactions degraded by the cosmic-ray muon removal process, MCParticle information is used to count the number of neutrino-induced hits and to classify the neutrino-induced, reconstructable particles in the visible final state. The results obtained by considering the collection of all hits, which form the input to PandoraCosmic, are then compared to those obtained, for the same events, by considering just the cosmic-removed hits. Events for which 10\% or more of the neutrino-induced hits are removed, or for which the classification of the neutrino-induced final state particles changes, are deemed to have been degraded. Table \ref{tab::CosmicRemovalDamage} shows the fraction of degraded events for a number of different neutrino interactions, with exclusive final states classified using the PandoraCosmic input hits. Between $5\%-18\%$ of events are degraded, with this fraction increasing with the number of final state particles, and increasing markedly with the presence of electromagnetic showers.

\begin{table}[!h]
  \begin{center}
    \begin{tabular}{ c | c  c  c }
      \toprule
      CCQE                        & $\mu$                    & $\mu + 1p$               & $\mu + 2p$                \\
                                  & $\,\,\,(5.9\pm0.2)\%$    & $\,\,\,(7.7\pm0.2)\%$    & $(10.5\pm1.2)\%$          \\ 
      \midrule
      CCRES $0\pi$ $0\gamma$      & $\mu$                    & $\mu + 1p$               & $\mu + 2p$                \\
                                  & $\,\,\,(5.1\pm0.6)\%$    & $\,\,\,(8.2\pm0.4)\%$    & $(10.1\pm0.4)\%$          \\ 
      \midrule
      CCRES $1\pi^{+}$ $0\gamma$  & $\mu + \pi^{+}$          & $\mu + \pi^{+} + 1p$     & $\mu + \pi^{+} + 2p$      \\
                                  & $\,\,\,(8.6\pm0.4)\%$    & $(10.3\pm0.3)\%$         & $(11.1\pm0.9)\%$          \\ 
      \midrule
      CCRES $1\pi^{0}$ $0\gamma$  & $\mu + \pi^{0}$          & $\mu + \pi^{0} + 1p$     & $\mu + \pi^{0} + 2p$      \\
                                  & $(15.0\pm0.5)\%$         & $(15.4\pm0.3)\%$         & $(17.8\pm1.0)\%$          \\ 
      \bottomrule
    \end{tabular}
  \end{center}
  \caption{The fraction of events deemed degraded by the cosmic-ray removal procedure, shown for a selection of BNB interactions with exclusive final states. In order to be deemed degraded, 10\% or more of the neutrino-induced hits must be removed, or sufficient hits removed to cause a change in the classification of the neutrino-induced final state particles.\label{tab::CosmicRemovalDamage}}
\end{table}

Visual scanning of the degraded events, examining the PandoraCosmic reconstruction output, reveals a number of challenging common issues. It is found that there is little mixing between neutrino-induced hits and cosmic-ray muon hits in the reconstruction; particles typically have either a very low or very high purity of neutrino-induced hits. Neutrino-induced muons are typically reconstructed as individual primary particles, which can be tagged as cosmic-ray muons. Protons can be lost if they are reconstructed as candidate delta rays and added as daughters of nearby true cosmic-ray muons, which are subsequently tagged and removed. The sparse showers from $\pi^{0}$ decays can, more frequently, be collected as daughter delta rays and so removed. In the subsequent analysis of pattern-recognition performance, any events deemed degraded are not assessed, as the performance metrics become ill-defined.

\begin{figure}[!h]
  \begin{center}
     \includegraphics[width=0.95\textwidth]{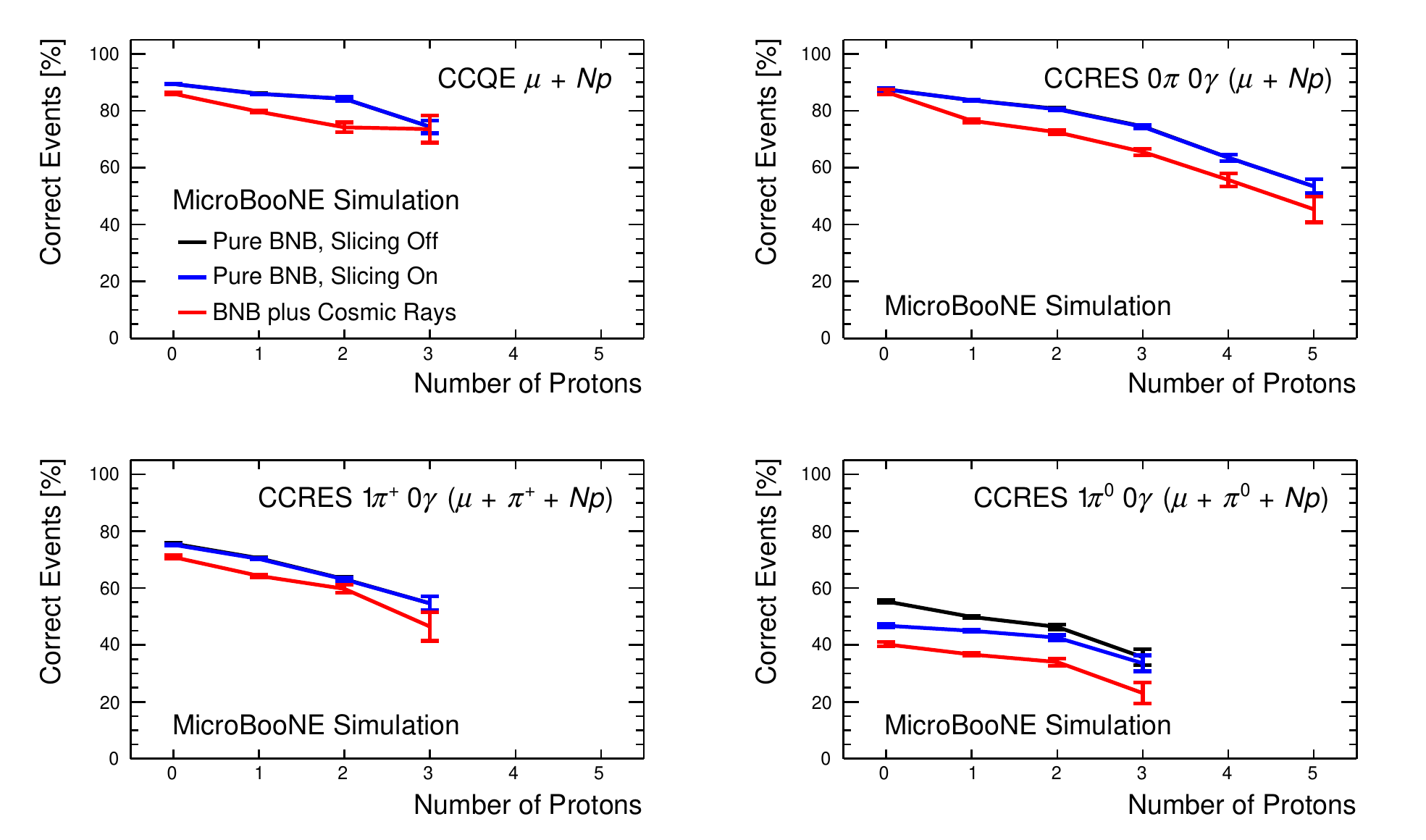}
     \caption{The fraction of events deemed to have correct pattern recognition, shown for a selection of different BNB interactions with exclusive final states. Results are shown separately for samples of pure BNB events, with and without the slicing algorithm enabled, and for BNB events in the presence of cosmic-ray backgrounds. Correct events are those deemed to have exactly one reconstructed particle matched to each target MCParticle. The slicing algorithm has little impact for events with only track-like particles in the final state, so the blue and black lines overlap in three of the plots.\label{fig::BNB_CR_Summary}}
  \end{center}
\end{figure}

Despite the degradation of the underlying neutrino interactions, is is found that the cosmic-ray muon tagging is conservative, with only 76\% of cosmic-ray muons (having at least 30 true hits) being tagged. A significant number of cosmic-ray muon remnants therefore enter the PandoraNu reconstruction and pose a challenge to the pattern recognition. Figure \ref{fig::BNB_CR_Summary} shows the fraction of BNB interactions deemed to have correct pattern recognition, for a number of different configurations: the results from Section \ref{sec::BNB_SUMMARY} are compared to those obtained for the very same interactions when the slicing algorithm (designed to address cosmic-ray muon remnants) is enabled; the results for corresponding interactions in the presence of cosmic-ray muon backgrounds are then shown (the slicing algorithm must, of course, be enabled for this configuration).

Results show that the slicing algorithm does not degrade the reconstruction of events with only track-like particles in the final state. For events with two showers in the final state, however, the slicing reduces the fraction of events deemed correct. This is because sparse shower topologies mean that the slicing can struggle to declare all the input hits as originating from a single interaction. Some shower elements are then placed in additional slices, which are reconstructed in isolation and produce separate reconstructed neutrino candidates. The single reconstructed neutrino particle containing the most neutrino-induced hits is assessed, so fragments of showers, sometime even entire showers, will be missing. The degradation in the correct event fraction is 5.8\%, averaging over all the ($\mu + \pi^{0} + Np$) final states.

The presence of cosmic-ray muon remnants causes additional degradation for all event types investigated. Averaging over all final states, and comparing the results to those in Section \ref{sec::BNB_SUMMARY}, the total degradation is 5.1\% for CC quasi-elastic events with ($\mu + Np$) final states. For CC resonance events, the total degradation is 7.2\% for ($\mu + Np$) final states, 5.5\% for ($\mu + \pi^{+} + Np$) final states and 13.7\% for ($\mu + \pi^{0} + Np$) final states. The degradation caused by removing the cosmic-ray muons should also be recalled when interpreting these figures. The challenge posed by the cosmic-ray muon background is substantial, but, with all effects considered, the pattern recognition is still functional. For neutrino interactions with purely track-like final state particles, the pattern recognition is typically deemed correct for 70\% of events. For interactions with two sparse showers, this fraction is typically 35\%.


\section{Concluding comments}
We have developed an innovative approach to pattern recognition in LArTPC detectors. The Pandora multi-algorithm approach uses large numbers of decoupled algorithms to gradually build up a picture of events and provide a fully-automated reconstruction of cosmic-ray muon and neutrino interactions. The pattern-recognition performance has been evaluated in detail for simulated events in the MicroBooNE detector. Strict metrics have been used to demonstrate the ability to match reconstructed particles to each true, generated particle in the visible final state. MicroBooNE is a surface-based experiment and substantial challenges are posed to the pattern recognition by the presence of the cosmic-ray muon background. The results presented in this paper provide a snapshot of the current performance. Significant improvements are expected, via the addition of new algorithms and refinements to the cosmic-ray muon removal.

\section*{Acknowledgements}
This material is based upon work supported by the following: the U.S. Department of Energy, Office of Science, Offices of High Energy Physics and Nuclear Physics; the U.S. National Science Foundation; the Swiss National Science Foundation; the Science and Technology Facilities Council of the United Kingdom; The Royal Society (United Kingdom); and the European Union's Horizon 2020 Research and Innovation programme under Grant Agreement No. 654168. Additional support for the laser calibration system and cosmic ray tagger was provided by the Albert Einstein Center for Fundamental Physics. Fermilab is operated by Fermi Research Alliance, LLC under Contract No. DE-AC02-07CH11359 with the United States Department of Energy.


\begin{thebibliography}{}

\bibitem{bib::Detector}
  R.~Acciarri {\it et al.} [MicroBooNE Collaboration],
  JINST {\bf 12}, no. 02, P02017 (2017)

\bibitem{bib::MiniBooNE} 
  A.~A.~Aguilar-Arevalo {\it et al.} [MiniBooNE Collaboration],
  Phys.\ Rev.\ Lett.\  {\bf 110}, 161801 (2013)

\bibitem{bib::PandoraSDK}
  J.~S.~Marshall and M.~A.~Thomson,
  Eur.\ Phys.\ J.\ C {\bf 75}, no. 9, 439 (2015)

\bibitem{bib::ILC}
  M.~A.~Thomson,
  Nucl.\ Instrum.\ Meth.\ A {\bf 611}, 25 (2009)

\bibitem{bib::CLIC}
  J.~S.~Marshall, A.~M\"{u}nnich and M.~A.~Thomson,
  Nucl.\ Instrum.\ Meth.\ A {\bf 700}, 153 (2013)

\bibitem{bib::Noise}
  R.~Acciarri {\it et al.} [MicroBooNE Collaboration],
  arXiv:1705.07341 [physics.ins-det].

\bibitem{bib::LArSoft}
  LArSoft Software: https://cdcvs.fnal.gov/redmine/projects/larsoft

\bibitem{bib::GENIE}
  C.~Andreopoulos {\it et al.},
  Nucl.\ Instrum.\ Meth.\ A {\bf 614}, 87 (2010)

\bibitem{bib::CORSIKA} 
  D.~Heck, G.~Schatz, T.~Thouw, J.~Knapp and J.~N.~Capdevielle,
  FZKA-6019.

\bibitem{bib::uboonecode}
    MicroBooNE Software: https://cdcvs.fnal.gov/redmine/projects/uboonecode

\end{thebibliography}
\end{document}